%% file: main.tex
\documentclass[amsmath,amssymb,floatfix,a4paper,h-physrev.bst,nofootinbib]{revtex4}%
\usepackage[utf8]{inputenc}
\usepackage{color}
\usepackage{epsf, array, color}
\usepackage{graphicx}
\usepackage{hyperref}
\usepackage{multirow}
\usepackage{xspace}
\usepackage{adjustbox}

\usepackage{lineno}
\begin{document}


\title{Report of the Topical Group on Higgs Physics for the Energy Frontier: The Case for Precision Higgs Physics}
\author{Sally Dawson$^a$, Patrick Meade$^b$, Isobel Ojalvo$^c$ and Caterina Vernieri$^d$}
\affiliation{
\vspace*{.5cm}
  \mbox{$^a$Department of Physics,\\
  Brookhaven National Laboratory, Upton, N.Y., 11973,  U.S.A.}\\
 \mbox{$^b$ C. N. Yang Institute for Theoretical Physics,}\\ 
 \mbox{Stony Brook University, Stony Brook,
 N.Y., 11794, U.S. A.}
 \mbox{$^c$Department of Physics,\\
 Princeton University, Princeton, NJ, 08554, U.S.A.}\\
 \mbox{$^d$SLAC National Accelerator Laboratory, Stanford, CA., 94309, U.S.A.}
 \vspace*{.5cm}}

\newcommand{\defaulteditor}{Editor}
\newcommand{\editnotecolor}{red}
\newcommand{\editnote}[2][\defaulteditor]{\noindent\textcolor{\editnotecolor}{\textit{[{\textbf{{#1}}}{{#2}}]}}}
\newcommand*{\klambda}{\ensuremath{\kappa_{\lambda}}\xspace}
\newcommand*{\mhh}{\ensuremath{m_{hh}}\xspace}
\newcommand*{\hh}{\ensuremath{hh}\xspace}
\newcommand{\IO}[1]{\editnote[IO]{#1}}
\newcommand{\SD}[1]{\editnote[SD]{#1}}
\newcommand{\PM}[1]{\editnote[PM]{#1}}
\newcommand{\CV}[1]{\editnote[CV]{#1}}
\mathchardef\mhyphen="2D
\def\hsm{h}
\def\ee{e^+e^-}
\def\mh{m_h}
\def\fbi{\mathrm{fb}^{-1}}
\def\abi{\mathrm{ab}^{-1}}
\def\CCC{C$^{3}$~}
\def\ee{e^+e^-}
\def\less{$<$}
\def\ye{Y_e}
\def\checked{$checkmark$}
\newcommand*{\hbb}{\ensuremath{h\rightarrow{b}\bar{b}}\xspace}
\newcommand*{\hcc}{\ensuremath{h\rightarrow{c}\bar{c}}\xspace}
\newcommand*{\bb}{\ensuremath{{b}\bar{b}}\xspace}
\newcommand*{\hyy}{\ensuremath{h\rightarrow\gamma\gamma}\xspace}
\newcommand*{\htt}{\ensuremath{h \rightarrow \tau^+ \tau^-}\xspace}
\newcommand*{\hww}{\ensuremath{h \rightarrow WW^*}\xspace}
\newcommand*{\hZZ}{\ensuremath{h \rightarrow ZZ^*}}
\newcommand*{\ttH}{\ensuremath{t\bar{t} h}}
\newcommand*{\hhbbbb}{\ensuremath{\hh\rightarrow b\bar{b}b\bar{b}}\xspace}
\newcommand*{\hhbbyy}{\ensuremath{\hh\rightarrow b\bar{b}\gamma\gamma}\xspace}
\newcommand*{\hhbbtt}{\ensuremath{\hh\rightarrow b\bar{b}\tau^{+}\tau^{-}}\xspace}
\newcommand*{\hhbbww}{\ensuremath{\hh\rightarrow b\bar{b}WW^{*}}\xspace}
\newcommand*{\hhbbvv}{\ensuremath{\hh\rightarrow b\bar{b}VV^*}\xspace}
\newcommand*{\hhbbVV}{\ensuremath{\hh\rightarrow b\bar{b}VV^*}\xspace}
\newcommand*{\hhbbWW}{\ensuremath{\hh\rightarrow b\bar{b}WW}\xspace}
\newcommand*{\hhbbZZ}{\ensuremath{\hh\rightarrow b\bar{b}ZZ(4\ell)}\xspace}
\newcommand*{\bbbb}{\ensuremath{b\bar{b}b\bar{b}}\xspace}
\newcommand*{\bbtautau}{\ensuremath{b\bar{b}\tau^+\tau^-}\xspace}
\newcommand*{\bbtt}{\ensuremath{b\bar{b}\tau^+\tau^-}\xspace}
\newcommand*{\bbgg}{\ensuremath{b\bar{b}\gamma\gamma}\xspace}
\newcommand*{\bbyy}{\ensuremath{b\bar{b}\gamma\gamma}\xspace}
\newcommand*{\bbzz}{\ensuremath{b\bar{b}ZZ^{*}}\xspace}
\newcommand*{\bbww}{\ensuremath{b\bar{b}WW^{*}}\xspace}
\newcommand*{\bbvv}{\ensuremath{b\bar{b}VV^*}\xspace}
\newcommand*{\hhwwyy}{\ensuremath{\hh\rightarrow WW^*\gamma\gamma}\xspace}
\newcommand*{\hhwwww}{\ensuremath{\hh\rightarrow WW^{*}WW^{*}}\xspace}
\newcommand*{\wwyy}{\ensuremath{WW^*\gamma\gamma}\xspace}
\newcommand*{\yyww}{\ensuremath{\gamma\gamma WW^*}\xspace}
\newcommand*{\wwww}{\ensuremath{WW^{*}WW^*}\xspace}

\maketitle


Contributers: S. Adhikari, F.  Abu-Ajamieh, A. Albert,
 H. Bahl, R. Barman, M.Basso,
 A. Beniwal,
 I. Bozovi-Jelisav,
 S. Bright-Thonney,
V. Cairo,
F. Celiberto
S. Chang, M. Chen, 
C. Damerell,
J. Davis, J. de Blas,
W. Dekens,  J. Duarte,
D. Ega˜na-Ugrinovic,
U. Einhaus,
Y. Gao,
 D. Gon\c{c}alves,
 A. Gritsan, H. Haber, 
U. Heintz,
S. Homiller,
S.-C. Hsu, 
D. Jean, 
S. Kawada,
E. Khoda, 
K. Kong,
N. Konstantinidis, 
A. Korytov,
S. Kyriacou, S. Lane, I.M. Lewis, 
K. Li, 
S. Li,
Z. Liu,
J.  Luo,
L. Mandacar-Guerra,
C. Mantel,
J.  Monroy,
M.  Narain,
R. Orr,
 R. Pan,
 A. Papaefstathiou, 
 M. Peskin, 
 M. T. Prim,
 F. Rajec,
 M. Ramsey-Musolf,
J. Reichert,
L. Reina
T. Robens, 
J. Roskes,
A. Ryd,
A. Schwartzman,
P. Scott
J. Strube,
D. Su,
W. Su,
M. Sullivan, 
T. Tanabe,
J. Tian,
A. Tricoli, 
E. Usai,
J. Va’vra,
Z. Wang,
G. White,
M. White,
A. G. Williams,
A. Woodcock,
Y. Wu, 
C. Young,
Y. Zhang, 
X. Zhu, 
R. Zou

\tableofcontents

\section{Abstract}

A future Higgs Factory will provide improved precision  on measurements of Higgs couplings beyond those obtained by the LHC,  and will enable a broad range of investigations across the fields of fundamental physics, including the mechanism of electroweak symmetry breaking, the origin of the masses and mixing of fundamental particles, the predominance of matter over antimatter, and the nature of dark matter.
Future colliders will measure Higgs couplings to a few per cent,
giving a window to  beyond the Standard Model (BSM)  physics in the 1-10 TeV range. In addition, they will make precise measurements of the Higgs width, and characterize the Higgs self-coupling.

\section{Why the Higgs is the Most Important Particle}
\input {Tex/intro.tex}

\section{Higgs Status}
\label{sec:Higgs}
\subsection{Experimental Status of SM Higgs}

\input{Tex/now.tex}

\subsection{Current status of theoretical Precision} 
\input{Tex/theory_now.tex}

\label{sec:theorynow}
\subsection{Multi Higgs production and Self Interactions}
\input{Tex/selfcoupling.tex}

\section{The future...}
\label{sec:future}
\subsection{Production Mechanisms at Future Colliders}
\input{Tex/lept_sigs.tex}
\subsection{Future mass and width measurements}~\label{sec:width}
\input{Tex/mass.tex}

\subsection{Couplings to Standard Model Particles} 
\input{Tex/coups.tex}

\subsection{Beyond the $\kappa$ framework}
\input{Tex/sme.tex}

\subsection{CP violating Higgs coupling measurements}

\input{Tex/cp.tex}

\subsection{Prospects for observing Double Higgs production and measuring Higgs self-couplings}

\input {Tex/hh_future.tex}
\newpage

\section{Learning about BSM Physics through Higgs measurements}~\label{sec:BSM}
\input{Tex/bsmintro.tex}

\subsection{Additional Higgs Singlets}~\label{sec:singlets}
\input{Tex/singlet.tex}

\subsection{Two Higgs Doublet Models}~\label{sec:doublets}
\input{Tex/2hdm.tex}

\subsubsection{Higgs and Flavor}\label{sec:flavor}
\input{Tex/flavor.tex}

\subsection{BSM in Higgs loops}~\label{sec:bsmloops}
\input{Tex/bsmloops.tex}

\subsection{Higgs Exotic Decays}~\label{sec:exoticdecays}
\input{Tex/exotic.tex}

\section{Detector/accelerator requirements to observe new physics}
\label{sec:Detector}
\input{Tex/detector.tex}

\section{Executive Summary}~\label{sec:conclusions}
\input{Tex/conclusions.tex}

\section*{Acknowledgements}
S.D.   is  supported by the U.S. Department of Energy under Grant Contract  de-sc0012704, P. M. is supported in part by the National Science Foundation grants  PHY-1915093 and PHY-2210533,
 I. O. is Supported by Department of Energy DE-SC0007968
and C. V.is supported  by Department of Energy Contract DE-AC02-76SF00515

\bibliographystyle{utphys}
\bibliography{refsef12}
\end{document}

%% file: Tex/intro.tex
Over the past decade, the LHC has fundamentally changed the landscape of high energy particle physics through the discovery of the Higgs boson and the first measurements of many of its properties.  As a result of this, and no discovery of new particles or new interactions at the LHC, the questions surrounding the Higgs have only become sharper and more pressing for planning the future of particle physics.  

The Standard Model (SM) is an extremely successful description of nature, with a basic structure dictated by symmetry.  However, symmetry alone is not sufficient to fully describe the microscopic world we explore: even after specifying the gauge and space-time symmetries, and number of generations, there are 19 parameters undetermined by the SM (not including neutrino masses).  Out of these parameters 4 are intrinsic to the gauge theory description, the gauge couplings and the QCD theta angle.  The other 15 parameters are intrinsic to the coupling of SM particles to the Higgs sector,
illustrating its paramount importance in the SM.  In particular, the masses of all fundamental  particles, their mixing, CP violation, and the basic vacuum structure are all undetermined and derived from experimental data. As simply a test of the validity of the SM, all these couplings must be measured experimentally. However, the centrality of the Higgs boson goes far beyond just dictating the parameters of the SM. 

The Higgs boson is connected to some of our most fundamental questions about the Universe.  Its most basic role in the SM is to provide a source of Electroweak Symmetry Breaking (EWSB). While the Higgs can describe EWSB, it is merely put in by hand in the Higgs potential.  Explaining {\it{why}} EWSB occurs is outside the realm of the Higgs boson, and yet at the same time by studying it we may finally understand its origin.  There are a variety of connected questions and observables tied to the origin of EWSB for the Higgs boson.  For example, is the Higgs mechanism actually due to dynamical symmetry breaking as observed elsewhere in nature?  Is the Higgs boson itself a fundamental particle or a composite of some other strongly coupled sector?  The answers to these questions have a number of ramifications beyond the origin of EWSB.  

\begin{figure}
\begin{centering}
\includegraphics[scale=0.3]{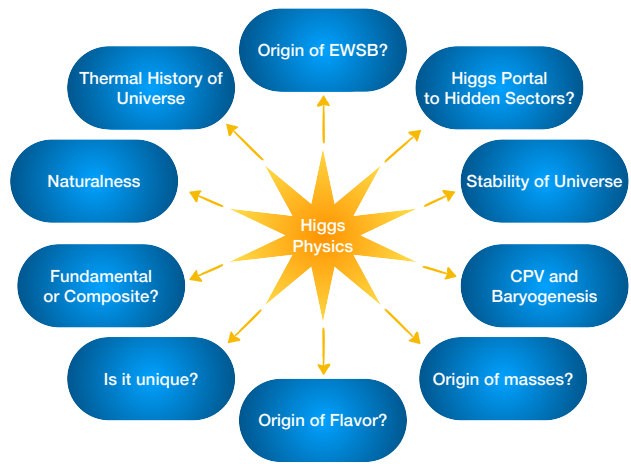} 
\par\end{centering}
\caption{The Higgs boson as the keystone of the Standard Model is connected to numerous fundamental questions that can be investigated by studying it in detail.}
\label{fig:higgscentral}
\end{figure}

If the Higgs boson is a fundamental particle, it represents the first fundamental scalar particle discovered in nature. This has profound consequences both theoretically and experimentally.  From our modern understanding of quantum field theory viewed through the lens of Wilsonian renormalization, fundamental scalars should not exist in the low energy spectrum without an ultra-violet  (UV)  sensitive fine tuning. This is known as the naturalness or hierarchy problem. From studying properties of the Higgs boson, one can hope to learn whether there is some larger symmetry principle at work such as supersymmetry, neutral naturalness, or if the correct theory is a composite Higgs model where the Higgs is a pseudo-Goldstone boson.

Experimentally, there are also a number of intriguing directions that open up if the Higgs boson is a fundamental particle.  The most straightforward question is whether the Higgs boson is unique as the only scalar field in our universe or is it just the first of many? From a field theoretic point of view, one can construct the lowest dimension gauge and Lorentz invariant operator in the SM from the Higgs boson alone.  This means that generically if there are other ``Hidden" sectors beyond the SM, at low energies the couplings of the Hidden sector particles to the Higgs boson are predicted to be the leading portal to the additional sectors.  Additionally, with a scalar particle the question remains as to whether the minimal Higgs potential is correct.  The form of the potential has repercussions for both our understanding of the early universe and its ultimate fate.  For the early universe, the SM predicts that the electroweak symmetry should be restored at high temperatures. However, depending on the actual form of the potential the question remains as to whether there even was a phase transition, let alone its strength.  Additionally, depending on the shape of the Higgs potential, it controls the future of our universe as our vacuum may only be metastable.

Finally, the Higgs boson is connected to some of the most puzzling questions in the universe: flavor, mass and CP violation.  While it is often stated that the Higgs boson gives mass to all elementary fermions, this is just the tip of the proverbial iceberg.  The Yukawa couplings determine not only the masses, but also the CKM matrix and its CP violating phase.  Thus, the Higgs boson is the only known direct connection to whatever is responsible for the origin of multiple generations, flavor and known CP violation.  By studying it with more precision, we may perhaps gain insight into these fundamental puzzles or, at the very least, test if this picture is correct.  Furthermore, these puzzles also extend to the neutrino sector.  Whatever form neutrino masses take, Majorana, Dirac or both, their mass still must connect to the SM Higgs boson or a new Higgs-like boson must  exist that also breaks the electroweak symmetry.

The fact that \emph{understanding the properties of the SM Higgs boson connects to so many fundamental questions illustrates how central it is to the HEP program}.  The connections briefly reviewed so far obviously can each be expanded in greater detail, but to collect the various themes in a simple to digest manner this is illustrated in Figure~\ref{fig:higgscentral}. 
The generality of the concepts and questions posed in Figure~\ref{fig:higgscentral} could even belie connections to additional fundamental mysteries.  For example, the Higgs portal could specifically connect to Dark Matter or other cosmological mysteries. 
\begin{figure}[ht]
\begin{centering}
\includegraphics[scale=0.2]{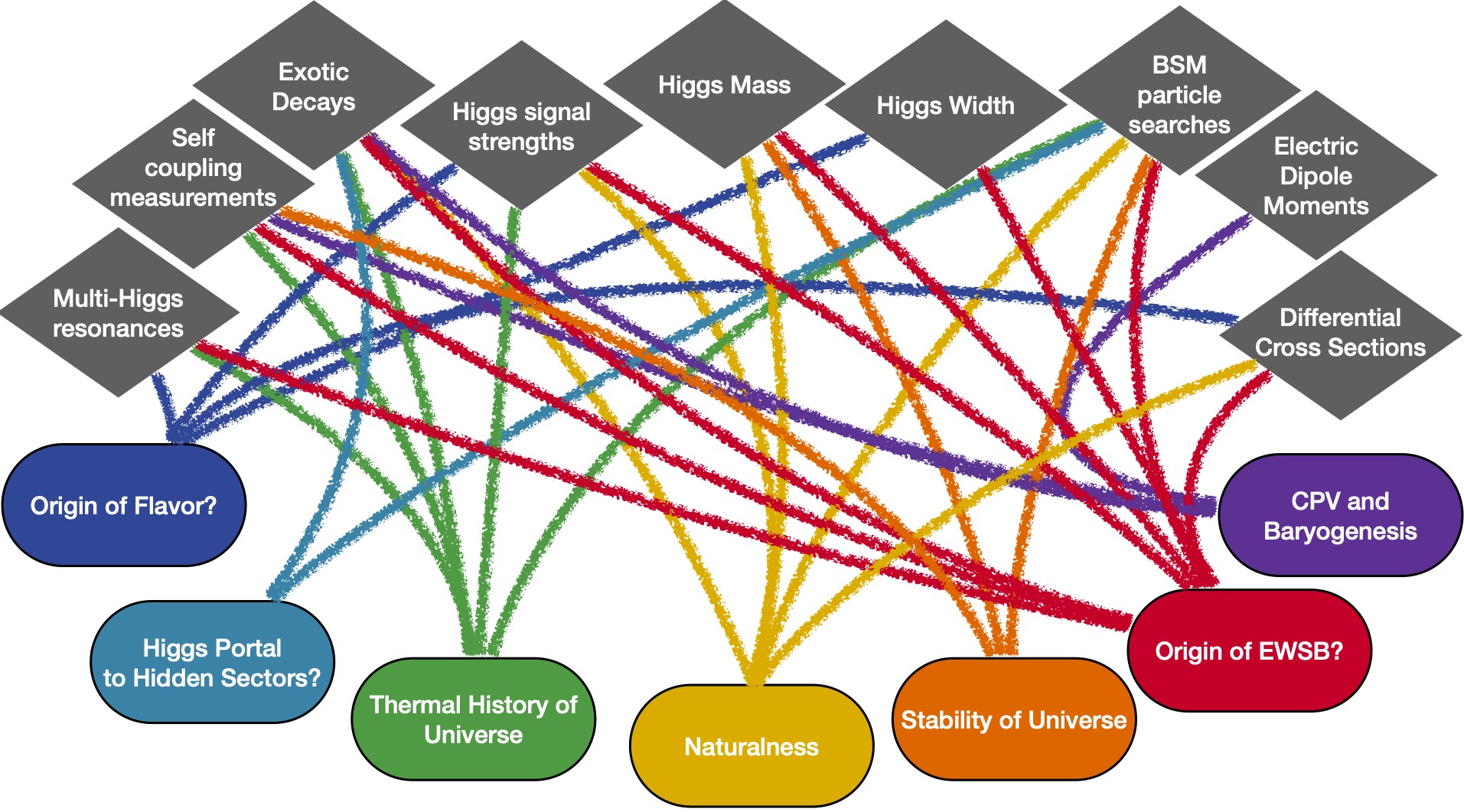} 
\par\end{centering}
\caption{Examples of the interplay between experimental observables and fundamental questions connected to the Higgs boson.}
\label{fig:higgscentral2}
\end{figure}

The goal of this topical report is to try to connect the many fundamental questions related to the Higgs boson to various observables and vice versa. The Higgs presents a challenge in HEP, because to 
test the consistency of the SM requires a dedicated experimental and theoretical program.
The previous Snowmass report\cite{Dawson:2013bba} advocated a bifurcation into Higgs factories and Energy frontier targets; however to understand the Higgs will require {\em both} directions as well as new theoretical concepts.  Therefore, understanding how to map various observables to the interesting questions is crucial as it helps enable a path to the future for deciphering what various collider projects can contribute.  In Figure~\ref{fig:higgscentral2} we give a suggestive visual representation of the types of observables and the deeply intertwined web that connects them to some of the fundamental ideas shown in Figure~\ref{fig:higgscentral}.

\begin{figure}[ht]
\begin{centering}
\includegraphics[scale=0.25]{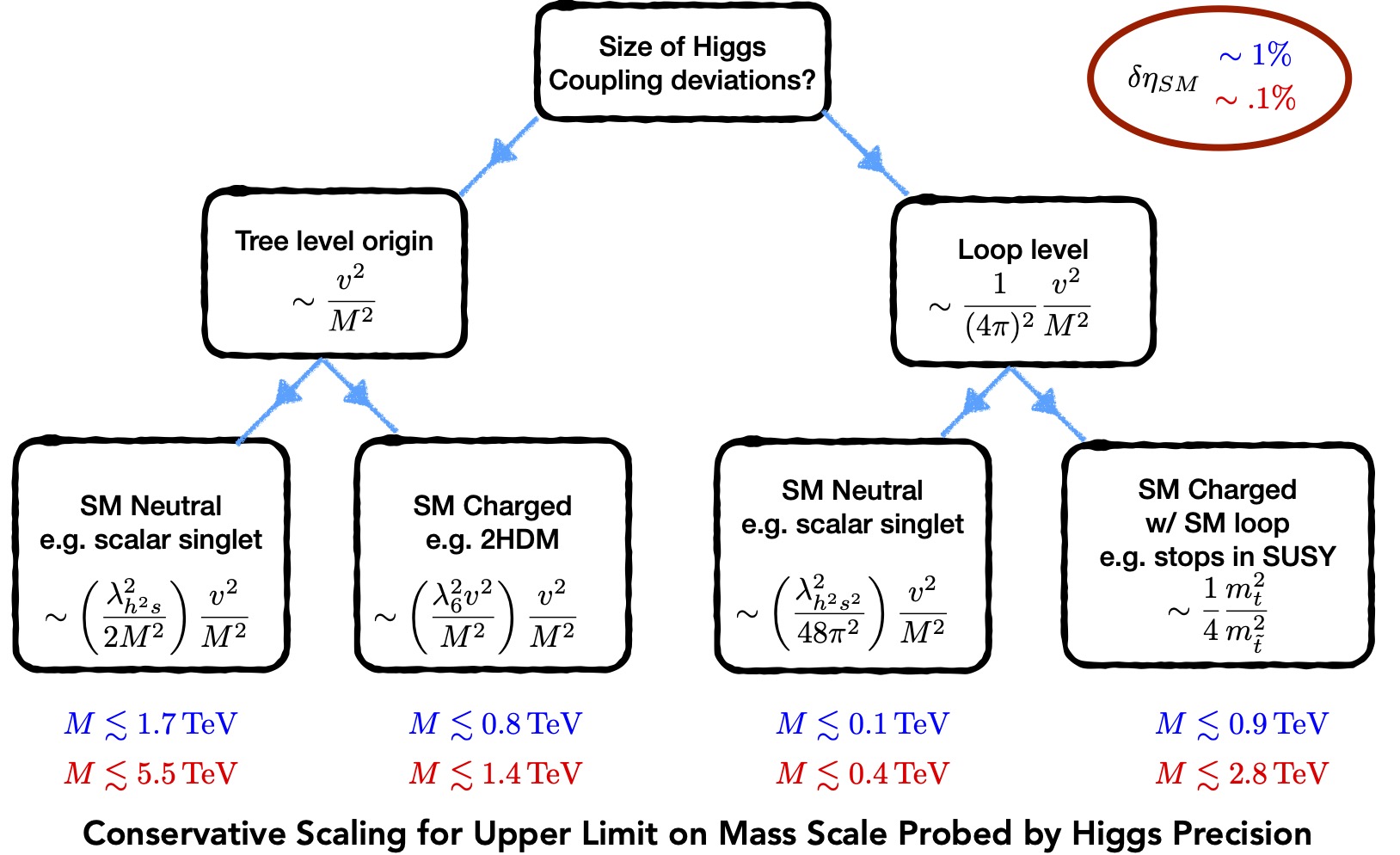} 
\par\end{centering}
\caption{Typical Higgs coupling deviations depending on whether the couplings are generated from new physics that generates tree level effects or loop level effects primarily.  Optimistically assuming all new physics couplings or ratios of new physics scales are $\mathcal{O}$(1) gives a conservative upper bound on the highest scales probed by Higgs coupling deviations. Based on assuming a precision for Higgs coupling deviations of $1\to.1\%$ this shows that Higgs couplings probe scales from as weak as $M\sim 100$ GeV to as strong as $M\sim 5.5$ TeV.  }
\label{fig:higgscentral3}
\end{figure}

While Figure~\ref{fig:higgscentral2} is qualitative, it does provide two important lessons:  The first is that many observables map to fundamentally different questions related to the Higgs boson. It is non-trivial to connect from observables related to Higgs physics with fundamental questions.
This has been referred to as the ``Higgs Inverse Problem", in analogy with the previously coined LHC inverse problem for BSM physics.  The second important lesson, alluded to in Figure~\ref{fig:higgscentral2}, is that Higgs related observables do not just fall into the standard $\kappa$ or effective field theory (EFT) fits.  While Higgs coupling deviations have become the gold-standard by which future collider projects are judged~\cite{deBlas:2019rxi}, they do not occur in isolation.  In particular, if there are any deviations in the Higgs couplings, or differential measurements etc., there {\em must} be new physics that couples to the Higgs boson which gives origin to it. In comparing various collider sensitives to new physics in the Higgs sector, one must also compare to other direct searches and indirect constraints on BSM physics simultaneously.  As an example of this, one could ask what is the meaning of achieving per cent or per mille level accuracy for Higgs couplings?  The standard approach to this question is to imagine that these deviations are caused by some higher dimension operator that arises from integrating out new BSM states.  To get a rough rule of thumb for this, one can imagine any gauge invariant operator in the SM that leads to some Higgs coupling, $\eta_{SM}$, being extended using the same trick as the Higgs portal, i.e. turning into a dimension 6 operator with the addition of a factor of $\hsm^\dagger \hsm$.   This, in turn, comes with a dimensional scale $M$ and a Wilson coefficient $c_\eta$, that when we expand around the vacuum expectation value  of the Higgs boson, leads to a predicted deviation of the corresponding SM Higgs coupling
\begin{equation}\label{eqn:EFT0}
\delta \eta_{SM}\sim c_\eta \frac{v^2}{M^2}.
\end{equation}
If one then categorizes the types of new physics contributions based on
whether they arise at tree or loop-level, and whether the new physics particles are charged under the SM then a more specific prediction can be made for $c_\eta$~\cite{deBlas:2017xtg}.  In Figure~\ref{fig:higgscentral3}, various possibilities are demonstrated, while also assuming a conservative scaling for the upper bound on the new physics mass scale $M$.  It is assumed that all new physics dimensionless couplings, or ratios of new physics scales are $\mathcal{O}(1)$.  In weakly coupled theories with valid EFT expansions one would expect a scaling with $c_\eta\ll 1$, and thus the upper bound on the scale $M$ would be even lower. This already demonstrates an important result for the interplay of BSM physics and Higgs physics: Depending on the type of new physics, reaching the per cent or per mille level accuracy for Higgs couplings corresponds to probing scales of $\mathcal{O}(.1\to 5.5~$ TeV). At the lower end, in the case of a SM gauge singlet scalar that affects Higgs precision measurements at loop level, the EFT formalism generically does not apply given the precision attainable at HL-LHC and future Higgs factories.  However, this does not mean that it is uninteresting from a Higgs precision point of view, rather it reflects that the effects on the Higgs sector must be considered broadly.  This is a generic lesson, as the scales generated are {\em all} within reach of the LHC or are in the few TeV range relevant for future discovery machines.  When planning for the future of HEP, it is crucial to consider the interplay of precision Higgs physics and direct searches to understand what is new territory, and what is complementary or ruled out by other experiments or analyses. 

The estimates coming from Eq.~\ref{eqn:EFT0} do not represent a no lose statement: this is impossible to make. For example, the scale of new physics could be slightly larger if the EFT description scaled differently due to strongly coupled dynamics, the canonical example being composite Higgs models~\cite{Kaplan:1983fs,Kaplan:1983sm,Georgi:1984ef,Georgi:1984af,Dugan:1984hq}.  Inherently there are not simple closed form predictions of arbitrary strongly coupled theories, and typically one relies upon guidance from large $N$ expansions.  In particular, there does not exist a calculable UV complete composite Higgs model that predicts a SM-like Higgs boson while satisfying all experimental constraints at this point. Instead, the phenomenology is often investigated in the context of some minimal symmetry based arguments of a low energy EFT where the Higgs arises as a pseudo Nambu-Goldstone boson (pNGB).  These models were more prevalent before the Higgs discovery, especially after the Little Higgs mechanism was introduced~\cite{Arkani-Hamed:2002iiv,Arkani-Hamed:2002ikv}. These models are more modest in scope and often fall under the Minimal Composite Higgs model~\cite{Agashe:2004rs} or Strongly Interacting Light Higgs (SILH) frameworks~\cite{Giudice:2007fh}. While many features are model dependent, there are some more model universal features that can be connected to Higgs physics~\cite{Liu:2018qtb}. In the Minimal Composite Higgs model, for example, if the Higgs couplings to gauge bosons were measured at the per-mille level without deviation, it would imply that the symmetry breaking scale $f$ would be probed to 5.5 TeV.  This is a larger scale than the weakly coupled assumptions shown in Figure~\ref{fig:higgscentral3} for SM charged states that would be composites of the strong dynamics.  This is not surprising as the states would strongly couple to the Higgs boson.  It is also important to note that $f$ is not necessarily the scale of the new composite states, they could in principle be higher or lower. Generic scaling arguments for composite mesons suggest $m_\rho \sim 4\pi f/\sqrt{N}$. In concrete models there often are top partners with $m\sim f$, or in Little Higgs constructions the gauge partners can be even lower $m\sim gf$.  Thus, it is difficult to draw concrete conclusions on the scales probed in strongly coupled theories via precision Higgs physics. Yet the lesson still persists that with precise Higgs measurements we are still generically exploring the TeV scale.  It is crucial to combine the myriad of related measurements to understand fully the Higgs sector. 

\begin{figure}
\begin{centering}
\includegraphics[scale=0.25]{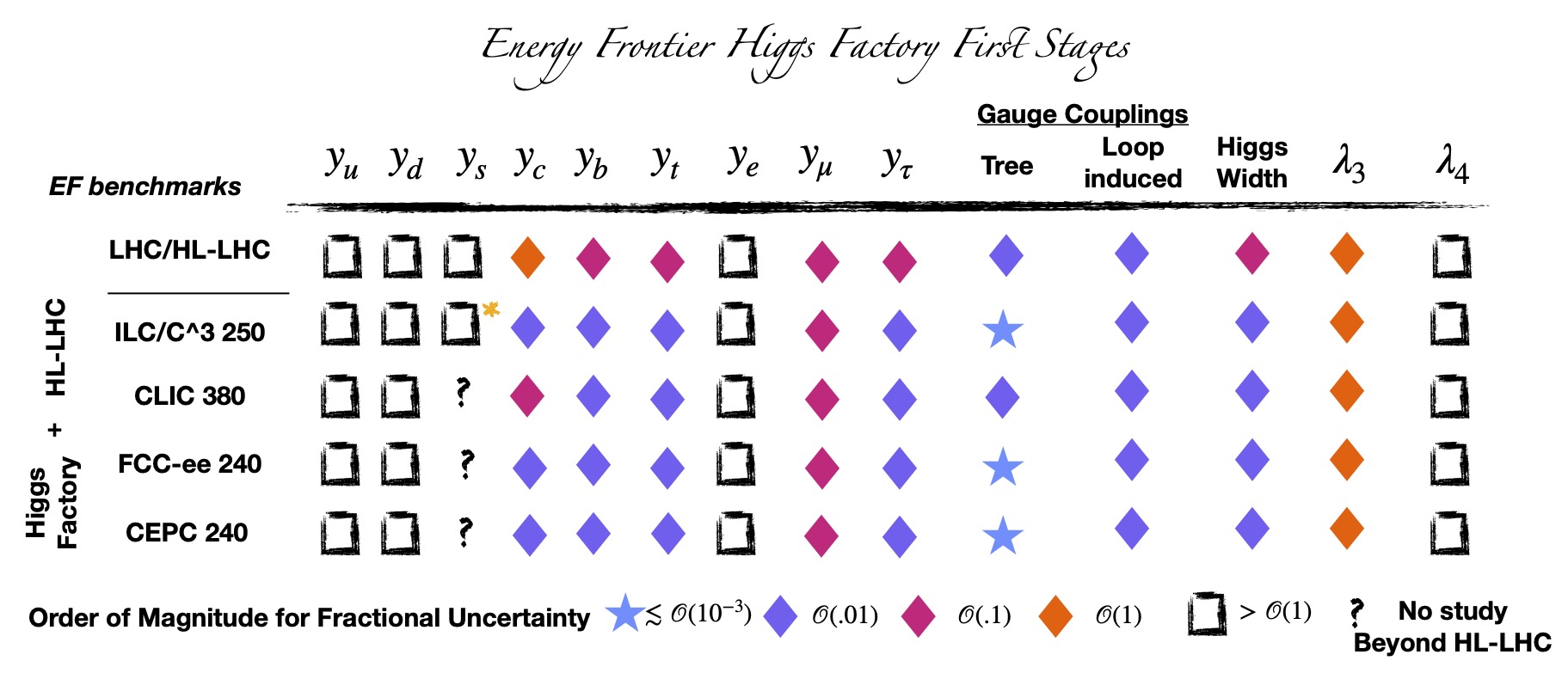} 
\par\end{centering}
\caption{A snapshot of future Higgs precision measurements of SM quantities based on the order of magnitude for the fractional uncertainties with the range defined through the geometric mean.  In this figure the first stages of each $e^+e^-$ Higgs factory are shown in combination with the HL-LHC, as well as the HL-LHC separately.  The Higgs factories are defined as those listed in Section 2.2 of the Energy Frontier Report, excluding the 125 GeV muon collider whose timescale is in principle longer term.  The specific precision associated to each coupling can be found in Section~\ref{sec:future} and references therein. A * is put on the ILC measurements for the strange Yukawa to single it out as a new measurement proposed during this Snowmass study, and is discussed further in Section~\ref{sec:flavor}.  The ? symbol is used in the case where an official study has not yet been performed, for example in the case of strange tagging for CLIC, FCC-ee, and CEPC.  This does not mean that they can not achieve a similar precision, but that it is yet to be demonstrated whether based on their detector concepts the measurement is worse or can be improved. 
}
\label{fig:higgssummarya}
\end{figure}

\begin{figure}
\begin{centering}
\includegraphics[scale=0.25]{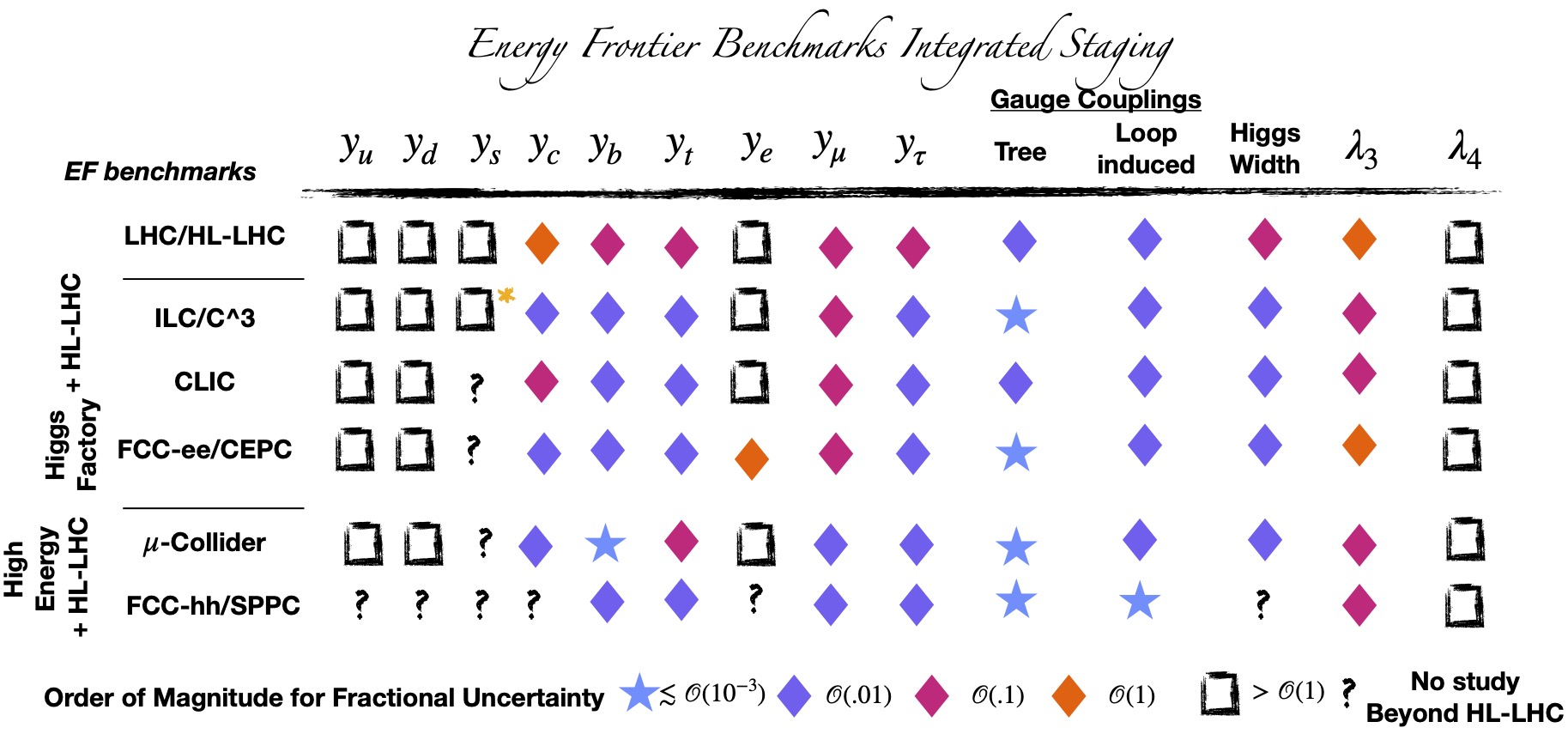} 
\par\end{centering}
\caption{A snapshot of future Higgs precision measurements of SM quantities based on the order of magnitude for the fractional uncertainties with the range defined through the geometric mean.  In this figure the ultimate reach of states of all Higgs factories and High Energy colliders are shown in combination with the HL-LHC results, as well as the HL-LHC separately.  All benchmarks and stages are defined in Section 2.2 of the Energy Frontier Report.  The specific precision associated to each coupling can be found in Section~\ref{sec:future} and references therein. A * is put on the ILC measurements for the strange Yukawa to single it out as a new measurement proposed during this Snowmass, and this is discussed further in Section~\ref{sec:flavor}.  The ? symbol is used in the case where an official study has not yet been performed, but does not connotate that it should be worse than similar colliders, simply that whether it is better or worse based on detector design has not been demonstrated.   Note that compared to Figure~\ref{fig:higgssummarya}, differences between Higgs Factories based on Linear Colliders and Circular Colliders can be seen.  Additionally for the High Energy Colliders such as FCC-hh and the Muon Collider, both offer extensions beyond the original Higgs factory proposals, of course on a longer timescale.
}
\label{fig:higgssummaryb}
\end{figure}

Given the basic link between the scale of new physics and the precision measurements of Higgs boson properties, it is useful to survey the proposed experiments to understand which options reach the per cent or per mille accuracy.  This is clearly one of the main highlights of this report, as well as the previous European Strategy Group report~\cite{deBlas:2019rxi}. In Section~\ref{sec:future}, the relevant inputs and specific projected sensitivities at various machines are shown. 
To give a more global perspective we illustrate schematically the outcome for precision Higgs physics in Figures~\ref{fig:higgssummarya} and \ref{fig:higgssummaryb}.

These snapshots differ from most in the literature in two key ways:  First, the more coarse grained approach to precision of the Higgs boson measurements, where we have delineated the capabilities based on the order of magnitude of the uncertainty achieved. While the usual fine grained approach is found in Section~\ref{sec:future}, based on the arguments about the scale of new physics probed, the difference between a 1\% and 2\% measurement is not particularly crucial compared to the order of magnitude.  This is especially true because the projected inputs to Snowmass and ESG~\cite{deBlas:2019rxi} were derived with different levels of rigor and assumptions.  As the LHC has demonstrated on numerous occasions, even in a difficult collider environment, experimental techniques can often surpass projections.  Second, there are numerous properties in the snapshot that are not typically listed in an EFT or "$\kappa$" fits such as first generation couplings, and the Higgs quartic coupling. This is to emphasize that the SM is far from being complete, and the Higgs boson, as its central figure, requires continued experimental effort to claim that the SM is ``complete".  Finally it also demonstrates where clearly more work is needed, including potentially new observables and ideas.

The summary of Higgs precision properties shown in Figures~\ref{fig:higgssummarya} and \ref{fig:higgssummaryb}, of course, contain numerous caveats, as the measurements of the various properties listed are done in very different ways.  As displayed, it can be thought of as akin to a ``kappa-0" or EFT fit.  Larger deviations in Higgs boson properties typically signify lower scale physics effects which are not  captured by EFT/$\kappa$ fits, and differential distributions or other observables may be key.  Moreover, with the Higgs portal motivation, there can be new decay modes of the Higgs which are not fully captured in Figs.~\ref{fig:higgssummarya}-\ref{fig:higgssummaryb}.  There is no possible way to model independently characterize all BSM effects on Higgs physics and going beyond this summary requires model interpretations as discussed further in Section~\ref{sec:BSM}. In this context, all EFT interpretations should also be thought of as models with thousands of parameters. What Fig.~\ref{fig:higgssummarya}-\ref{fig:higgssummaryb} do show is that {\em all} of the currently proposed colliders that are Energy Frontier  benchmarks offer exciting windows into understanding the Higgs.  To further differentiate amongst collider options requires understanding the differences in the types of BSM physics that these Higgs precision measurements correlate with, that we attempt to address more in Section~\ref{sec:BSM}, as well as how useful they are in the context of other Topical Group measurements.  Additionally, one must ask the question what is the precision goal for these properties? This question requires an understanding of the interplay shown in Figure~\ref{fig:higgscentral2} and how to prioritize various measurements. 

The rest of the report is organized as follows:  Section~\ref{sec:Higgs} contains a description of current measurements of Higgs properties, Section~\ref{sec:future} discusses future projections of measurements of Higgs properties, and Section~\ref{sec:BSM} contains a brief overview of the information gained from the measurements of Higgs properties. 
Section~\ref{sec:Detector} discusses detector and accelerator requirements for the observation of new physics in Higgs measurements.

%% file: Tex/now.tex
\begin{figure}[h!]
\begin{centering}
\includegraphics[scale=0.45]{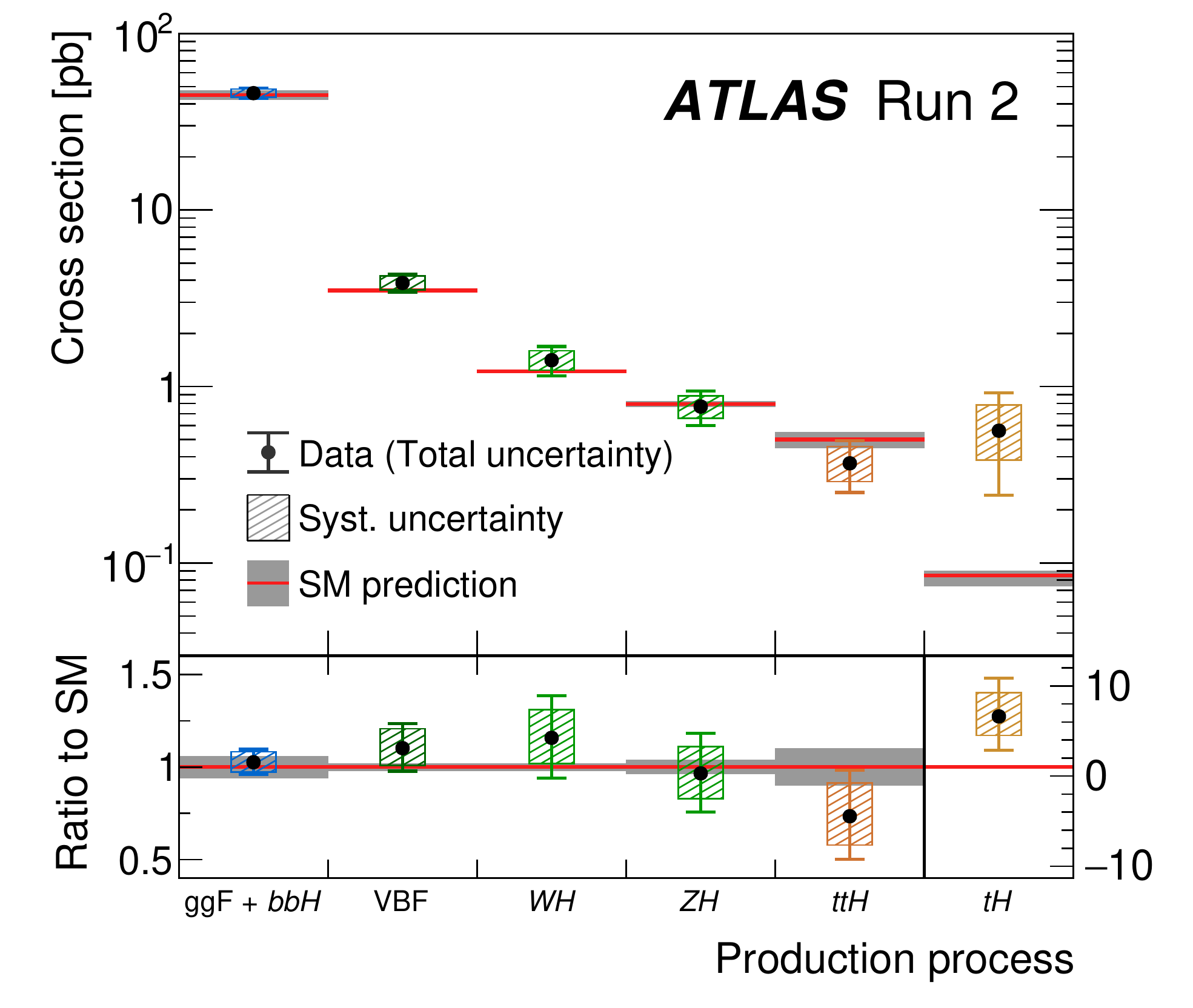} 
\includegraphics[scale=0.4]{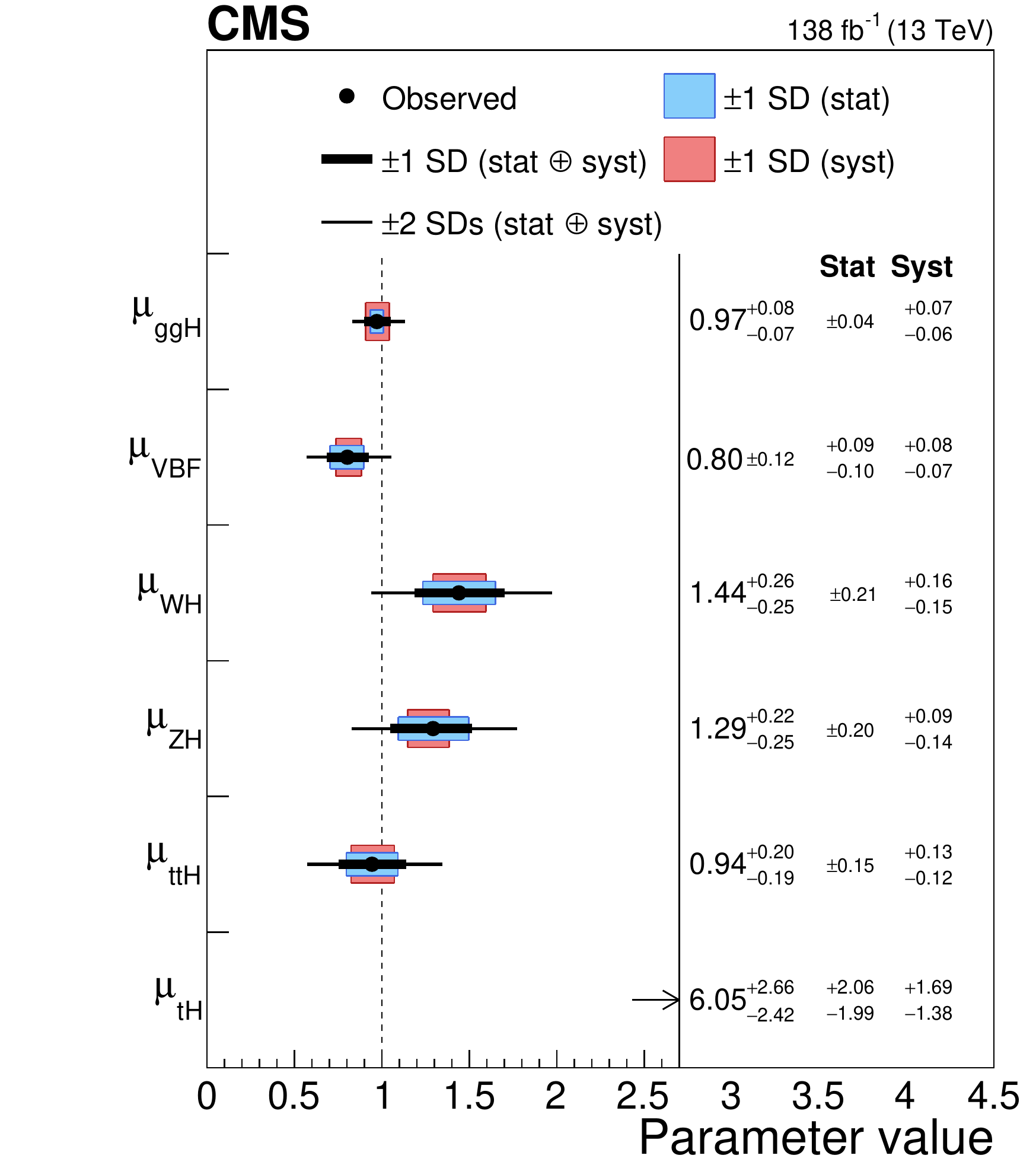} 
\par\end{centering}
\caption{Measured cross sections for ggF, VBF, W$\hsm$, Z$\hsm$, \ttH, and t$\hsm$ normalized to their SM predictions, assuming SM values for the decay branching fractions for ATLAS (left) and CMS (right)~\cite{ATLAS:2022vkf,CMS:2022dwd}. 
}
\label{fig:sigs_now}
\end{figure}
\begin{figure}[h!]
\begin{centering}
\includegraphics[scale=0.45]{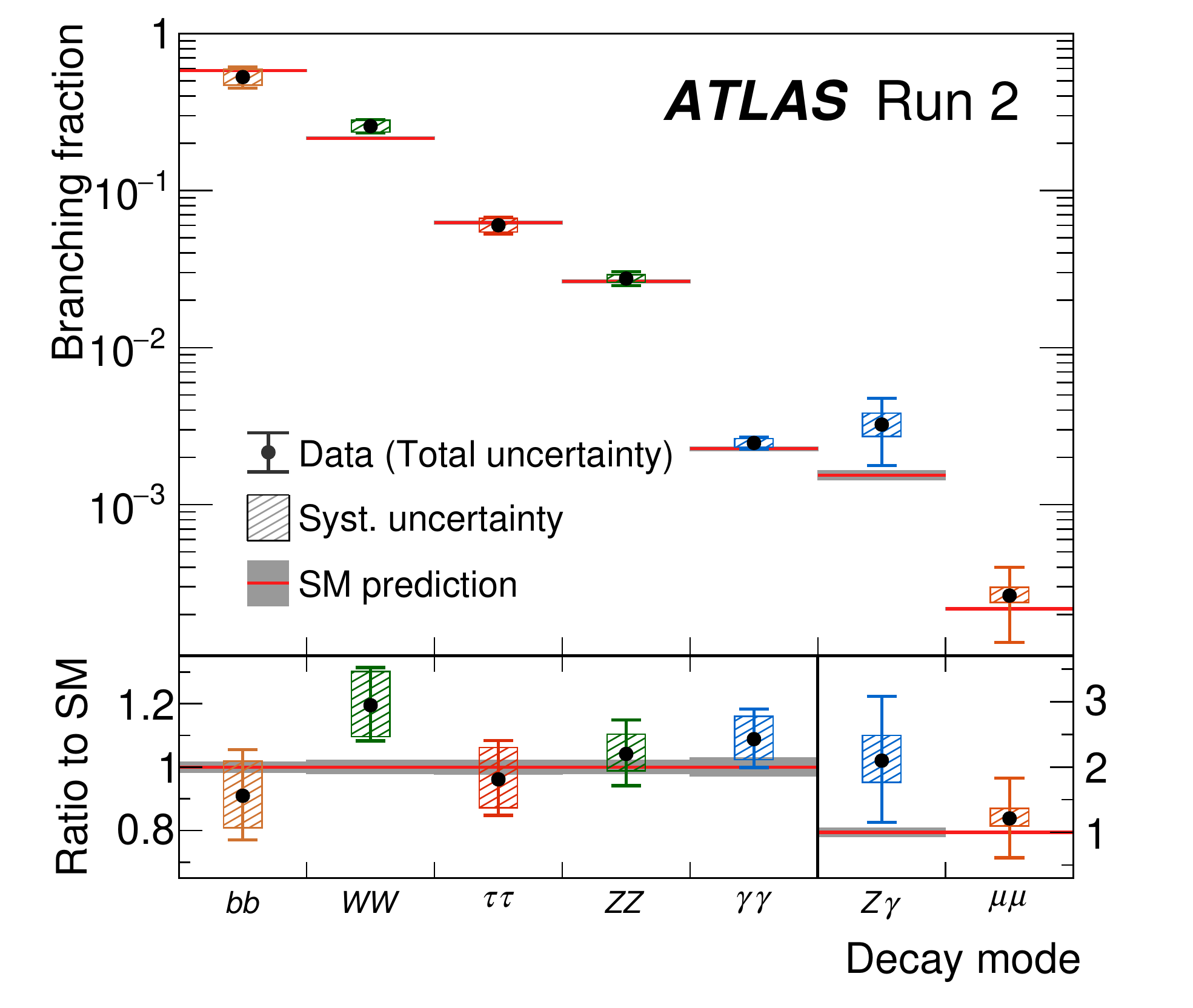} 
\includegraphics[scale=0.4]{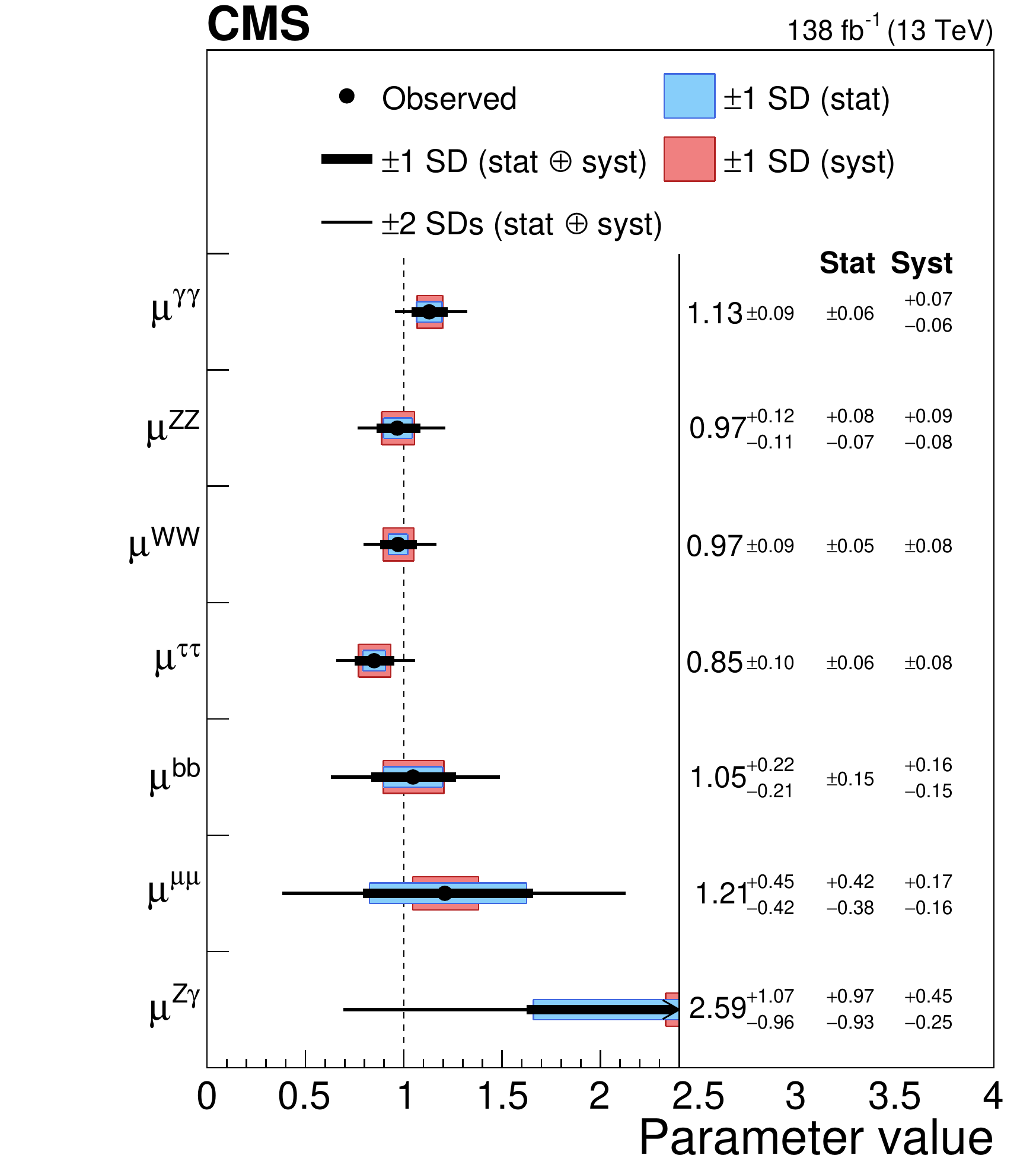} 
\par\end{centering}
\caption{Left, observed and predicted Higgs boson branching fractions for different Higgs boson decay modes~\cite{ATLAS:2022vkf}. Right, CMS signal strength modifiers for the various decay modes~\cite{CMS:2022dwd}.}
\label{fig:brs_now}
\end{figure}
\begin{figure}[ht!]
\begin{centering}
\includegraphics[scale=0.4]{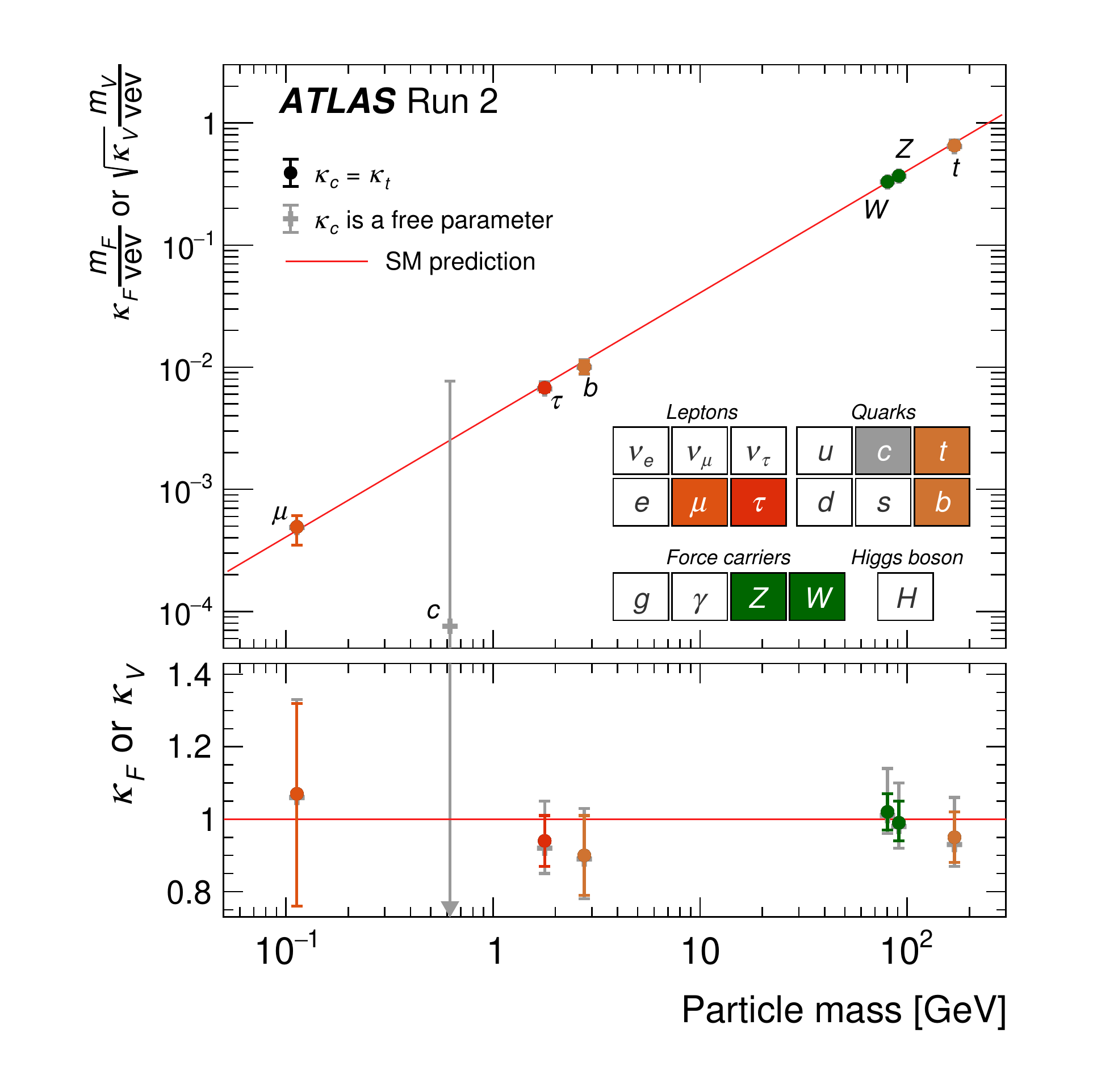}
\includegraphics[trim={0 -0.6cm 0 0},clip,scale=0.276]{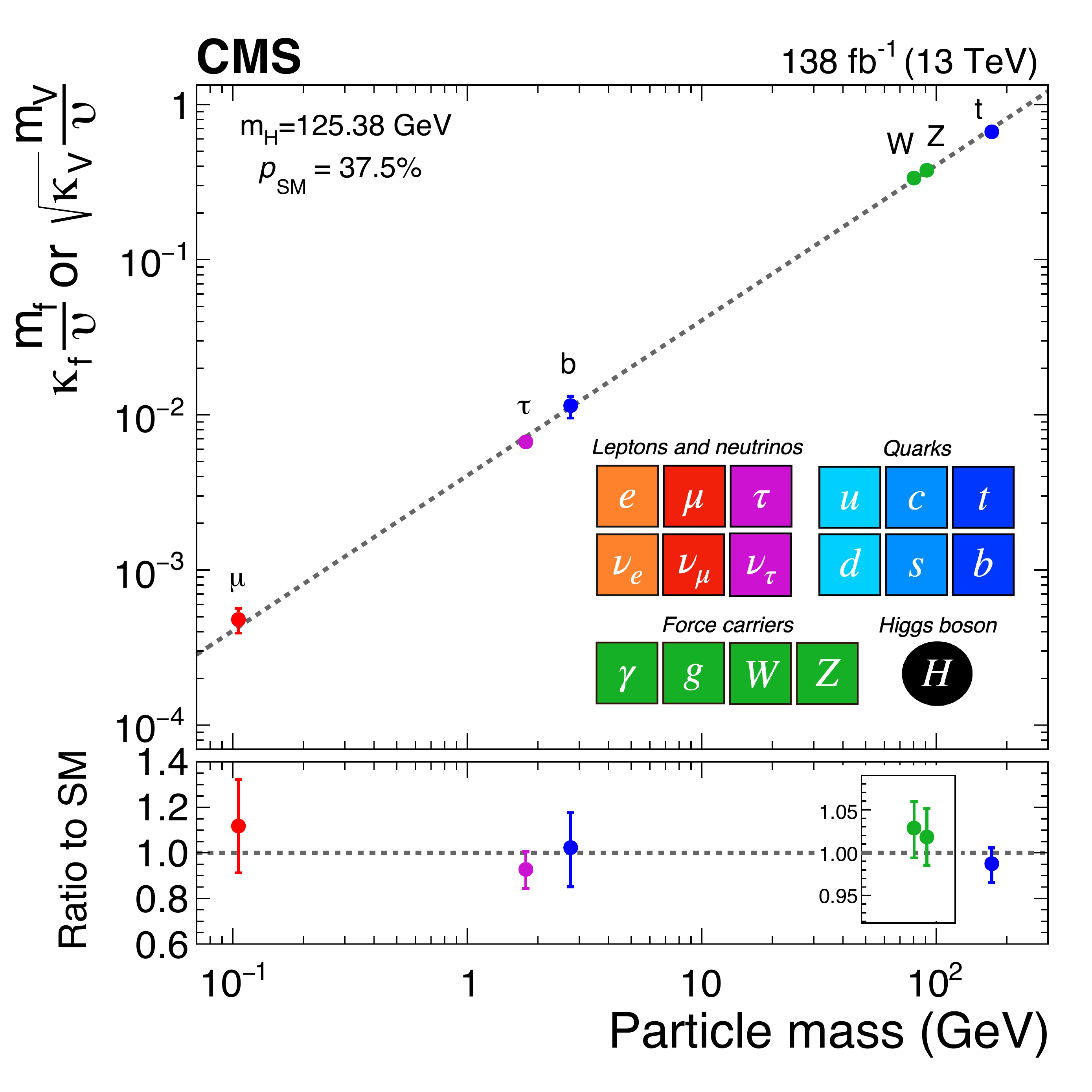} 
\par\end{centering}
\caption{The best-fit estimates for the reduced coupling modifiers extracted for fermions and weak bosons compared to their corresponding predictions from the SM. The associated error bars represent 68\% CL intervals for the measured parameters for ATLAS (right)~\cite{ATLAS:2022vkf} and CMS (left)~\cite{CMS:2022dwd}. ATLAS considers two fit scenarios with $\kappa_c =\kappa_t$ (coloured circle markers) and $\kappa_c$ left free-floating in the fit (grey cross markers).}
\label{fig:couplings}
\end{figure}

LHC Run 2 with $\sim$140 $\fbi$ of data analyzed is providing a wealth of new measurements for the Higgs sector. The mass is a free parameter in the SM and it is now known to per mille accuracy.
The most recent Higgs boson mass measurements, from CMS and ATLAS set its to value to be 125.38$\pm$0.14 GeV~\cite{CMS:2020xrn} and 124.92$\pm$0.21 GeV~\cite{ATLAS:2020coj} respectively, using both the $\gamma\gamma$ and
ZZ decay channels.  With some of the Higgs boson coupling measurements approaching $\mathcal{O}$(5-10)\% precision, we are entering the era of precision Higgs physics. 
All of the major production mechanisms of the Higgs boson have been observed at the LHC: gluon fusion (ggF), vector-boson fusion (VBF), the associated production with a W or Z boson (W\hsm, Z\hsm), and the associated production with top quarks (\ttH, t\hsm), as shown in Figure~\ref{fig:sigs_now}.
The most updated measurements of Higgs decay modes are shown in Figure~\ref{fig:brs_now}.  The experimental sensitivity of some production and decay modes are nearing the precision of state-of-the-art theory predictions.
The values of the Higgs boson couplings  to the elementary particles that are extracted from the measured cross sections and branching ratios are given in Figure~\ref{fig:couplings}; it is seen that the strengths of the couplings increase with the masses of the elementary particles, in good agreement with the SM predictions, within the systematic uncertainties.  

The couplings to the first and second generations have not yet been measured. 
Probing the charm Yukawa in the high-pile up environment at the LHC is very challenging. Novel jet reconstruction, identification tools and analysis techniques have been developed to look for \hcc in the V\hsm~production mode, leveraging also the expertise developed for \hbb in the same topology. The most stringent constraint to date is set by CMS using 138~$\fbi$ of Run 2 data. The observed 95\% CL interval (expected upper limit) is 1.1 $< |\kappa_c| < $ 5.5 ($|\kappa_c| <$ 3.4)~\cite{HIG-21-008-pas}\footnote{ The $\kappa$'s are defined as the ratio of the measured Higgs couplings to the SM predictions.}.  This should be compared to indirect bounds on the charm Yukawa, since if $\kappa_c \sim 5$, this would already be ruled out by contributions to the Higgs width if $\kappa_c$ were the only parameter that was modified in the SM, see for example~\cite{Delaunay:2013pja,Coyle:2019hvs}. 
CMS has reported the first evidence of Higgs decay to $\mu\mu$ with 137 $\fbi$ at 13 TeV~\cite{CMS:2020xwi}, but the measurement of the Higgs coupling to the $\mu$ will require the additional dataset of the HL-LHC.

In the SM, the branching fraction to invisible final states, BR(\hsm$\rightarrow$ inv), is only about 0.1\%,
from the decay of the Higgs boson via ZZ$^{*} \rightarrow 4\nu$.  Observation of an invisible decay, would be a clear signal of new physics beyond the Standard Model. 
The most stringent constraint currently is set by CMS exploiting the VBF topology and with 101~$\fbi$ at 13 TeV. The observed (expected) upper limit on the invisible branching fraction of the Higgs boson is found to be 18\% (10\%) at the 95\% CL, assuming the SM production cross section~\cite{CMS:2022qva}. ATLAS with 139~$\fbi$ at 13~TeV in the same final state has set an observed (expected) limit of 14.5\% (10.3\%) at 95\% CL~\cite{ATLAS:2022yvh}.

\begin{figure}[ht!]
\begin{centering}
\includegraphics[trim={0 1.6cm 0 0},clip,scale=0.4]{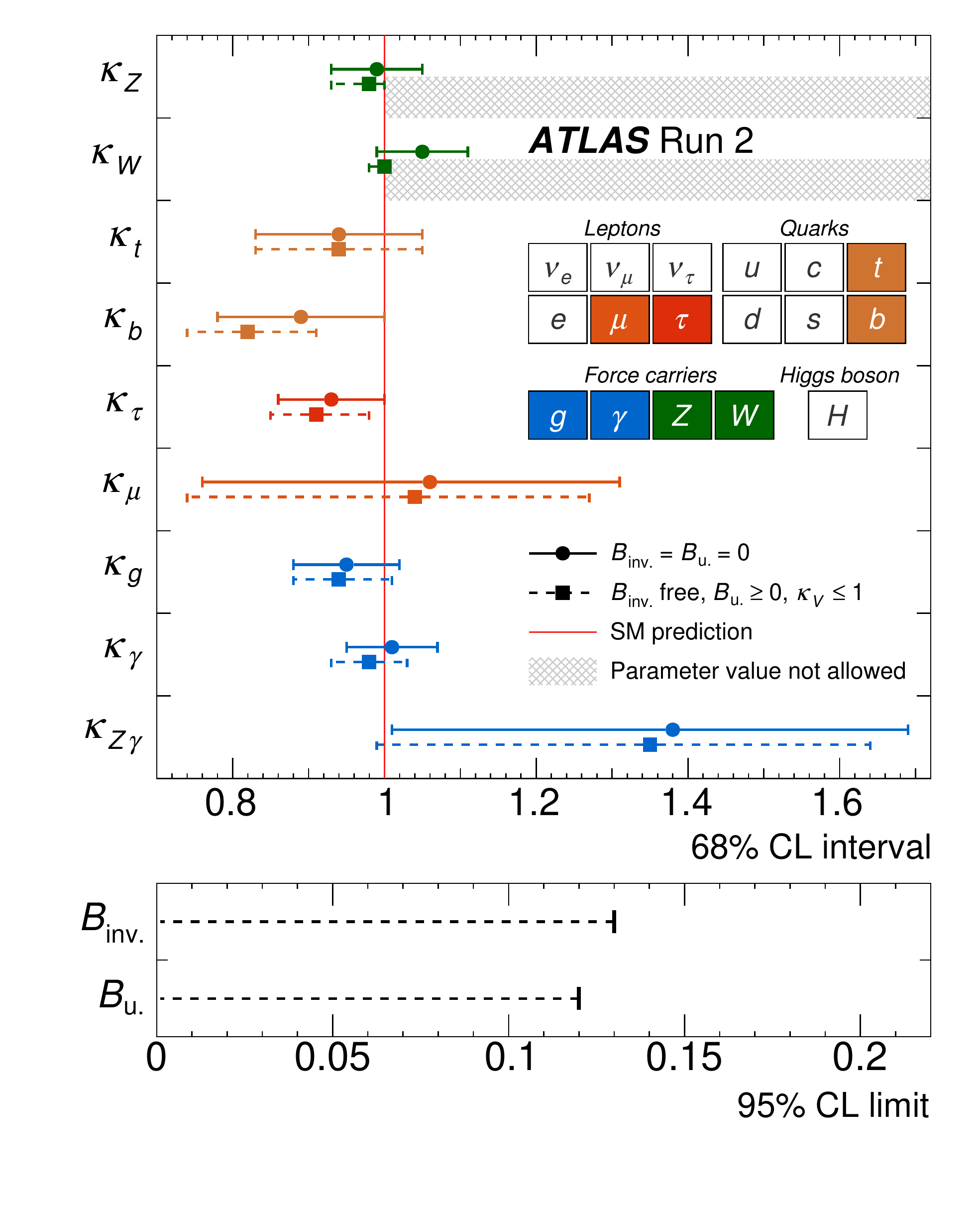} 
\includegraphics[trim={0 -0.01cm 0 0},clip,scale=0.45]{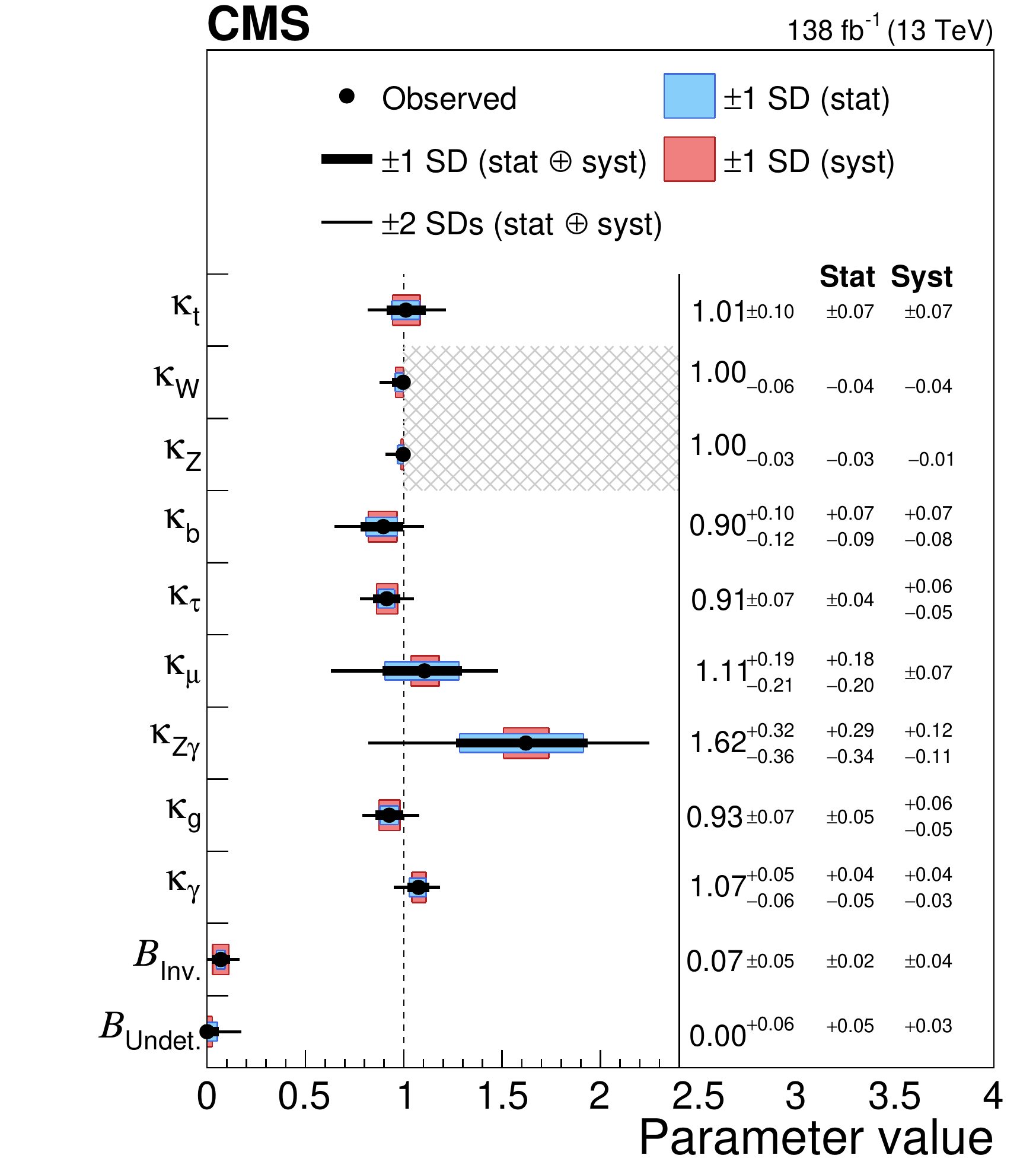} 
\par\end{centering}
\caption{Left, ATLAS best-fit values and uncertainties for Higgs boson coupling modifiers per particle type with effective photon and gluon couplings and the branching fraction to invisible (B$_i$) and undetected decays (B$_u$) included as free parameters and the measurement of the Higgs boson decay rate into invisible final states included in the combination~\cite{ATLAS:2020qdt}. Right, CMS summary of the couplings modifiers $\kappa$. The thick (thin) black lines report the 1$\sigma$ (2$\sigma$) confidence intervals\cite{CMS:2022dwd}.}
\label{fig:coups_now}
\end{figure}

In addition to the previously mentioned, channel-independent measurements, a simultaneous fit of many individual production times branching fraction measurements is performed to determine the values of the Higgs boson coupling strengths. The $\kappa$-framework defines a set of parameters that affect the Higgs boson coupling strengths without altering any kinematic distributions of a given process. SM values are assumed for the coupling strength modifiers of first-generation fermions, the other coupling strength modifiers are treated
independently. The results are shown in Figure~\ref{fig:coups_now} for ATLAS and CMS. In this particular fit, the presence of non-SM particles in the loop-induced processes is parameterized by introducing additional modifiers for the effective coupling of the Higgs boson to gluons, photons and Z$\gamma$, instead of propagating modifications of the SM particle couplings through the loop calculations. In these results, it also assumed that any potential effect beyond the SM does not substantially affect the kinematic properties of the Higgs boson decay products. The coupling modifiers are probed at a level of uncertainty of 10\%, except for $\kappa_b$ and $\kappa_{\mu}$ ($\approx 20\%$), and $\kappa_{Z\gamma}$ ($\approx 40\%$).

\newpage

%% file: Tex/theory_now.tex
\begin{figure}
\begin{centering}
\includegraphics[scale=0.3]{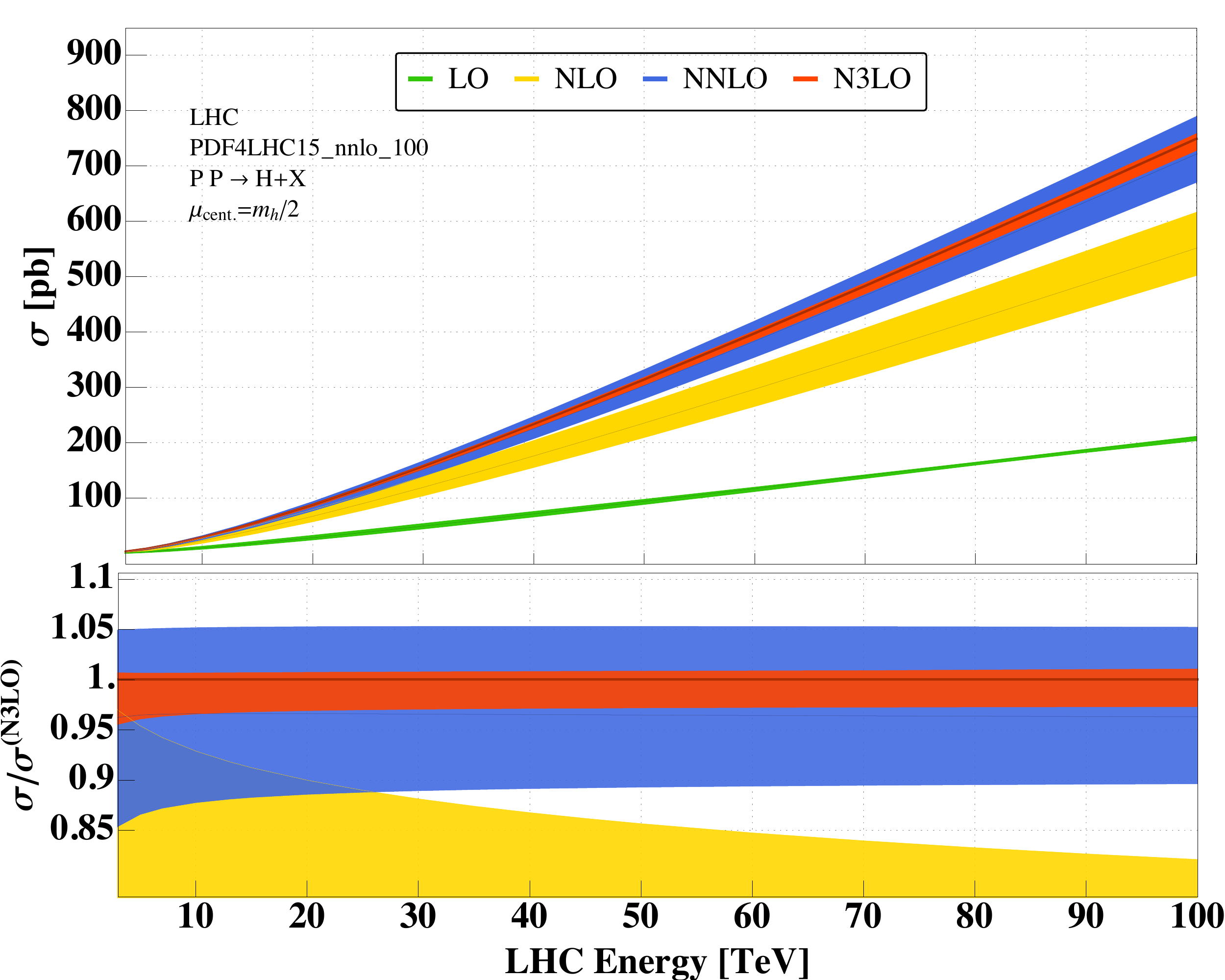}
\par\end{centering}
\caption{Gluon fusion contribution to the Higgs boson cross section at the LHC as a function of the p-p collision energy at LO, NLO, NNLO, and N$^3$LO~\cite{Caola:2022ay}. The bands represent an estimate of the theoretical uncertainty.}
\label{fig:ggn3}
\end{figure}
\begin{figure}
\begin{centering}
\includegraphics[scale=0.15]{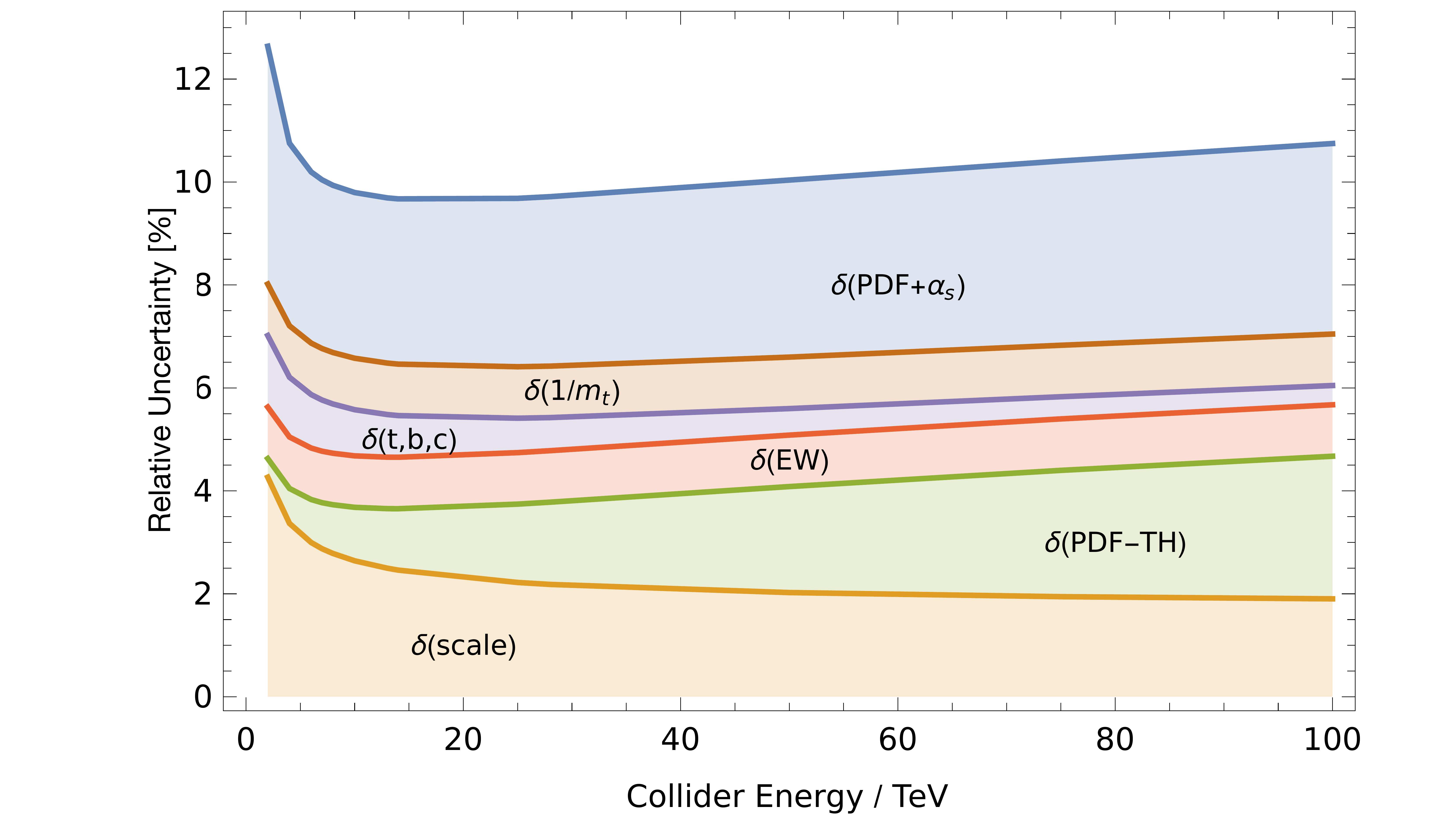}
\par\end{centering}
\caption{Linear sum of relative uncertainties on gluon fusion Higgs boson production as a function of the p-p collision energy~\cite{Cepeda:2019klc}.}
\label{fig:gguncert}
\end{figure}

The large number of Higgs boson events at the LHC offers the opportunity for precision measurements of Higgs cross sections and the extraction of the Higgs couplings to fermions and gauge bosons, requiring correspondingly precise theory calculations. 
Predictions for the inclusive cross sections at 14~TeV and 27~ TeV including higher order QCD and electroweak corrections are given in Table \ref{tab:higgs_sig}. It is apparent that the uncertainties rise with the machine energy.  
The total rates for all important Higgs production channels at the LHC are known to NNLO QCD, with N$^3$LO results available for the gluon fusion channel, as seen in Figure~\ref{fig:ggn3}.  Nevertheless, a major source of uncertainty on the Higgs boson couplings is expected to arise from theory as shown schematically in Figure \ref{fig:gguncert}, with the theory uncertainty expected to be comparable to the expected statistical and systematic uncertainties of the measurements. The theory uncertainties arise from unknown higher order QCD and electroweak corrections, effects of fermion masses, and uncertainties in the knowledge of the PDFs.  Impressive theoretical progress has been, and is continuing,  to be achieved, leaving theorists optimistic that the theory uncertainties can be reduced by a factor of two in the future~\cite{Caola:2022ayt}.  Meeting this necessary theoretical accuracy will  require a dedicated effort with significant computational resources~\cite{Cordero:2022gsh}.

\begin{table}[t]
\centering
\renewcommand{\arraystretch}{1.5}
\begin{tabular}{||c|c|c||}
\hline\hline
Process & $\sigma$(pb)$@14$ TeV & $\sigma$(pb)$@27$
TeV\\ \hline \hline
$ggF$ (N$^3$LO QCD + NLO EW) & $54.72^{+4.28\%({\text{theory}})+1.85\%({\text{PDF}})+2.60\% (\alpha_s)} 
_{-6.46\%({\text{theory}})-1.85\%({\text{PDF}})+2.62\% (\alpha_s)} 
$ & $146.65^{+4.53\%({\text{theory}})+1.95\%({\text{PDF}})+2.69\% (\alpha_s)} 
_{-6.43\%({\text{theory}})-1.95\%({\text{PDF}})+2.64\% (\alpha_s)} 
$
\\
\hline 
$VBF$ (NNLO QCD) &  $4.260^{+.45\% ({\text{scale}})+2.1\%({\text{PDF}}+\alpha_s)} 
_{-.34\%({\text{scale}})-2.1\%({\text{PDF}}+\alpha_s)} 
$ & $11.858 ^{+.66\% ({\text{scale}})+2.1\%({\text{PDF}}+\alpha_s)} 
_{-.36\%({\text{scale}})-2.1\%({\text{PDF}}+\alpha_s)} $ \\
\hline
$W\hsm$ (NNLO QCD+NLO EW) & $1.498 \pm .51\% ({\text{scale}})\pm 1.35\%({\text{PDF}}+\alpha_s) $ & $3.397^{+.29\% ({\text{scale}})+1.37\%({\text{PDF}}+\alpha_s)}
_{-.72\% ({\text{scale}})-1.37\%({\text{PDF}}+\alpha_s)} $\\
\hline
$Z\hsm$ (NNLO QCD+NLO EW) & $.981^{
 +3.61\% ({\text{scale}})+ 1.90\%({\text{PDF}}+\alpha_s)}_{-2.94\% ({\text{scale}})- 1.90\%({\text{PDF}}+\alpha_s)} $ 
 & $2.463^{+5.42\% ({\text{scale}})+2.24\%({\text{PDF}}+\alpha_s)}
_{-4.00\% ({\text{scale}})-2.24\%({\text{PDF}}+\alpha_s)} $
\\
\hline 
$t{\overline{t}}\hsm$ (NLO QCD + NLO EW) & $.6128 ^{
 +6.0\% ({\text{scale}})+ 3.5\%({\text{PDF}}+\alpha_s)}_{
 -9.2\% ({\text{scale}})- 3.5\%({\text{PDF}}+\alpha_s)}$ & 
 $2.860^{
 +7.8\% ({\text{scale}})+ 2.8\%({\text{PDF}}+\alpha_s)}_{
 -9.0\% ({\text{scale}})- 2.8\%({\text{PDF}}+\alpha_s)}$\\
\hline \hline
\end{tabular}
\caption{Predicted Higgs boson cross sections in p-p interactions with $\mh=125$ GeV from~\cite{Cepeda:2019klc}. }
\label{tab:higgs_sig}
\end{table}

Comparisons of theory and data, however, involve fiducial cross sections and theoretical progress has been made in extending these calculations to NNLO QCD and higher and thereby reducing the theory uncertainties.   In gluon fusion, for example, the decay $\hsm\rightarrow \gamma \gamma$ with fiducial cuts is known to N$^3$LO QCD, along with N$^3$LL' resummation, with a resulting theory uncertainty 
of ${\cal{O}}(3\%)$\cite{Billis:2021ecs,Chen:2021isd}.  Along with the need for higher order calculations including fiducial cuts comes the requirement to match the theory to higher order parton shower calculations which contributes to further theoretical uncertainties~\cite{Darvishi:2022gqt,Campbell:2022qmc}.   

The theoretical predictions for Higgs branching ratios given the Higgs boson mass and SM inputs give targets for future experimental measurements on the ${\cal{O}}(1-3\%)$ accuracy.  A few of the branching ratios are shown in Table~\ref{tab:higgs_br_th} for $\mh=125$ GeV and a complete set of SM branching rates (including known higher order corrections) can be found in 
\cite{Cepeda:2019klc}.

\begin{table}[t]
\centering
\renewcommand{\arraystretch}{1.5}
\begin{tabular}{||c|c||}
\hline\hline
Decay & Branching Ratio\\
\hline\hline
$\hsm\rightarrow b{\overline{b}}$
&
$.582^{+.65\%(\text{Theory})
+.72\% (m_q)+.78\%(\alpha_s)}_
{-.65\%(\text{Theory})
-.74\% (m_q)-.80\%(\alpha_s)}$\\
\hline
$\hsm\rightarrow c{\overline{c}}$ & 
$.02891^{+1.20\%(\text{Theory})
+5.26\% (m_q)+1.25\%(\alpha_s)}_
{-1.20\%(\text{Theory})
-.98\% (m_q)-1.25\%(\alpha_s)}$
\\
\hline
$\hsm\rightarrow \tau^+{\overline{\tau}}^-$
&
$
.06272^ {+1.17\% (\text{Theory}) + .98\% (m_q)+.62\%(\alpha_s)}_
{-1.16\%(\text{Theory})
-.99\% (m_q)-.62\%(\alpha_s)}$ \\
\hline
$\hsm\rightarrow \gamma\gamma$&
$
.00227 ^ {+1.73\% (\text{Theory}) + .93\% (m_q)+.61\%(\alpha_s)}_
{-1.72\%(\text{Theory})
-.99\% (m_q)-.62\%(\alpha_s)}$
\\
\hline
$ \hsm\rightarrow ZZ\rightarrow 4l (l=e,\mu,\tau)$
&
$.0002745 \pm 2.18\%$ \\
\hline
$ \hsm\rightarrow WW\rightarrow l^+l^- \nu {\overline{\nu}} (l=e,\mu,\tau)$
&
$.02338 \pm 2.18\%$ \\
\hline\hline
\end{tabular}
\caption{Higgs branching ratios with $\mh=125$ GeV from \cite{Cepeda:2019klc}.The $m_q$ uncertainty is the parametric dependence on the quark masses. }
\label{tab:higgs_br_th}
\end{table}

%% file: Tex/selfcoupling.tex
The scalar potential of the Higgs boson field, responsible for the EWSB mechanism, is still very far from being probed. After EWSB, the Higgs boson potential gives rise to cubic and quartic terms in the Higgs boson field, inducing a self-coupling term. The Higgs boson self-coupling, within the SM, is fully predicted in terms of the Fermi coupling constant and the Higgs boson mass, which has been measured at per-mille level accuracy by the ATLAS and CMS experiments~\cite{ATLAS:2020coj, CMS:2020xrn}. The Higgs self-coupling is accessible through Higgs boson pair production ($\hsm\hsm$) and inferred from radiative corrections to single Higgs measurements. Measuring this coupling is essential to shed light on the structure of the Higgs potential, whose exact shape can have deep theoretical consequences.

\begin{table}[ht!]
\begin{center}
{
\begin{tabular}{|l|c|c|cc|}
\hline
\hline
Search channel & Collaboration & Luminosity ($\fbi$) &  \multicolumn{2}{|c}{95\% CL Upper Limit}  \\
& & & expected & observed \\
\hline
\multirow{2}{*}{\bbbb} & 
ATLAS~\cite{ATLAS-CONF-2022-035} & 126 &
8.1 & 5.4   \\
                        & 
CMS~\cite{CMS:2022cpr} & 138 &
4.0 & 6.4 \\
\hline
\multirow{2}{*}{\bbgg} & 
ATLAS~\cite{ATLAS:2021ifb} & 139 &
5.5 & 4.0 \\
                                  & 
CMS~\cite{CMS:2020tkr} & 137 &
5.5 & 8.4 \\
\hline
\multirow{2}{*}{\bbtautau} & 
ATLAS~\cite{ATLAS-CONF-2021-030} & 139 & 
3.4 & 4.2 \\
                              & 
CMS~\cite{hig-20-010} & 138 &
5.2 & 3.3 \\
\hline
\multirow{2}{*}{\bbvv} & 
ATLAS~{\cite{Aaboud:2018zhh}} & 36.1 &
40 & 29 \\                            & 
CMS~\cite{hig-20-004} & 137 & 40 & 32  \\
\hline
\multirow{2}{*}{\wwyy} & 
ATLAS~{\cite{Aaboud:2018ewm} } & 36.1 &
230 & 160 \\
                              & 
CMS &  -- & -- & \\
\hline
\multirow{2}{*}{\wwww} & 
ATLAS~{\cite{Aaboud:2018ksn}} & 36.1 &
160 & 120 \\

& CMS~{\cite{hig-21-002}} & 138 &
19 & 21 \\
\hline
       \multirow{2}{*}{comb} & 
ATLAS~{\cite{ATLAS-CONF-2022-052}} & 126-139 &
2.2 & 2.2 \\

& CMS~{\cite{CMS:2022dwd}} & 138 &
2.5 & 3.4 \\                   
\hline
\hline
\end{tabular}
}
\end{center}
\vspace*{-0.4cm}
\caption{List of $\hsm\hsm$ searches at the LHC based on the p-p data collected by ATLAS and CMS at 13 TeV and corresponding to up to 126-139$~\fbi$ of integrated luminosity. Observed and expected upper limits on the SM \hh production cross section are normalised to the SM prediction~\cite{deFlorian:2016spz}. The combination of the three main channels, \bbbb, \bbtt and \bbyy, is also shown. }\label{exp-summary-table}
\end{table}

The 95\% CL expected and observed upper limits on the signal strength $\mu = \sigma_{\hsm \hsm} /\sigma_{\hsm \hsm}^{SM}$ are reported in Table~\ref{exp-summary-table} for each individual $\hsm \hsm$ final state. 
The best final state for the non-resonant $\hsm \hsm$ production is $b{\overline{b}}\tau^+\tau^- $ in ATLAS, and \bbyy in CMS. For each experiment the combined sensitivity of all channels together is improved by about 40\% with respect to the best channel~\cite{ATLAS-CONF-2022-052}. This can be easily explained by a relatively comparable sensitivity of the \hhbbyy, \hhbbtt and \hhbbbb~final states.

\begin{figure}[ht!]
\begin{center}
\includegraphics[width=0.50\textwidth]{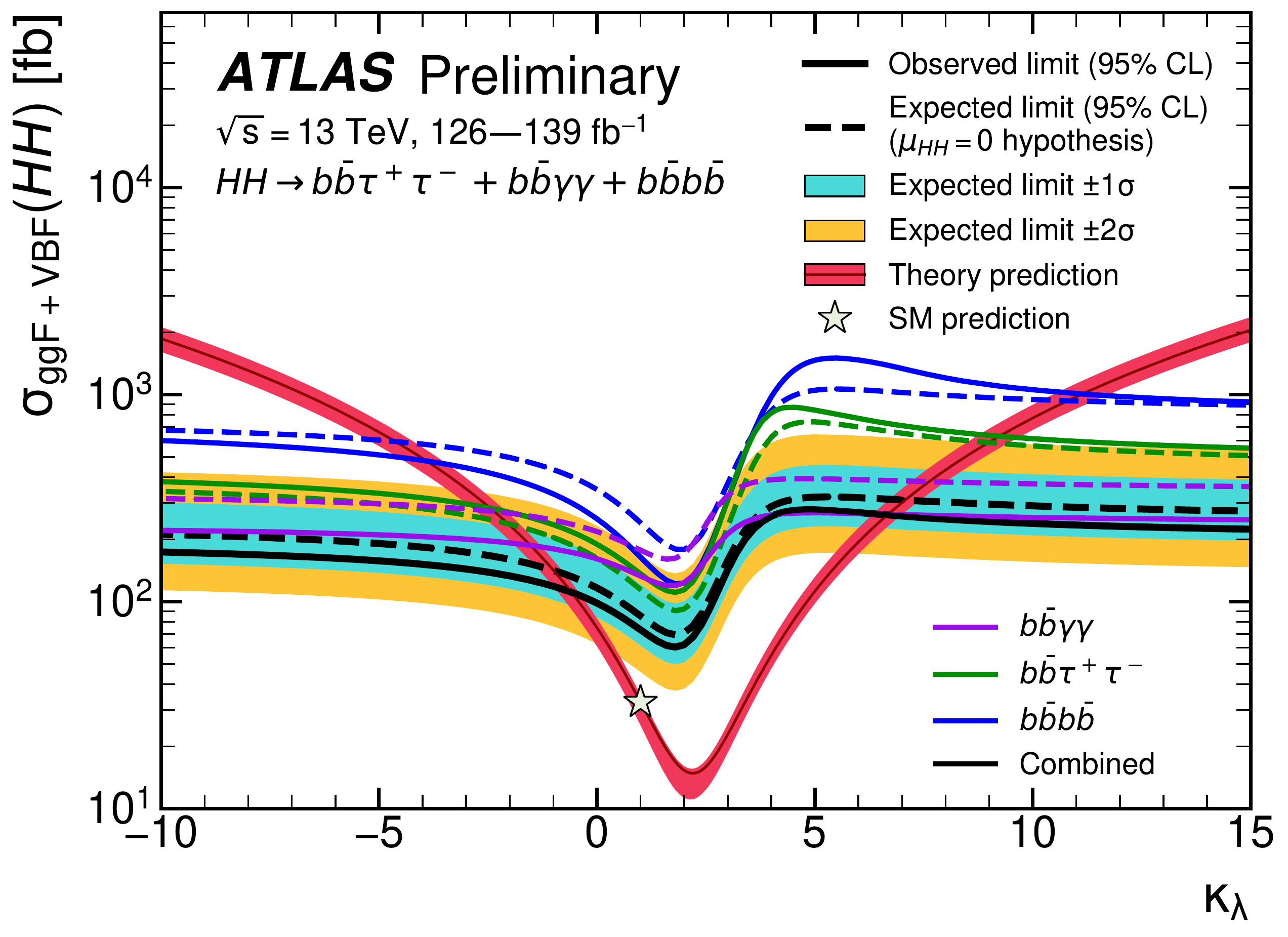}
\includegraphics[width=0.48\textwidth]{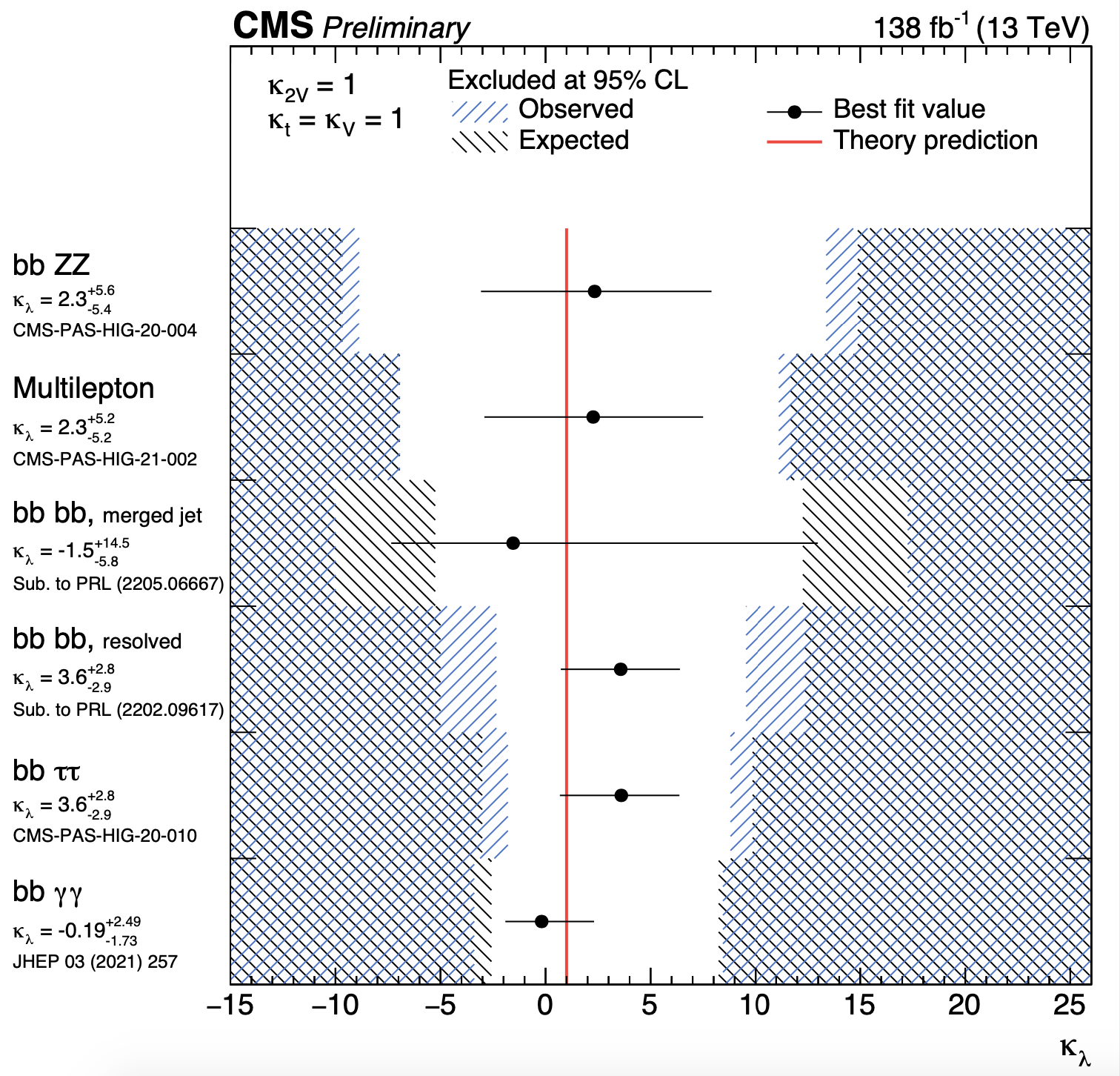}
\caption{Expected and observed 95\% CL upper limits on the \hh production cross section as a function of $\klambda$ for ATLAS (left)~\cite{ATLAS-CONF-2022-052}. 95\% confidence intervals on \klambda superimposed by the best fit value on this parameter~\cite{TwikiCMS}.}
\label{fig:comb:lambda}
\end{center}
\end{figure}

\begin{figure}[ht!]
\begin{center}
\includegraphics[width=0.44\textwidth]{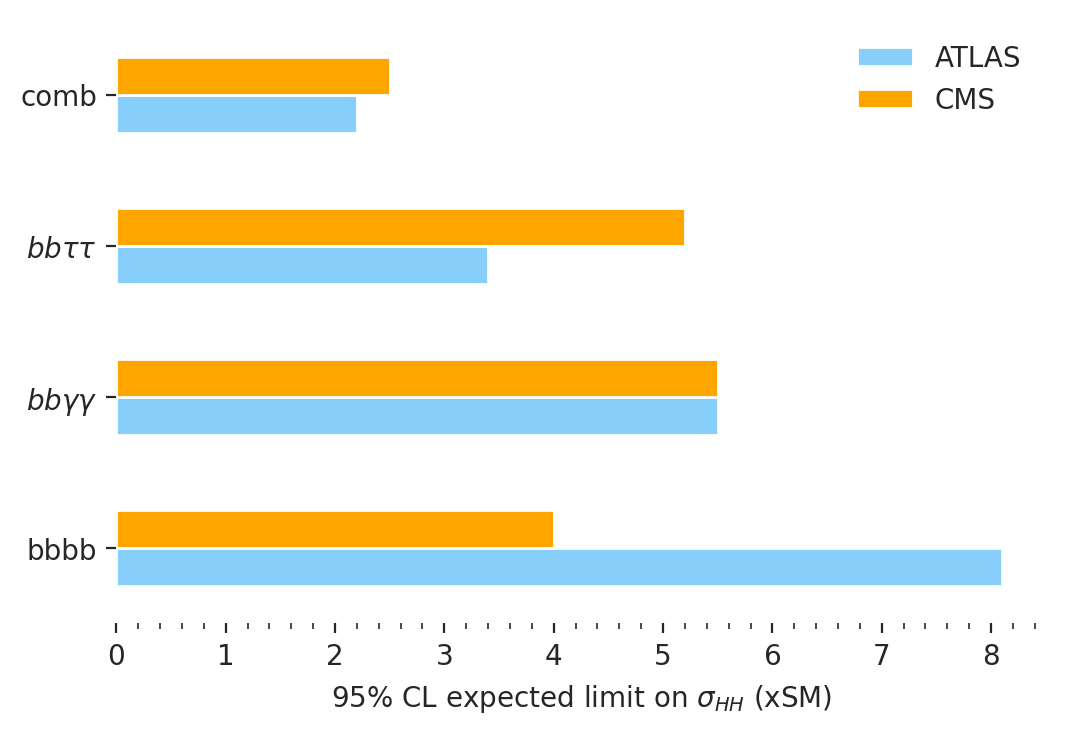}
\includegraphics[width=0.48\textwidth]{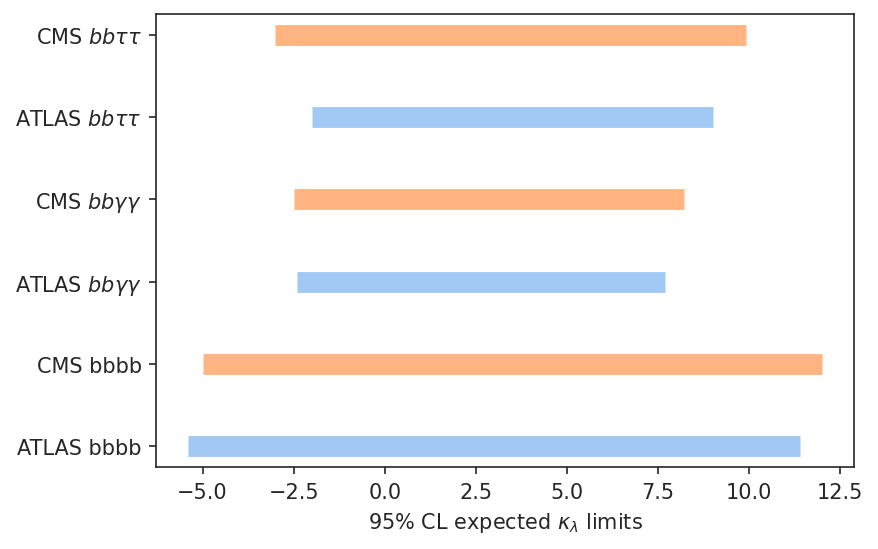}
\caption{Expected 95\% CL upper limits on the \hh production cross section (left) and $\klambda$ (right) for ATLAS and CMS. Results are based on the full Run 2 dataset corresponding to a luminosity of 126-138~$\fbi$.}
\label{fig:comb2}
\end{center}
\end{figure}

Assuming all the other couplings are set to their SM value, any modification to the self-coupling value would affect both the \hh production cross section and decay kinematics. These effects are fully simulated for each \klambda$=\lambda/\lambda_{\mathrm{SM}}$ value considered in the scan performed by the ATLAS and CMS collaborations, where 
$\lambda_{\mathrm{SM}}\equiv \mh^2 G_F/\sqrt{2}$ is the SM tri-linear Higgs coupling. Modifications to the Higgs boson decay branching fractions through one loop electroweak corrections are not considered in the analyses of ATLAS and CMS, although they can modify the results up to O(10\%). 
Figure~\ref{fig:comb:lambda} shows the upper limit on $\sigma(pp \to \hh)$ for a given value of \klambda published by  ATLAS and CMS . Figure~\ref{fig:comb2} shows the expected upper limit on $\sigma(pp \to \hh)$ and \klambda for the most significant channels analyzed by ATLAS and CMS with full Run 2 data.

The maximum value of the acceptance is obtained for $\klambda \sim 2$, where the cross section obtains its minimum. This $\klambda$ value corresponds to the maximum destructive interference between the box and the triangle contributions to the $gg\rightarrow \hsm\hsm$ sub-process, resulting in a harder $\mhh$ spectrum that increases the signal acceptance. For $|\klambda|>10$, the triangle diagram becomes dominant and the upper limit becomes symmetric in $\klambda$.
The corresponding intervals where  \klambda is observed (expected) to be constrained at 95\% CL are listed in Table ~\ref{tab:comb:summary} for the main channels.

\begin{table}[ht!]
\begin{center}
{
\begin{tabular}{|l|c|c|c|}
\hline
\hline
Final state & Collaboration & \multicolumn{2}{c}{allowed \klambda interval at 95\% CL} \\
               &             & observed & expected \\
\hline
\multirow{2}{*}{\bbbb} & ATLAS & -3.5 -- 11.3 & -5.4 -- 11.4 \\
                        & CMS   & -2.3 -- 9.4  & -5.0 -- 12.0  \\
\hline
\multirow{2}{*}{\bbtautau} & ATLAS & -2.4 -- 9.2 & -2.0 -- 9.0 \\
                              & CMS   & -1.7 -- 8.7  & -2.9 -- 9.8  \\
\hline
\multirow{2}{*}{\bbyy} & ATLAS & -1.6 -- 6.7 & -2.4 -- 7.7 \\
                                  & CMS   & -3.3 -- 8.5 & -2.5 -- 8.2 \\

\hline
\multirow{2}{*}{comb} & ATLAS & -0.6 -- 6.6 & -1.0 -- 7.1 \\
                                  & CMS   & -1.2 -- 6.8 & -0.9 -- 7.1 \\
 
                                
\hline
\hline
\end{tabular}
}
\end{center}
\vspace*{-0.5cm}
\caption{\label{tab:comb:summary} The The observed and expected 95\% CL intervals on $\kappa_{\lambda}$ for the most sensitive individual final states analyzed for non-resonant $hh$ production at 13 TeV with about 126-139~fb$^{-1}$ under the assumption of no $hh$ production. All other Higgs boson couplings are set to their SM values. Constraints derived under different statistical assumptions are also available in~\cite{CMS:2020tkr,hig-21-002,CMS:2022cpr,ATLAS-CONF-2021-052,ATLAS-CONF-2022-035,ATLAS:2022jtk,CMS:2022dwd}. }
\end{table}


In addition to the direct determination of the Higgs self-coupling through the study of Higgs boson pair production, an indirect measurement is also possible utilizing the NLO electroweak corrections to single Higgs measurements\cite{Degrassi:2021uik}. 
We note that the uncertainties are quite different in the indirect fit from in the direct $\hsm\hsm$ measurement.
The first experimental constraint on $\klambda$ from single Higgs measurements has been determined by the ATLAS experiment~\cite{ATLAS-CONF-2022-052}, by fitting data from single Higgs boson analyses taking into account the NLO $\klambda$ dependence of the cross sections of the ggF, VBF, $Vh$ and $\ttH$ production modes and the $\gamma\gamma$, $WW$, $ZZ$, $\tau\tau$ and $\bb$ decay modes, including differential information with STXS. These single Higgs analyses use data collected at 13 TeV with an integrated luminosity of 126-138~$\fbi$.
A likelihood fit is performed to constrain the value of the Higgs boson self-coupling $\klambda$, while all other Higgs boson couplings are set to their SM values. Thus assuming the new physics modifies only the Higgs boson self-coupling, the constraints on $\klambda$ derived through the combination of single Higgs measurements can be directly compared to the constraints set by double Higgs production measurements. The 95\% CL allowed interval for $\klambda$ from single Higgs production is $-4.0 < \klambda < 10.3$ (observed) and $-5.2 < \klambda < 11.5$ (expected). This interval is competitive with the one obtained from the direct \hh searches using an integrated luminosity up to 138~$\fbi$, which is  $-0.6 < \klambda < 6.6$ (observed) and  $-2.1 < \klambda < 7.8$ (expected)~\cite{ATLAS-CONF-2022-052}.

The sensitivity on $\klambda$ derived from single Higgs processes in an exclusive fit is comparable to those from \hh direct searches, but the constraints become significantly weaker when non-Standard Model like Higgs couplings are allowed in the indirect fit~\cite{ATLAS-CONF-2022-052}.
A  combination of single Higgs analyses and double Higgs analyses (\bbbb, \bbyy, \bbtautau with an integrated luminosity of up to 138 $\fbi$) has been performed by ATLAS under the assumption that new physics affects only the Higgs boson self-coupling, excluding values outside the interval -0.4 $< \klambda < $6.3 at 95\% CL
while the expected excluded range assuming the SM predictions is -1.9$< \klambda < $ 7.5~\cite{ATLAS-CONF-2022-052}.
A preliminary CMS result~\cite{CMS:2020gsy}, based on single analyses using part of the Run 2 dataset is available. Similarly to the ATLAS combination, all the most sensitive decay modes were included: $\gamma\gamma$, $WW$, $ZZ$, $\tau\tau$, \bb and $\mu\mu$. Most of the results are based on the dataset collected in 2016, with the exception of $ZZ$ which exploits the full Run 2 data sample (138~$\fbi$). The 95\% CL interval, assuming all other couplings fixed to their SM values, is observed to be $-3.5 < \klambda < 14.5$. 


However, the sensitivity to $\klambda$ from double Higgs measurements is reduced if the coupling to the top quark ($\kappa_t$) is left free to float, due to a $\kappa_t^4$ dependence of the total $pp \to \hh$ cross section~\cite{CMS:2020tkr}. Therefore a determination of $\klambda$ which would take into account beyond the Standard Model contributions affecting $\kappa_t$, would be possible only through a simultaneous analysis of both single and double Higgs measurements. As the experimental sensitivity increases, the addition of more differential information, in particular for $\ttH$ and ggF, would allow for a more general EFT interpretation of these measurements.

%% file: Tex/lept_sigs.tex
The collider scenarios studied by the Energy Frontier working group are given in Table \ref{tab:efscen}.

\begin{table}[h]
\begin{center}
\begin{tabular}[c]{||l l|c|c|c||}
\hline
 \hline
Collider	&	Type	&	$\sqrt{s}$	&	$\mathcal{P} [\%]$	&	$\mathcal{L}_{\rm int}$	\\
	&		&		&	$e^-/e^+$	&	${\rm ab}^{-1}$	/IP\\ \hline\hline
HL-LHC	&	pp	&	14 TeV	&		&	3 	\\ 
\hline
ILC and C$^3$	&	ee	&	250 GeV	&	$\pm80/\pm30$	&	2	\\
c.o.m almost &		&	350 GeV	&	$\pm80/\pm30$	&	0.2	\\
similar	&		&	500$^{\ast}$
GeV	&	$\pm80/\pm30$	&	4	\\
	&		&	1 TeV	&	$\pm80/\pm20$	&	8	\\
	\hline
CLIC	&	ee	&	380 GeV	&	$\pm80/0$	&	1	\\
\hline
CEPC	&	ee	&	$M_Z$	&		&	60	\\
	&		&	2$M_W$	&		&	3.6	\\
	&		&	240 GeV	&		&	20	\\ 
	&		&	360 GeV	&		&	1 \\
	\hline
FCC-ee	&	ee	&	$M_Z$	&		&	150	\\
	&		&	2$M_W$	&		&	10	\\
	&		&	240 GeV	&
	&	5	\\
	&		&	2 $M_{top}$	&		&	1.5	\\
\hline
muon-collider (higgs)	&	$\mu\mu$	&	125 GeV &		&	0.02\\		
\hline \hline
\end{tabular}
\label{tab:efscen}
\end{center}
\caption{Benchmark scenarios for Snowmass 2021 Higgs factory studies, ($^{\ast}$) for C$^3$ 550 GeV is being considered instead of 500 GeV. }
\end{table}

Before detailing the specific precision achievable at future colliders, it is useful to review the new production mechanisms available.  Clearly, production at the HL-LHC will be through the same mechanisms as given in the previous section, and this holds for FCC-hh as well. The obvious strengths for both the HL-LHC and FCC-hh programs are increased energy for multi-Higgs productions and differential measurements, as well as the largest number of single Higgs bosons that can be produced at any proposed collider.  The downside is the large SM backgrounds, primarily QCD, which is unavoidable at hadron colliders. 

Lepton colliders provide alternative methods of production compared to hadron colliders {\em and} reduced SM backgrounds.  The production mechanism at lepton colliders often depends on the energy scale and the type of lepton collider.  Colliders that are ready to be built in the near future (5-10~years) with minimal R\&D, are the $e^+e^-$ Higgs factories, of which, 5 are currently proposed ---ILC~\cite{ILCInternationalDevelopmentTeam:2022izu}, C$^{3}$\cite{Dasu:2022nux}, CEPC~\cite{CEPCPhysicsStudyGroup:2022uwl}, CLIC~\cite{Robson:2018zje}, and FCC-ee~\cite{Bernardi:2022hny}.  All of these proposals start with low-energy Higgs production stages which are dominated by $\ee\rightarrow Z\hsm$.  The s-channel resonant production of the Higgs could also be available at a circular $\ee\rightarrow \hsm$, given increased luminosity, or at a $125$ GeV muon collider $\mu^+\mu^-\rightarrow \hsm$, due to the larger value of the muon Yukawa, $y_\mu$,
compared to $y_e$. Both of these ``low energy" options are much further in the future and not part of the first stage of the Higgs factory plans.

For C$^3$, 550 GeV is being considered instead of 500 GeV, which would mainly affect the top-Higgs prediction, as shown in Figure~\ref{fig:coups_tth}. In this report, we assume the same physics reach for ILC and C$^3$.
An additional possibility is a photon-photon collider, XCC, at the Higgs resonance (125 GeV) based on X-ray FEL beams \cite{Barklow:2022vkl}.  A part of the program produces  $e \gamma$ collisions at $\sqrt{S}$=140 GeV, to observe the process $e \gamma\rightarrow 
e \hsm$.  This gives tagged Higgs decays, similar to  $e^+e^-\rightarrow Z\hsm$, allowing one to absolutely normalize the Higgs couplings.

\begin{figure}[h]
\begin{minipage}{6in}
  \centering
  $\vcenter{\hbox{\includegraphics[height=2.25in]{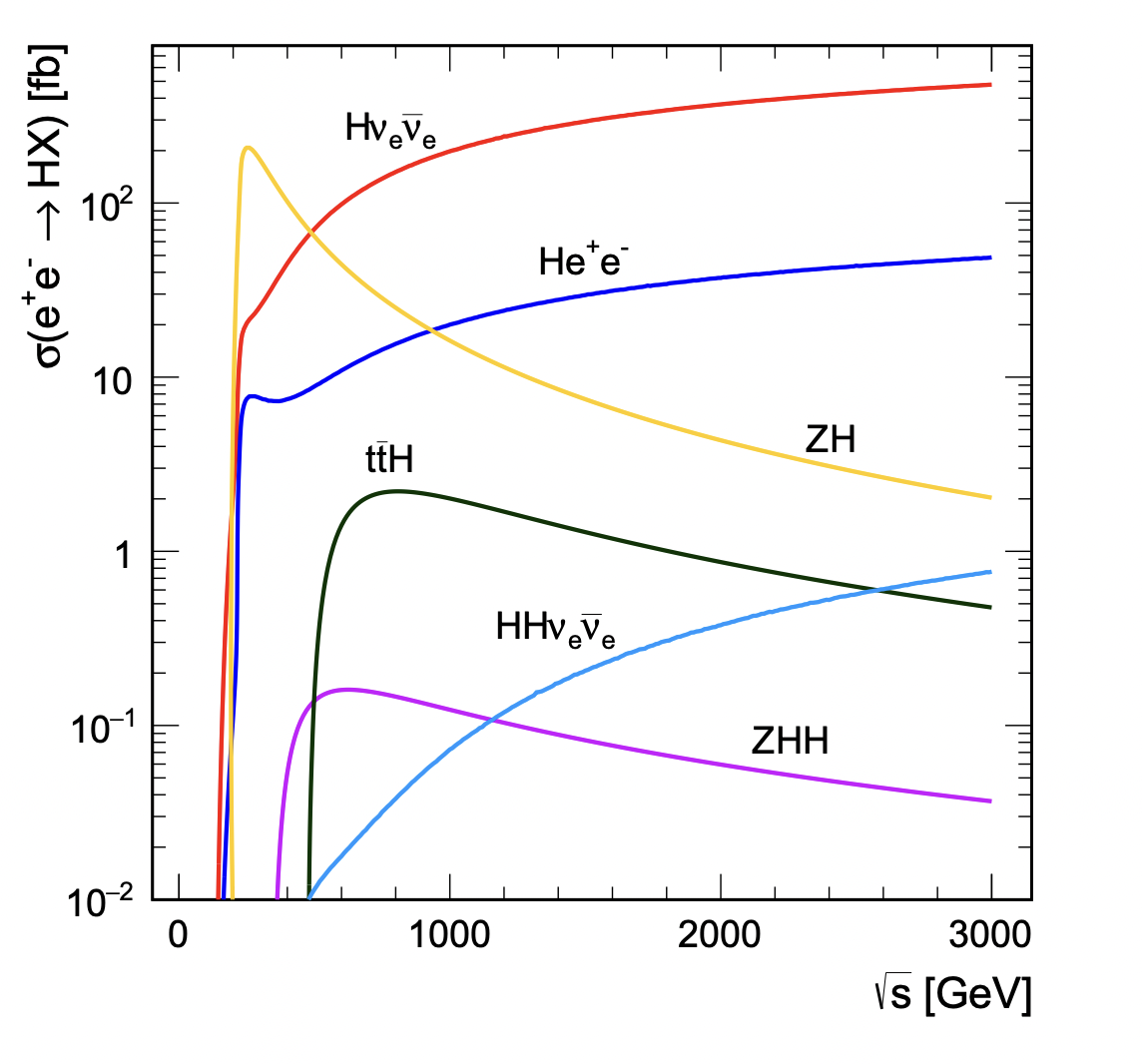}}}$
  \hspace*{.2in}
  $\vcenter{\hbox{\includegraphics[height=1.75in]{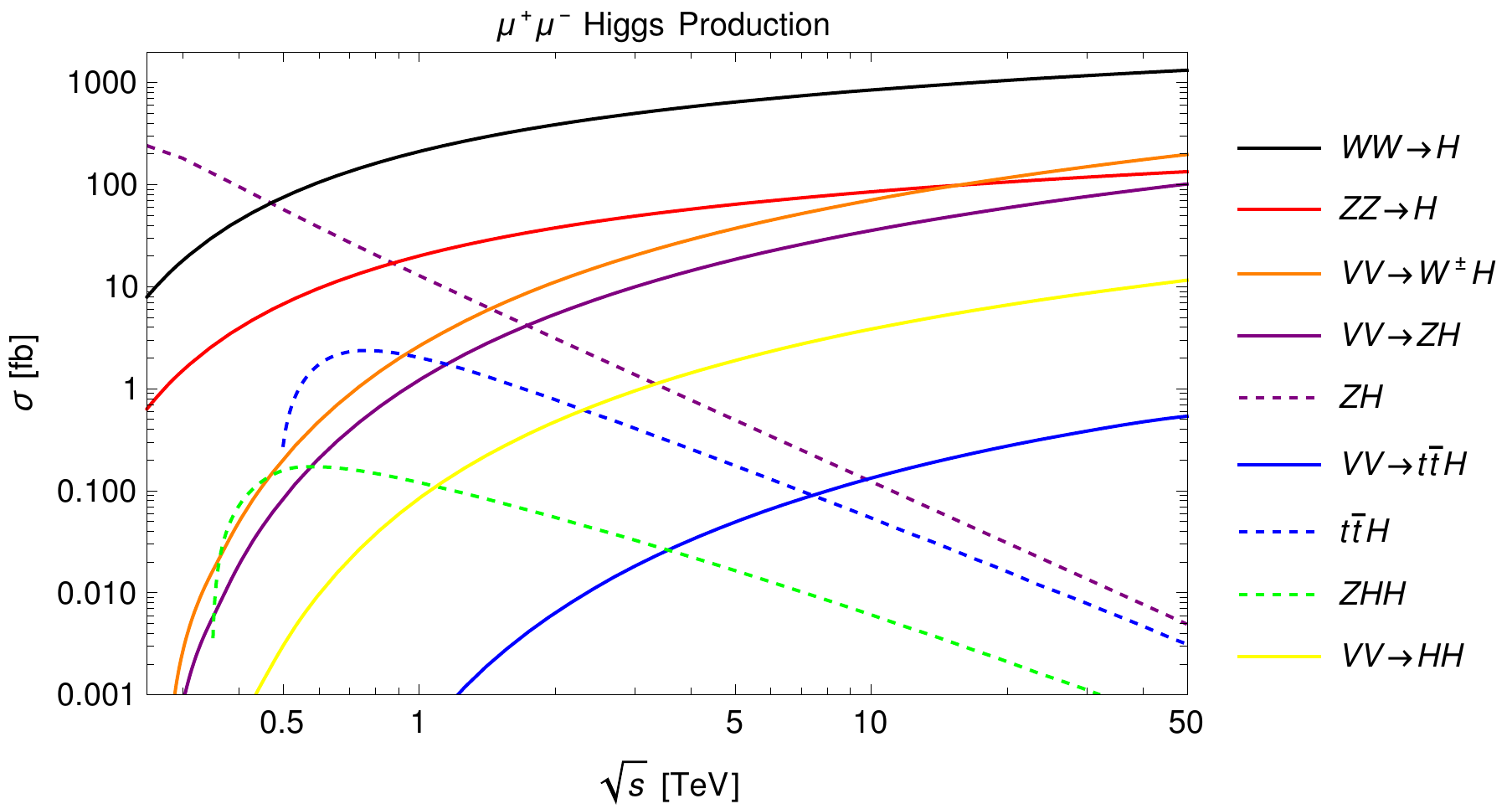}}}$
\end{minipage}
\par
\caption{
Left, Higgs boson production rates for $\ee$ interactions as a function of collision energy~\cite{CLICdp:2018cto}. Right: production rates for $\mu^+\mu^-$ Higgs production as a function of collision energy, where the dashed curves correspond to annihilation cross sections.}
\label{fig:lepxsecs}
\end{figure}

For the second stage of linear $\ee$ colliders running at higher energies, the vector boson fusion processes takes over, e.g. $\ee\rightarrow \nu {\overline{\nu}}\hsm$, and $\ee\rightarrow \ee\hsm$.  For high energy muon colliders, $\mu^+\mu^-\rightarrow \nu {\overline{\nu}}\hsm$ and $\mu^+\mu^-\rightarrow \mu^+\mu^-\hsm$ are always the primary production mechanisms, and above 7 TeV VBF even dominates for $t\bar{t}\hsm$ production.  The cross-section dependence for lepton colliders is illustrated in Fig~\ref{fig:lepxsecs} where the range of $\ee$ colliders is shown on the left and for muon colliders on the right.

\begin{figure}[h]
\begin{centering}
\includegraphics[scale=0.4]{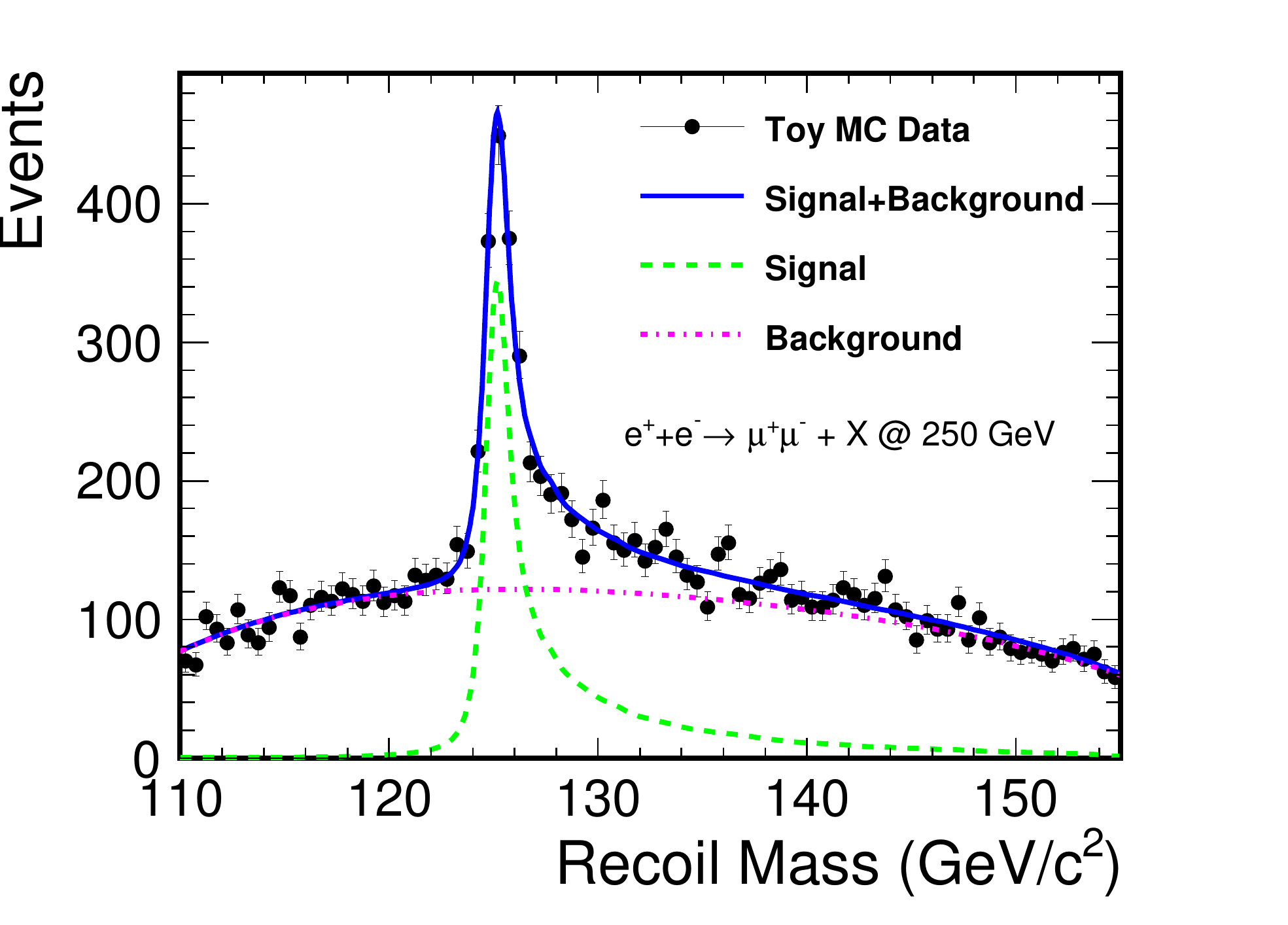}
\par\end{centering}
\caption{
Example of the recoil mass for 250~GeV $\ee$ collision energy at ILC~\cite{ILCInternationalDevelopmentTeam:2022izu}.}
\label{fig:ilcsigs}
\end{figure}

Lower energy Higgs factories offer advantages in terms of the absolute measurements of production cross sections, whether a $\mathcal{O}(250$~GeV) $\ee$ collider or an eventual Higgs resonance collider.  For example, at a 250 GeV $\ee$ collider, the dominant production mechanism is Z$\hsm$. The total Z$\hsm$ cross section can be extracted independently of the Higgs boson's detailed properties by counting events with an identified Z boson, and for which the mass recoiling against the Z clusters around the Higgs boson mass. This model-independent measurement of the $\hsm ZZ$ coupling is unique to $\ee$ colliders. By using the recoil mass distribution (shown in Figure ~\ref{fig:ilcsigs}), the Z$\hsm$ total cross section can be measured from the area of the signal peak to $\sim \mathcal{O}(1\%)$ precision. At higher center of mass energies for $\ee$, $\mu^+\mu^-$ and $p\mhyphen p$, there are a larger number of Higgs bosons produced, new types of observables, new production modes with top quarks, and multi-Higgs bosons which will be further discussed in the rest of this section.  

%% file: Tex/mass.tex
The Higgs mass is a fundamental parameter of theory and has implications for our understanding of the meta-stability of the universe.  In addition, it is a predicted quantity in certain BSM models, such as the MSSM\cite{Slavich:2020zjv}.
The ILC projects a measurement of $\mh$ from the position of the recoil mass peak in $\ee\rightarrow Z \hsm$ with a precision of $\pm 14$~MeV\cite{Yan_2016}.  Similarly, the FCC-ee projects 
a mass measurement of 6-9~MeV statistical error with the potential to improve this measurement further by including the $Z\rightarrow \ee$ decay. This would lead to an ultimate precision of $\delta \mh \sim$ 4~MeV with FCC-ee.  At the HL-LHC 
CMS projects a measurement $\Delta \mh\sim 30$~MeV in the $\hsm\rightarrow ZZ\rightarrow 4l$ and $\hsm \rightarrow \gamma \gamma$ channels\cite{ATL-PHYS-PUB-2022-018},
assuming detector upgrades give a $25\%$ improvement in the $4\mu$
resolution and a $17\% $ increase in the $4\mu$ and $4e$ channels.

It was long thought that it was impossible to measure the Higgs width at the LHC, due to the smallness of the SM Higgs width.  However, it was realized in Refs.~\cite{Caola:2013yja,Kauer:2012hd}  that
 the interference of the off-shell Higgs boson with the full amplitude in the
 $ZZ\rightarrow 4l$ channel is sensitive to the Higgs width.  By comparing measurements above the Higgs resonance and on the Higgs resonance, a measurable sensitivity
 to the width can be observed and CMS has recently used
 this technique to obtain the first measurement
 $\Gamma_\hsm=3.2^{+2.4}_{-1.7} ~MeV$\cite{https://doi.org/10.48550/arxiv.2202.06923}. The HL-LHC projects a combined ATLAS-CMS width measurement of $\Gamma_\hsm=4.1^{+0.7}_{-0.8}$~MeV,
 corresponding to roughly a $17\%$ accuracy using this technique\cite{ATL-PHYS-PUB-2022-018}.   If non-Standard Model Higgs interactions exist, the resulting limits on the width are altered.  

\begin{figure}[h]
\begin{centering}
\includegraphics[scale=0.5]{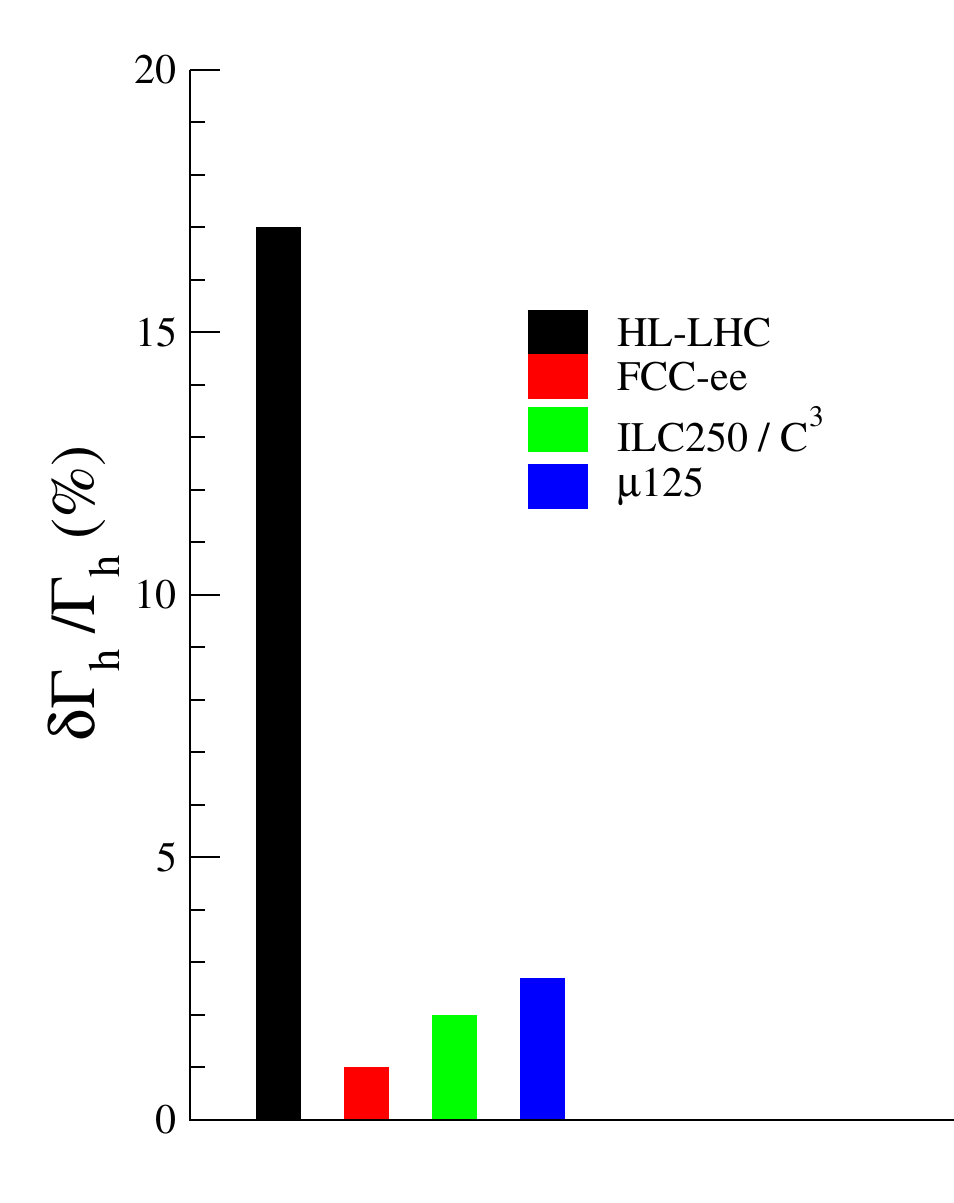}
\par\end{centering}
\caption{Summary of width measurements at future colliders.  The LHC measurement uses interference effects in the off-shell decay of $ZZ\rightarrow 4l$, while the lepton colliders use the recoil mass in $e^+e^-\rightarrow Z\hsm$.}
\label{fig:width}
\end{figure}

Lepton colliders offer the opportunity to obtain a fit to the Higgs width using the Z$\hsm$ kinematic distributions.  The fully reconstructed Z boson
in the final state along with the well-determined 4-momenta of the initial state leptons in the $Z\hsm$ process allows for a clean determination of the
Higgs boson kinematics regardless of the Higgs decay channel. The full FCC-ee program (combined with HL-LHC) allows for a $1\%$ measurement of the Higgs
width~\cite{Bernardi:2022hny}.  Using a SMEFT fit, the ILC finds similar results for the full program, but with just the initial center of mass energy 250~GeV run, a $2\%$ measurement on the total width is projected~\cite{ILCInternationalDevelopmentTeam:2022izu}. 
A muon collider running at $\sqrt{S}=125$~GeV can obtain a model-independent measurement of the Higgs total width at the 68$\%$ level of $2.7\% ~(1.7\%)$
with $5~ \fbi (20~ \fbi)$  by using a line-shape measurement~\cite{MuonCollider:2022xlm}.  A high-energy muon collider should obtain a similar order of
magnitude precision using the indirect methods employed at the LHC with the same theoretical assumptions.  The width measurements at future colliders are summarized in Fig.
\ref{fig:width}
It is important to note that the width measurements shown in Table \ref{tab:higgs_sig_comb} are obtained assuming that there is no contribution from BSM physics and no unobserved decay channels and,
therefore, do not represent model-independent measurements of the width.  .

%% file: Tex/coups.tex
\begin{figure}
\begin{centering}
\includegraphics[scale=0.35]{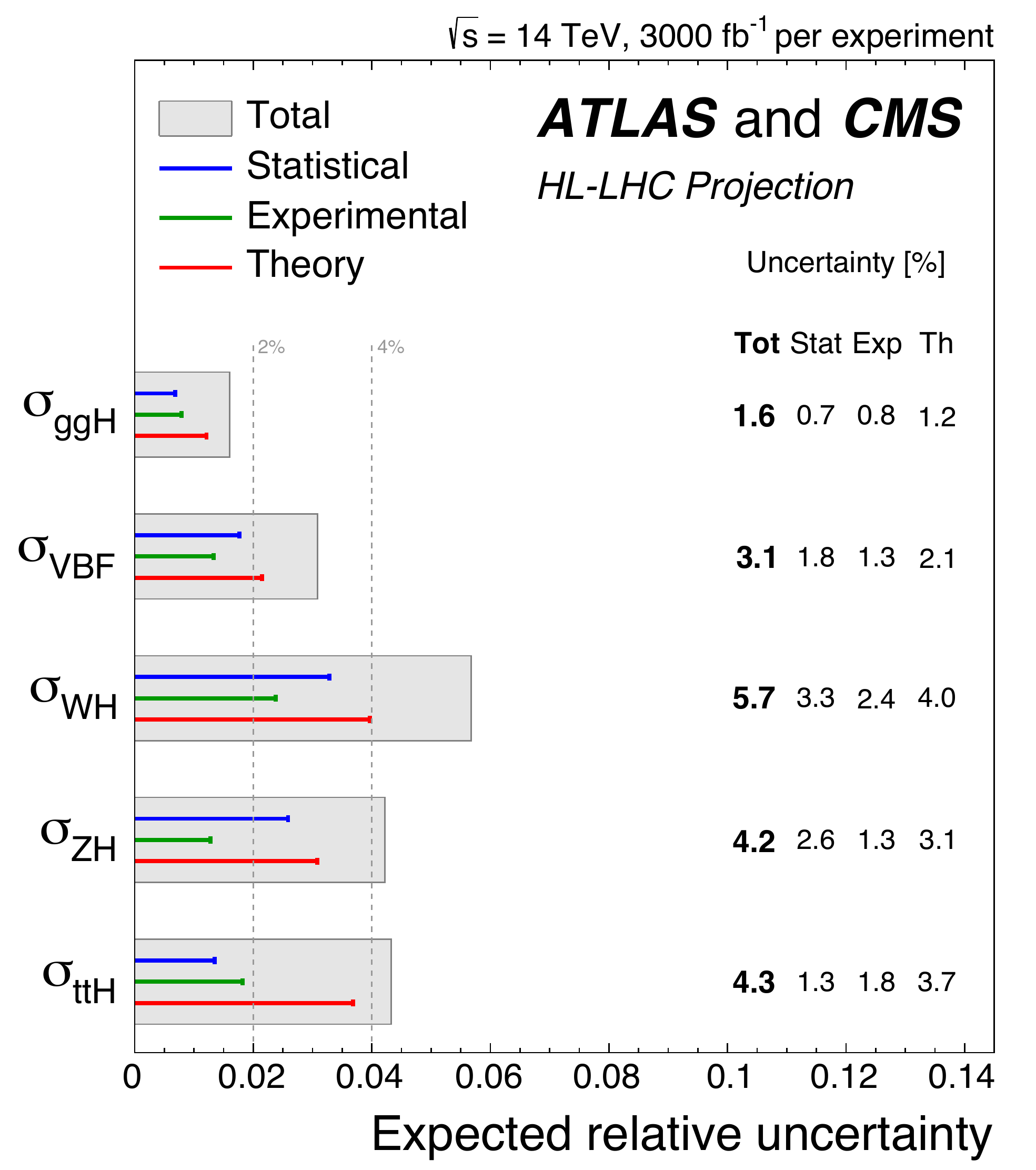} 
\includegraphics[scale=0.4]{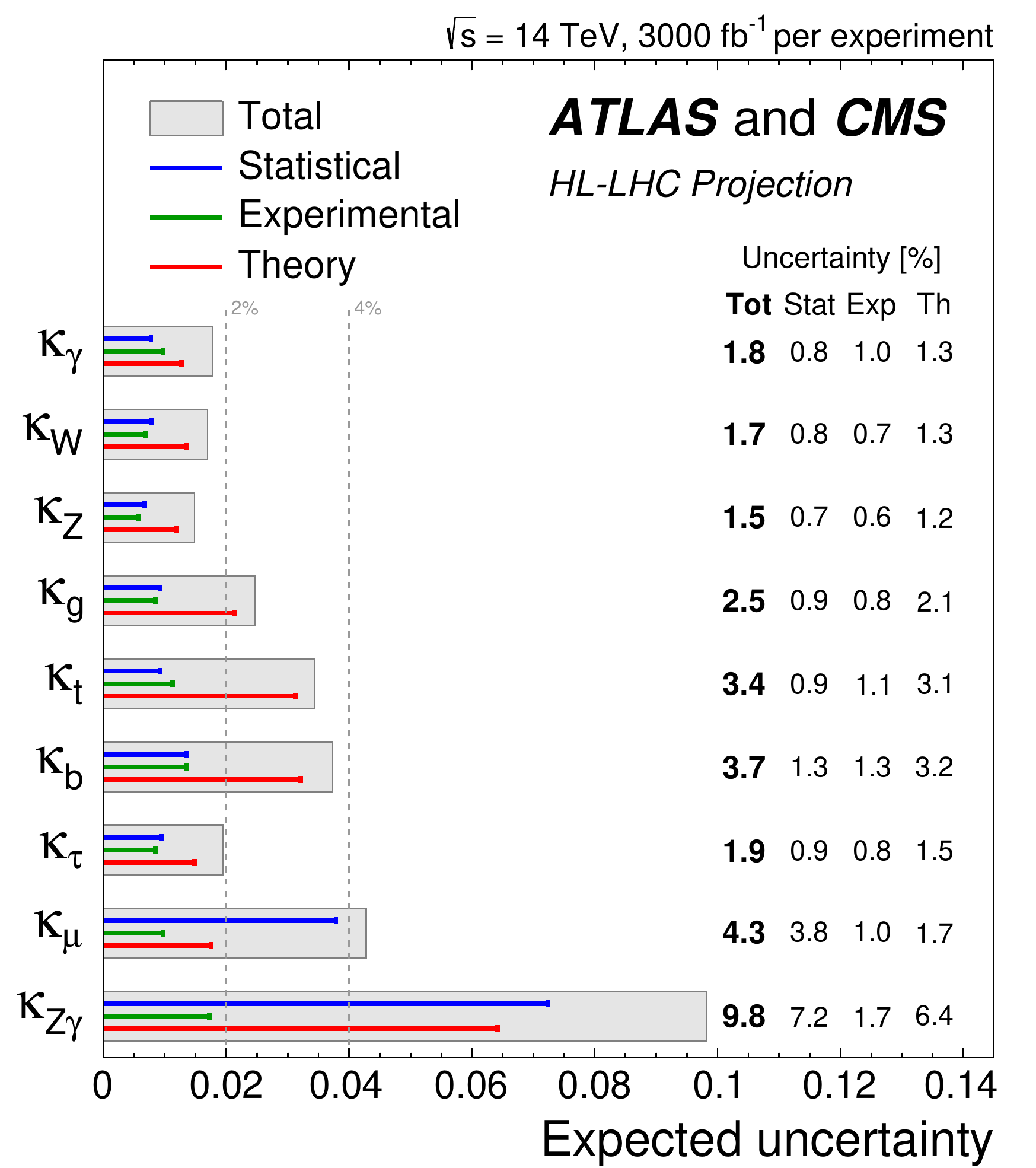} 
\par\end{centering}
\caption{Projections for HL-LHC Higgs coupling measurements. \cite{ATL-PHYS-PUB-2022-018}.}
\label{fig:HL-snow}
\end{figure}

The planned High Luminosity era of the LHC (HL-LHC), starting in 2029\footnote{This refers to the updated schedule presented in January 2022~\cite{LHCschedule}} will extend the LHC  Higgs dataset by a factor of $\mathcal{O}(10)$, and produce about 170 million Higgs bosons and 120 thousand Higgs boson pairs. This would allow an increase in the precision for most of the Higgs boson coupling measurements. The conditions of the data-taking will present challenges of higher data rates, larger radiation doses, and unprecedented levels of pileup - with about 200 collisions on average per bunch crossing - and both the ATLAS and CMS experiments are going through major upgrades to ensure robust performance.

HL-LHC will dramatically expand the physics reach for Higgs physics. Current projections are based on the Run 2 results and some basic assumptions that systematic uncertainties will scale with luminosity and that improved reconstruction and analysis techniques will be able to mitigate pileup effects.  Studies based on the 3000$~\fbi$ HL-LHC dataset estimate that we could achieve $\mathcal{O}(2-4\%)$ precision on the couplings to W, Z and third generation fermions.  But the couplings to first and second generation quarks will still not be accessible at the LHC and the self-coupling will only be probed with $\mathcal{O}$(50\%) precision. The projected sensitivity to the muon coupling has of $\mathcal{O}$(8.5-7\%) uncertainty on the signal strength modifier depending on the assumptions for the systematic uncertainties~\cite{ATL-PHYS-PUB-2022-018}.  We will, however,  be able to exclude the hypothesis corresponding to the absence of self-coupling at the 95\% CL in these projections for HL-LHC, but not to test the SM prediction.  

It is clear that to gain a complete and precise understanding of the Higgs boson properties and measure new physics effects we will need to go beyond the  LHC and HL-LHC\cite{deBlas:2019rxi}. Gaining access to the very high energy regime could potentially enable the production of on-shell new physics particles, if they exist,  that are related to new forces. If the new particles are too heavy to be produced at the HL-LHC, precision measurements of the Higgs boson couplings will give a hint about modifications of the SM. Precision of $\mathcal{O}$(few~\%) level or below requires collider experiments designed for high precision. The complementarity between leptonic and hadronic initial states will eventually lead to the most precise and comprehensive understanding of the Higgs couplings to gather insight on where new physics lies. 
Future machines are charged with the challenging tasks of
improving the HL-LHC measurements of Higgs couplings, of 
testing the SM predictions of  measurements of the Higgs boson Yukawa couplings to light flavor quarks and measuring the Higgs self-coupling. The latter demands access to high energy center-of-mass collisions to benefit from  the larger dataset of $\hsm\hsm$ pairs.

The projections for the extraction of the Higgs boson production cross sections at the HL-LHC with a combined CMS and ATLAS analysis are shown in
Figure~\ref{fig:HL-snow}\cite{ATL-PHYS-PUB-2022-018}.
The expected precision with $3~\abi$ is $\sim 1.6\%$ for ggF and rises to $\sim 5.7\%$ for $W\hsm$ production, while the major decay channels can be determined with an accuracy of a few percent: 
$\gamma \gamma\sim 2.6\%$, 
$ZZ\sim 2.9\%$, 
$WW\sim 2.8\%$, 
$\tau^+\tau^-\sim 2.9\%$,
and 
$b {\overline{b}} \sim 4.4\%$. 
These projections can be used to determine the Higgs boson couplings to fermions and gauge bosons that is a fundamental goal of all future colliders. ATLAS and CMS have significantly improved the precision of the predictions for couplings based on the full Run 2 analyses, and dedicated HL-LHC simulations as shown in
Figure~\ref{fig:HL-snow}\cite{ATL-PHYS-PUB-2022-018}. As discussed in Section~\ref{sec:theorynow}, the interpretation of the measurements is dominated by theory uncertainty.  We note that the theory uncertainties assumed in Figure~\ref{fig:HL-snow} represent a significant improvement over the current theory uncertainties shown in Figure ~\ref{fig:gguncert}.

\begin{figure}
\begin{centering}
\includegraphics[scale=0.4]{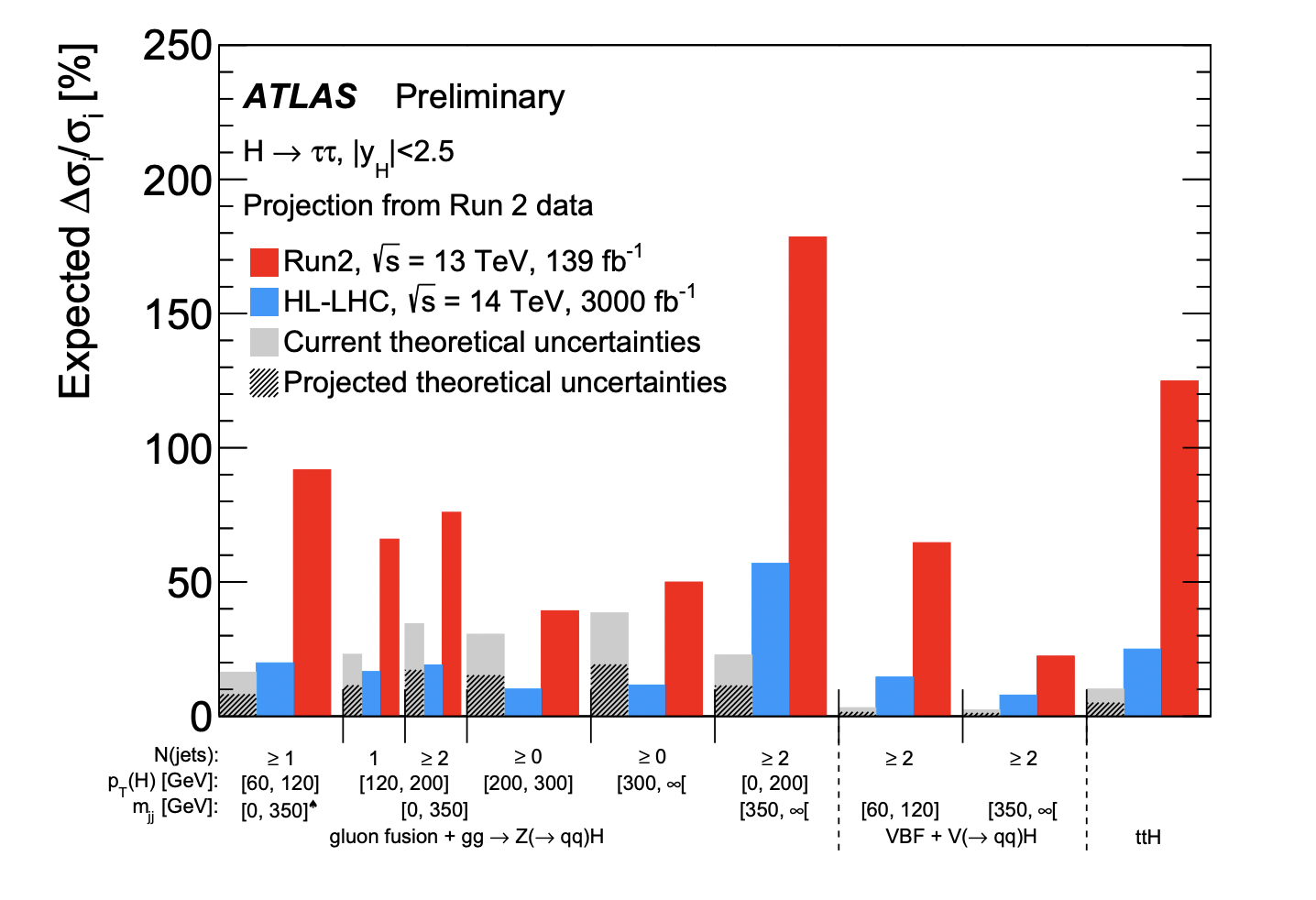} 
\includegraphics[scale=0.5]{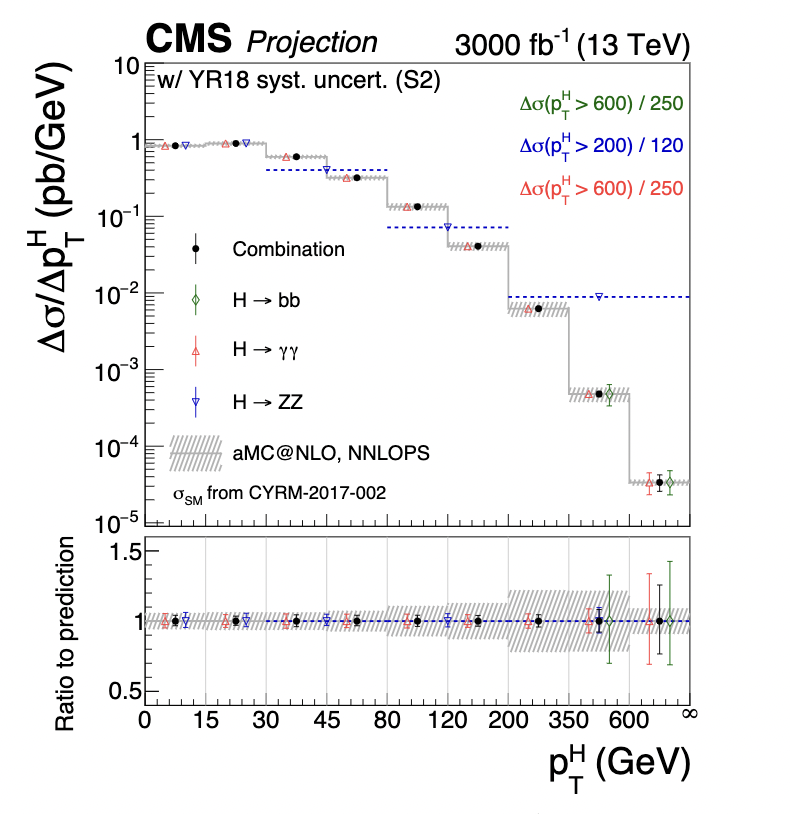} 
\par\end{centering}
\caption{Left, Run 2 (red) and projected HL-LHC (blue) expected precision of the $\htt$ measurement scaled to the cross-section expectation value in various STXS bins labelled i~\cite{ATL-PHYS-PUB-2022-018}.
Right, Projected differential cross sections for ggF $\hyy$, $\hZZ$ and $\hbb$ decay channels at an integrated luminosity of 3000 $\fbi$~\cite{Cepeda:2019klc}.}
\label{fig:highmomentum}
\end{figure}

With the increased luminosity at HL-LHC, differential cross sections will play an important role in determining Higgs properties. 
At the HL-LHC, the inclusive $\htt$ cross-section measurement is projected to have a precision of 5\%. The projected precision of the four dominant production mode measurements are 11\%, 7\%, 14\% and 24\% for ggF, VBF, V\hsm, \ttH~respectively. Theoretical uncertainties on the signal prediction dominate the uncertainty for the ggF and VBF projections, while in the V\hsm~ projection there are similar contributions from experimental uncertainties and uncertainties coming from the data sample size. For the \ttH~projection, the largest impact is from the various experimental uncertainties, although closely followed by theoretical uncertainties on the signal prediction and from the data sample size. In all cases systematic uncertainties have a larger contribution than the statistical ones from the data sample size. 
In the simplified template cross section (STXS) framework, the most sensitive projected measurements are the VBF + V\hsm~ cross-section in events with at least two jets and a di-jet invariant mass of at least 350 GeV (VBF topology), with an uncertainty of 7\%, and the ggF + ggZ\hsm~ cross-section in events with a Higgs boson transverse momentum between 200 and 300 GeV, with an uncertainty of 10\%, and above 300 GeV, with an uncertainty of 11\%\cite{ATL-PHYS-PUB-2022-018}. 
The differential rate for the $\hbb$~ final state will also be important at the HL-LHC for the momentum range $> 300$ GeV. Figure~\ref{fig:highmomentum}-left shows a comparison to the Run 2 $\htt$ measurements and to the current and projected theoretical uncertainties for the STXS study of differential rates at the HL-LHC and the RHS of  Figure~\ref{fig:highmomentum} shows the projected sensitivity for the combined ggF cross-section measurement with the $\hyy$, $\hZZ$ and $\hbb$ decay channels, based on a preliminary Run 2 analysis with 35.9 $\fbi$\cite{Cepeda:2019klc}. With respect to the uncertainties affecting the measurement based on an integrated luminosity of 35.9 $\fbi$, the uncertainties at 3000 $\fbi$  in the higher momentum region are about a factor of ten smaller. This is expected, as the uncertainties in this region remain statistically dominated.

With the basic Higgs mechanism for mass generation now demonstrated, the next task for Higgs studies is to search for the influence of new interactions that can explain why the Higgs field has the properties required in the SM. If the new particles associated with these interactions are too heavy to be produced at the HL-LHC, they can still cause measurable deviations in the pattern of Higgs boson couplings from the SM predictions.

An $\ee$ Higgs factory will lead to insight on the Higgs Yukawa couplings at the next level beyond the third generation fermions and with more precision than the HL-LHC. Indeed, at an $e^+e^-$ Higgs factory the precision can be enhanced by the availability of precise calculations combined with much more democratic production rates: Higgs production is roughly of the same order as other processes in $e^+e^-$ collisions, whereas the LHC must trigger and select Higgs events among backgrounds that are multiple orders of magnitude larger.  In the SM, the Higgs Yukawa couplings are exactly proportional to mass and this clearly makes the observation of the Higgs couplings to the first and second generation fermions difficult.   Tagging of charm and strange quarks, as previously demonstrated at SLC/LEP, gives effective probes for advancing this program. The cleaner $\ee$ environment aided by beam polarization would be a sensitive probe that could reveal more subtle phenomena~\cite{Dasu:2022nux}. 

Studies for the five current $\ee$ Higgs factory proposals---ILC, \CCC, CEPC, CLIC, and FCC-ee---demonstrate that experiments at these facilities can meet and even exceed these requirements for high precision.  Actually, despite their different strategies, all these proposals lead to very similar projected uncertainties on the Higgs boson couplings.  The higher luminosity proposed for circular $\ee$ machines is compensated by the advantages of polarization at linear colliders, yielding very similar projected sensitivity for the precision of Higgs couplings~\cite{Fujii:2018mli,Barklow:2017suo}. 
In combination with the measurement of the rate of $Z\hsm $ events with an $\hsm \rightarrow ZZ$ decay, a model-independent determination of the Higgs total width can be obtained at an $\ee$ collider. The analysis of the other Higgs decays similarly provides a set of model-independent Higgs partial width and coupling measurements~\cite{Blondel:2021ema}.

\begin{figure}
\begin{centering}
\includegraphics[scale=0.8]{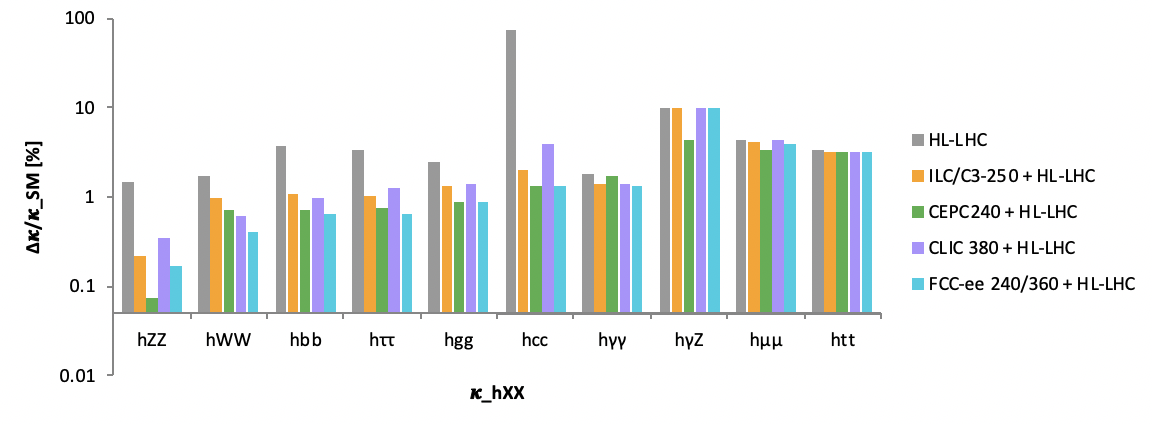} 
\par\end{centering}
\caption{Projected relative Higgs coupling measurements in $\%$  when combined with HL-LHC results. All values assume no beyond the Standard Model  decay modes. In addition, only initial stages are shown for near-term colliders: 
This corresponds to $3~ab^{-1}$ and two interaction points (IPs), ATLAS and CMS, for the HL-LHC at 14~TeV, $2~ab^{-1}$ and 1~IP at 250~GeV for ILC/C$^3$, $20~ab^{-1}$ and 2~IP at 240~GeV for CEPC, $1~ab^{-1}$ and 1~IP at 380~GeV for CLIC, and $5~ab^{-1}$ and 4~IP at 240~GeV for FCC-ee. Note that the HL-LHC $\kappa_{hcc}$ projection uses only the CMS detector and is an upper bound \cite{ATL-PHYS-PUB-2022-018}. }
\label{fig:higgs_sig_comb_initial_fig}
\end{figure}

\begin{figure}
\begin{centering}
\includegraphics[scale=0.6]{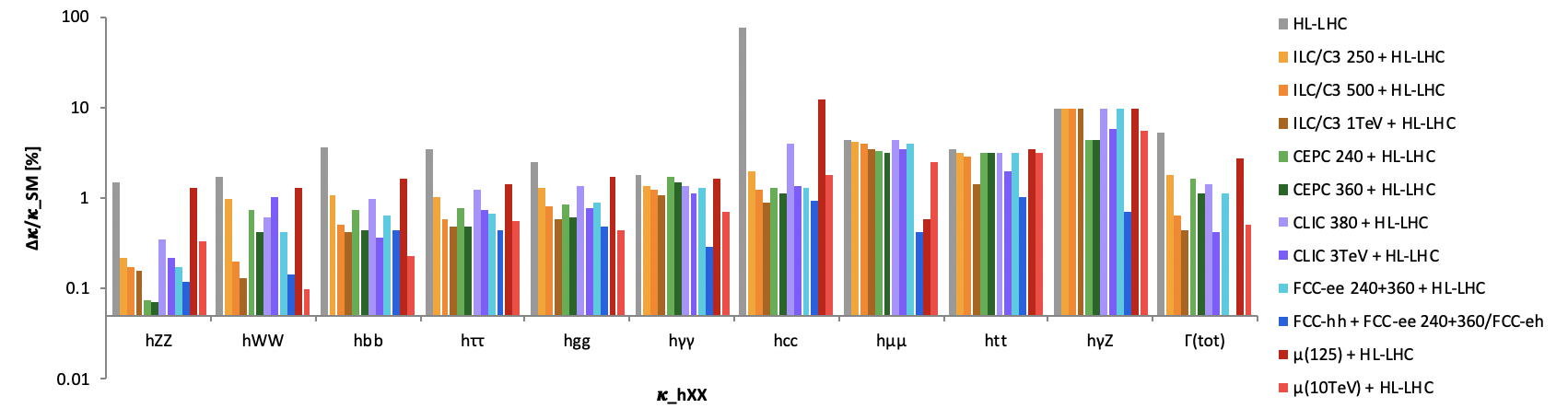}
\par\end{centering}
\caption{Relative Higgs coupling measurements in $\%$  when combined with HL-LHC results. All values assume no beyond the Standard Model  decay modes of the Higgs boson. The energies and luminosities are those defined in Table \ref{tab:efscen}.}
\label{fig:higgs_sig_comb_fig}
\end{figure}

We show the projected sensitivity for the first stages of possible lepton colliders combined with HL-LHC projections in Figure~\ref{fig:higgs_sig_comb_initial_fig}. It is clear that the dominant improvement from HL-LHC results is in the couplings to $b$'s, $\tau's$ and the $W$ and $Z$ gauge bosons. We note that since no beyond the Standard Model physics is allowed in this fit, the width measurements are just a result of summing the various channels and do not represent independent measurements. 
In Figure~\ref{fig:higgs_sig_comb_fig}, we show the potential improvements from higher energy runs of the $\ee$ colliders along with possible muon collider and FCC-hh input
and observe the significant gain in our understanding of the Higgs couplings.  The $Z\gamma$ interaction remains difficult to measure at all of these machines and the measurement of the coupling to top is not significantly improved from the HL-LHC results in the initial stages of the proposed $\ee$ machines.  These results are based on the $\kappa_0$ scenario of the ESG\cite{Cepeda:2019klc}
(combined with projections for HL-LHC results) which does not allow for beyond the Standard Model decays of the Higgs boson. 
A muon collider running on the Higgs resonance has very similar reach as the 
lepton colliders except for the $\hsm \mu \mu$ coupling which can be measured 
with $\sim {\cal{O}}(.1 \%)$ precision. 
Exact results are given in Table \ref{tab:higgs_sig_comb}, where the figure caption references the sources of the various numbers. 

\begin{table}[t]
\centering
\renewcommand{\arraystretch}{1.5}
\begin{adjustbox}{angle=90}
\begin{tabular}{||c|c|c|c|c|c|c|c|c|c|c|c|c||}
\hline\hline
Higgs Coupling &HL-LHC &  ILC250  & ILC500  &ILC1000 & FCC-ee&CEPC240  &CEPC360&CLIC380&CLIC3000&$\mu$(10TeV)& $\mu$125& FCC-hh\\
 ($\%$)& &  + HL-LHC  &+HL-LHC &+ HL-LHC & + HL-LHC &+ HL-LHC&+HL-LHC&+ HL-LHC&+HL-LHC &+ HL-LHC&+HL-LHC &+FCCee/FCCeh\\
\hline \hline 
$\hsm ZZ$               &$1.5$ &$.22$  &$.17$  &.16 &$.17$ &.074 &.072&$.34$  &$.22$ &.33 &1.3   &.12 \\
\hline
$\hsm WW$               &$1.7$ &$.98$  &$.20$  &.13 &$.41$ &.73  &.41 &$.62$  &$1$   &.1  &1.3   &.14 \\
\hline
$\hsm b {\overline{b}}$ &$3.7$ &$1.06$ &$.50$  &.41 &$.64$ &.73  &.44 &$.98$  &$.36$ &.23 &1.6   &.43 \\
\hline
$\hsm \tau^+\tau^-$     &$3.4$ &$1.03$ &$.58$  &.48 &$.66$ &.77  &.49 &$1.26$ &$.74$ &.55 &1.4   &.44\\
\hline 
$\hsm gg$.              &$2.5$ &$1.32$ &$.82$  &.59 &$.89$ &.86  &.61 &$1.36$ &$.78$ &.44 &1.7   &.49\\
\hline
$\hsm c {\overline{c}}$ &-     &$1.95$ &$1.22$ &.87 &$1.3$ &1.3  &1.1 &$3.95$ &$1.37$&1.8 &12    &.95\\
\hline
$\hsm \gamma\gamma$     &$1.8$ &$1.36$ &$1.22$ &1.07&$1.3$ &1.68 &1.5 &$1.37$ &$1.13$&.71 &1.6   &.29\\
\hline
$\hsm \gamma Z$         &$9.8$ &$10.2$ &$10.2$ &10.2&$10$  &4.28 &4.17&$10.26$&$5.67$&5.5 &$9.8$ &.69\\
\hline 
 $\hsm \mu^+\mu^-$      &$4.3$ &$4.14$ &$3.9$  &3.53&$3.9$ &3.3  &3.2 &$4.36$ &$3.47$&2.5 &.6    &.41 \\
\hline
$\hsm t {\overline{t}}$ &$3.4$ &$3.12$ &$2.82$ &1.4 &$3.1$ &3.1  &3.1 &$3.14$ &$2.01$&3.2 &$3.4$ &1.0\\
\hline
$\Gamma_{tot}$          &$5.3$ &$1.8$  &$.63$  &.45 &$1.1$ &1.65 &1.1 &$1.44$ &$.41$ &.5  &2.7   & \\
\hline \hline 
\end{tabular}
\end{adjustbox}
\caption{Relative Higgs coupling measurements in $\%$ at future colliders when combined with HL-LHC results.  The FCC-ee numbers are from \cite{Bernardi:2022hny} and assume 8 years running split between $\sqrt{S}=240$ GeV and  $\sqrt{S}=365$ GeV. 
 The $\mu$(125) numbers assume $5$ fb$^{-1}$ and the $\mu$(10TeV) numbers assume 10 ab$^{-1}$ and are from \cite{MuonCollider:2022xlm} and ~\cite{Forslund:2022xjq}.
 All numbers assume no BSM decay modes. 
 The FCC-hh numbers are from 
 \cite{deBlas:2019rxi} and the ILC and  CLIC numbers are updated from \cite{deBlas:2019rxi} by J. de Blas.   The CEPC numbers are from
 \cite{CEPCPhysicsStudyGroup:2022uwl} and assume $\sqrt{S}=$240 GeV and 20 ab$^{-1}$ and are combined with HL-LHC. Note that since there are no beyond the Standard Model decays allowed in this table, the width is  constrained by the sum of the SM contributions.  }
\label{tab:higgs_sig_comb}
\end{table}

\begin{table}[h!]
\centering
\begin{tabular}{||c|cc|cc||}
\hline\hline
 & 2/ab-250 & +4/ab-500 &5/ab-250 & +1.5/ab-350\\
coupling & pol. & pol. & unpol. & unpol. \\
\hline \hline 
\hsm ZZ & 0.50 & 0.35 & 0.41 & 0.34 \\
\hsm WW & 0.50 & 0.35 & 0.42 & 0.35 \\
\hsm $b\bar{b}$ & 0.99 & 0.59 & 0.72 & 0.62 \\
\hsm $\tau\tau$ & 1.1 & 0.75 & 0.81 & 0.71 \\
\hsm $gg$ & 1.6 & 0.96 &1.1 & 0.96  \\
\hsm $c\bar{c}$ & 1.8 & 1.2 &1.2 & 1.1 \\
\hsm $\gamma\gamma$ & 1.1 & 1.0 & 1.0 & 1.0 \\
\hsm $\gamma Z$ & 9.1 & 6.6 & 9.5 & 8.1 \\
\hsm $\mu\mu$ & 4.0 & 3.8 & 3.8 & 3.7 \\
\hsm $tt$ & - & 6.3 & - & - \\
\hsm\hsm\hsm  & - & 20 & - & - \\
\hline
$\Gamma_{tot}$ & 2.3 & 1.6 &1.6 & 1.4 \\
$\Gamma_{inv}$ & 0.36 & 0.32 & 0.34 & 0.30 \\
$\Gamma_{other}$ & 1.6 & 1.2 & 1.1 & 0.94 \\

\hline \hline 
\end{tabular}
\caption{Projected uncertainties in the Higgs boson couplings computed within the SMEFT framework and including projected improvements in precision electroweak measurements, as described in the ILC reports and the FCC-ee CDR~\cite{Bambade:2019fyw,ILCInternationalDevelopmentTeam:2022izu, FCC:2018evy}. }
\label{tab:polarization}
\end{table}

There are extensive comparisons between the FCC-ee/CEPC and the ILC/C$^3$ run plans that indicate they offer rather similar precision to study the Higgs Boson. When analyzing Higgs couplings with SMEFT, 2$\abi$ with polarized beams yields similar sensitivity to 5$\abi$ with unpolarized beams. Electron polarization is essential for this. Positron polarization does not add precision, but it offers cross-checks on sources of systematic error. Positron polarization becomes more relevant at high energy ($>$ TeV) where the most important cross sections are initiated from $e^{-_L}e^{+_R}$.
This is shown in Table~\ref{tab:polarization}, which also takes account of the different levels of improvement in precision electroweak measurements expected in the ILC and FCC-ee programs~\cite{Bambade:2019fyw,ILCInternationalDevelopmentTeam:2022izu, FCC:2018evy}.

\newpage
\subsubsection{Top Yukawa}
\begin{figure}[h!]
\begin{centering}
\includegraphics[scale=0.4]{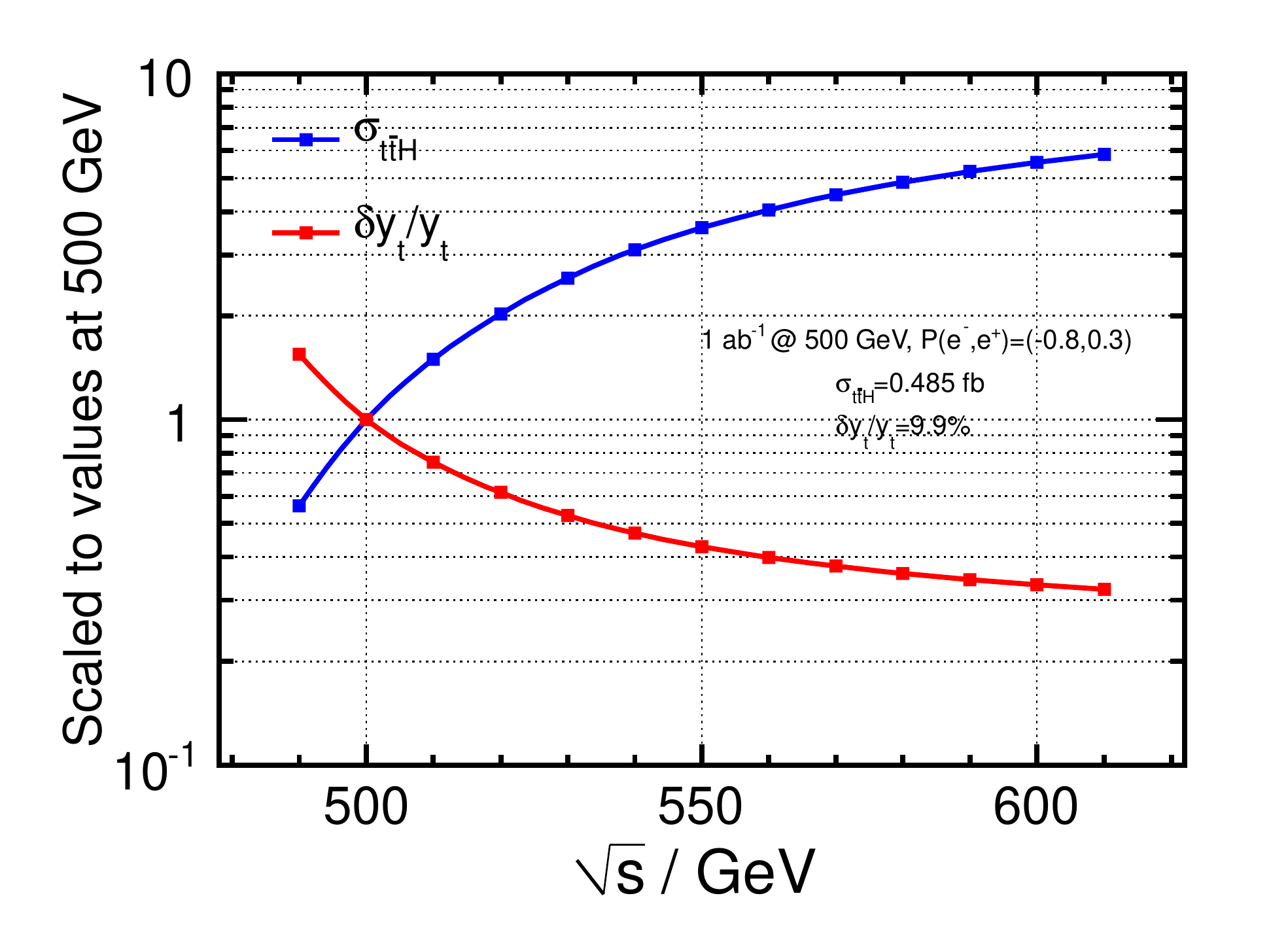} 
\par\end{centering}
\caption{Uncertainty on the top quark Yukawa measurement at an $\ee$ collider as a function of collider energy, showing the improvement at higher center of mass energy\cite{Barklow:2015tja}.}
\label{fig:coups_tth}
\end{figure}

Many models of BSM physics have large effects on the top quark Yukawa.  The gluon fusion rate at the HL-LHC measures a combination of the top quark Yukawa and an effective $gg\hsm$ coupling, while  
the \ttH~and $t\hsm$ channels provide a theoretically cleaner determination of the top quark Yukawa.  The full program of future $\ee$ colliders can  reduce the uncertainty on the top quark Yukawa coupling from that of the HL-LHC, and the uncertainty decreases rapidly as the energy of the $\ee$ collider
is increased, as seen in Figure~\ref{fig:coups_tth}\cite{Barklow:2015tja}, which is an important motivation for higher energy lepton colliders. 

\subsubsection{Charm Yukawa}
\begin{figure}
\begin{centering}
\includegraphics[scale=0.34]{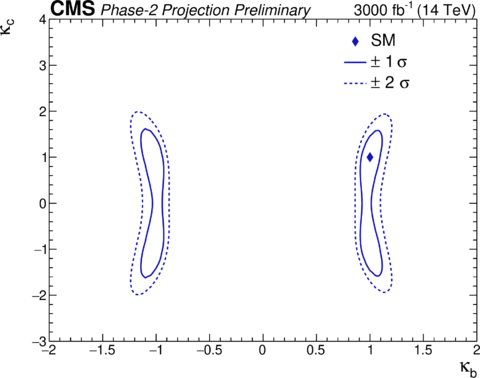}
\includegraphics[scale=0.4]{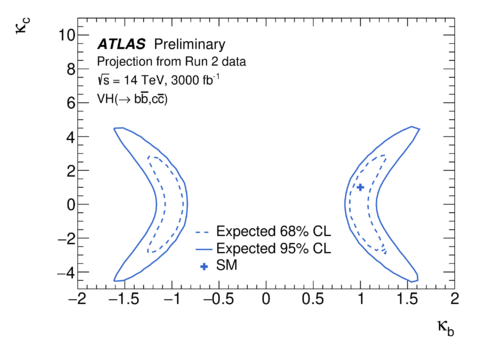} 
\par\end{centering}
\caption{Projected simultaneous sensitivity to $b$ and $c$ quark Yukawa couplings at the HL-LHC~\cite{ATL-PHYS-PUB-2022-018}.}
\label{fig:coups_kc_kb}
\end{figure}

There has been significant progress in the understanding of the sensitivity of the HL-LHC to the charm quark Yukawa.  
CMS and ATLAS have studied the charm quark Yukawa using the associated W$\hsm$ and Z$\hsm$ channels, with an expected limit, $\mid \kappa_c\mid  < 3.4$ at 95$\%$ CL based on the full Run 2 dataset.  The CMS constraint is a factor of 4 better than the ATLAS result, which is attributed the the use of multi-variate techniques and the inclusion of a boosted analysis using substructure techniques.
A combined fit to $\kappa_b-\kappa_c$ results in a projected constraint of
$\mid \kappa_b/\kappa_c\mid <$ 2.6 at 95$\%$ CL at the HL-LHC. The HL-LHC projections for a 2-parameter fit 
to $\kappa_c$ and $\kappa_b$ from 
$V\hsm$ production are shown in Figure \ref{fig:coups_kc_kb}. 
From Table~\ref{tab:higgs_sig_comb},
we see that the lepton colliders have a clear advantage in determining the charm Yukawa coupling over the HL-LHC, with projected uncertainties of ${\cal{O}}(1-2~\%)$.

\subsubsection{Strange and light Yukawa}
\begin{figure}
\begin{centering}
\includegraphics[scale=0.16]{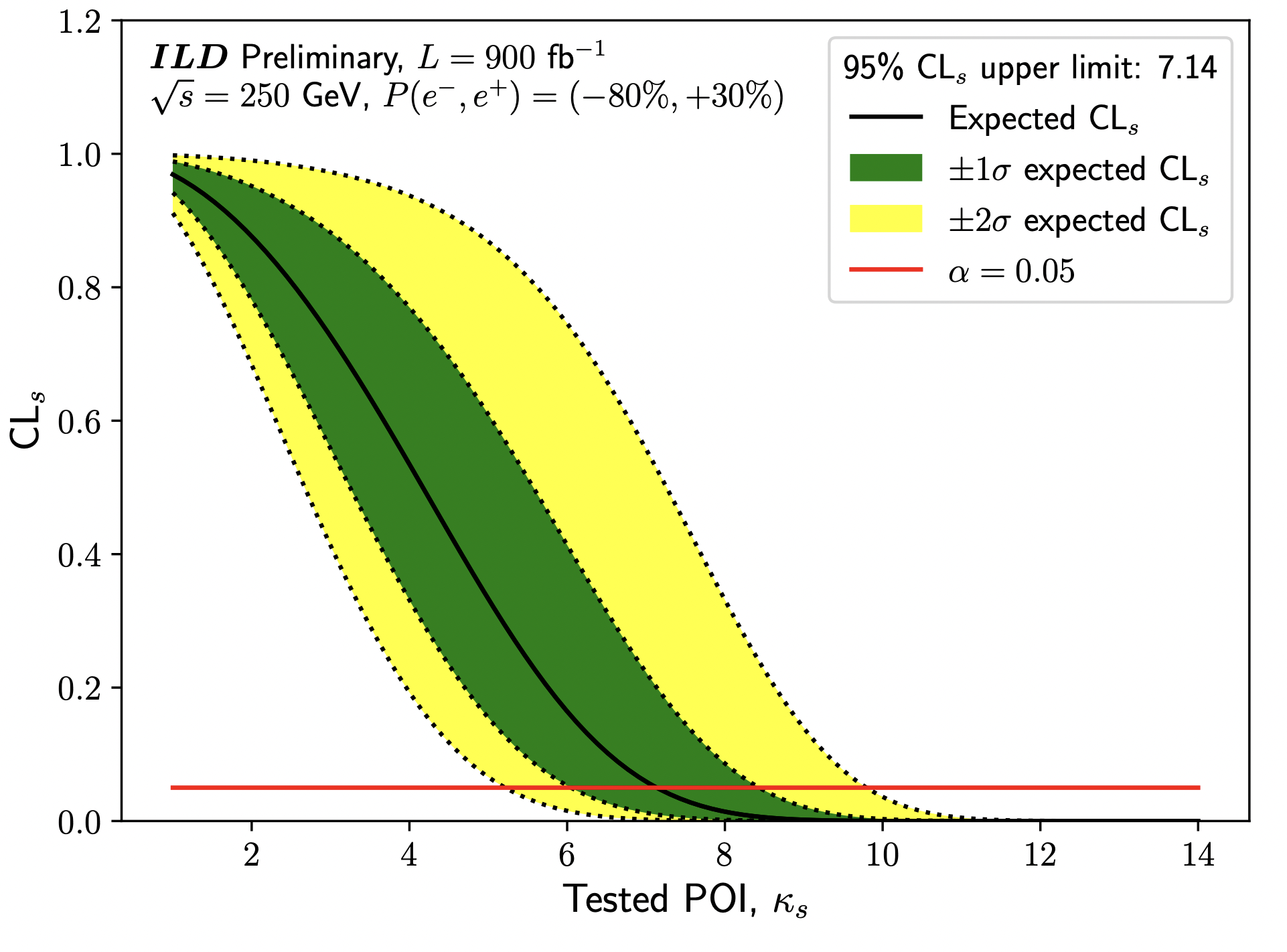}
\includegraphics[scale=0.225]{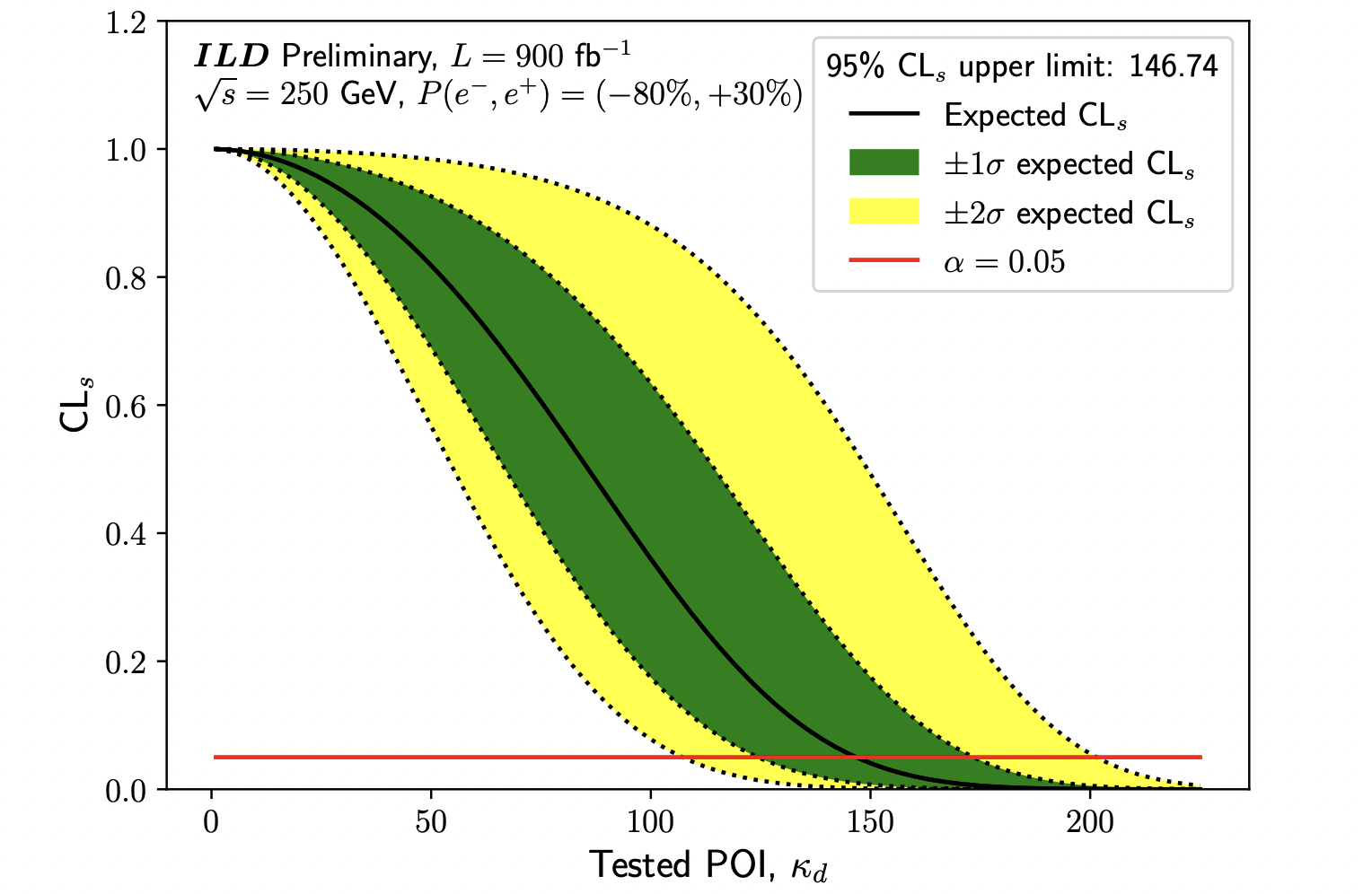} 
\includegraphics[scale=0.222]{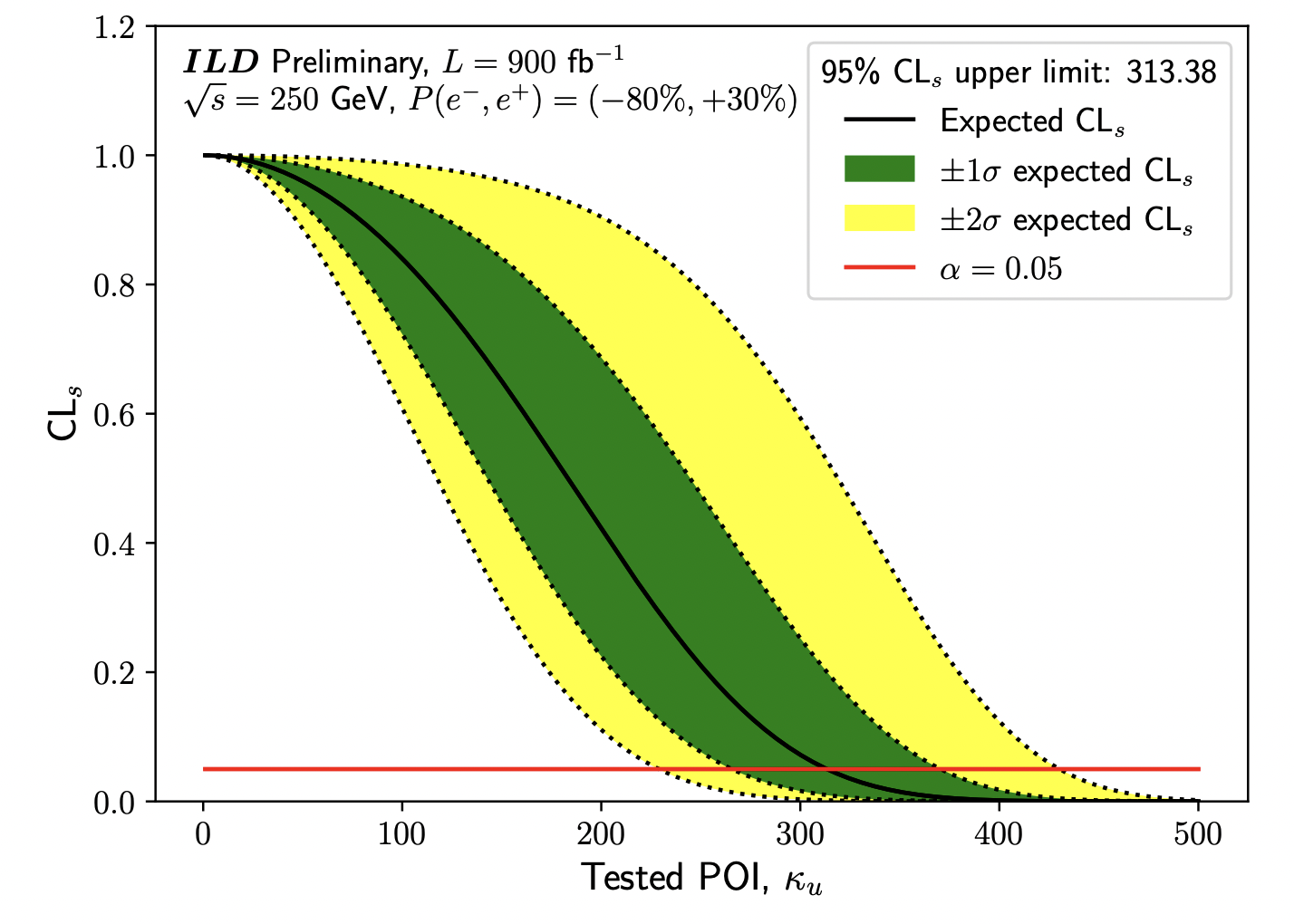} 
\par\end{centering}
\caption{Upper limit plots for the Higgs coupling strength modifiers, $\kappa_s$ (left), $\kappa_d$ (center), $\kappa_u$ (right) using strange tagging at the ILC.  The crossing of the black and red lines indicates the 95\% confidence level~\cite{Albert:2022mpk}.}
\label{fig:coups_ks_ku}
\end{figure}

The prospects for strange quark Yukawa measurements at HL-LHC are not promising, although there have been suggestions that it may be accessible through $\phi$ or $\rho$ meson measurements. This measurement is more promising at lepton colliders. The measurement of $\hsm\rightarrow s\bar{s}$ is performed using the associated Z$\hsm$ production mode in two channels based on
the decay of the Z: neutrinos and leptons. Using the jet flavor tagging algorithm described in~\cite{Albert:2022mpk}, based on the ILD detector, the projected sensitivity to the strange quark Yukawa coupling is $\kappa_s < 7.14$ at $95\%$ CL with 900 $\fbi$ at the $250$ GeV ILC with polarization $P(e^+, e^-)=(-80\%, +30\%)$. Limit plot for $\kappa_s$ is shown in the left side of Figure~\ref{fig:coups_ks_ku} for the combined results of Z decaying to leptons and neutrinos.

This jet flavor algorithm is a multi-classifier that combines information of the jet-level variables and the 10 leading momentum particles contained within the jet. Their kinematics is redefined relative to the jet’s axis and their momentum and mass scaled by the momentum of the jet. The ILD detector will provide particle ID information for each particle in the jet, including electron, muon, pion, kaon/strange hadron and proton likelihoods. The reconstructed likelihoods based on the dE/dx and TOF information have been replaced with the truth likelihoods, resulting in a best-case scenario for tagging performance. Compared to the combined limit achieved using a jet flavour tagger with PID, $\kappa_s <7.14$, there is a $\approx$8$\%$ degradation in the expected performance without PID.

Without modifying the signal region selections for the $\hsm\rightarrow s\bar{s}$ analysis and exploiting the same multi-classifier algorithm, the 95\% CL upper limits on the Higgs-down quark Yukawa coupling,
$\kappa_d$, and the Higgs-up quark Yukawa coupling, $\kappa_u$ have been derived and shown in Figure~\ref{fig:coups_ks_ku}. 

These ILD bounds, based only on 900 $\fbi$ of the data foreseen
at the ILC, compare favorably with current and future indirect LHC limits and would provide the strongest limits for a second Higgs doublet in the spontaneous flavor violating 2HDM model described in Sec. V.B.1 with masses between approximately 80 and 200 GeV within the model\cite{Albert:2022mpk}.

\subsubsection{Electron Yukawa}
\begin{figure}
\begin{centering}
\includegraphics[scale=0.2]{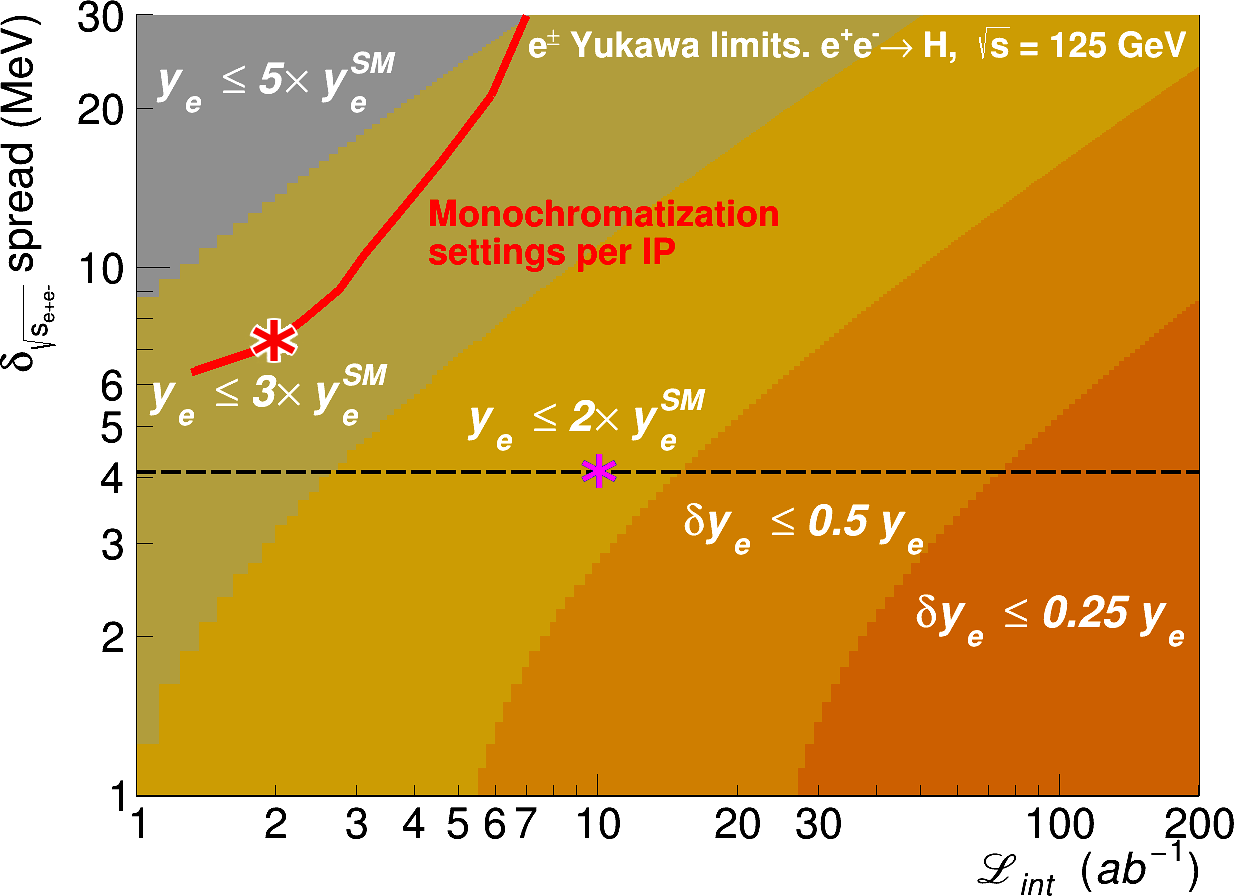} 
\par\end{centering}
\caption{Prospects for measuring the electron Yukawa in a dedicated FCC-ee run with $\sqrt{s}=m_\hsm$~\cite{Bernardi:2022hny}.}
\label{fig:coups_ee}
\end{figure}

Measuring the electron Yukawa coupling would give deep insight into  the Higgs boson interactions with the first generation fermions, since it is the smallest Yukawa coupling in the SM, $\ye\sim 3\times 10^{-6}$. At  hadron colliders,
this measurement is assumed to be impossible, since the $\hsm\rightarrow \ee$ signal is dwarfed by the immense Drell-Yan background.  
A proposal to run the FCC-ee on the $s$- channel Higgs resonance offers the first glimmer of hope
that this measurement could be accomplished\cite{d_Enterria_2022}. The measurement requires that the $\ee$ beams have a very small energy spread, and the current best estimate
of this spread is shown in Figure \ref{fig:coups_ee}, where it appears that with $2~\abi$, a measurement of $\ye$  within a factor of $3$ of the SM prediction might 
be achievable.

%% file: Tex/sme.tex
A consistent theoretical framework requires the use of effective field theory (SMEFT) techniques in place of the $\kappa$ approach.The $\kappa$ approach is never the less of value, since it offers a figure of merit to compare different collider sensitivities to Higgs physics.  The SMEFT approach allows for the combination of Higgs data with data from electroweak precision observables, diboson production and top quark physics for a more comprehensive understanding of high scale physics\cite{deBlas:2022ofj}.  The SMEFT fit assumes that there are no new light particles other than the SM particles and that the Higgs boson is in an $SU(2)$ doublet. 

\begin{figure}
\begin{centering}
\includegraphics[scale=0.44]{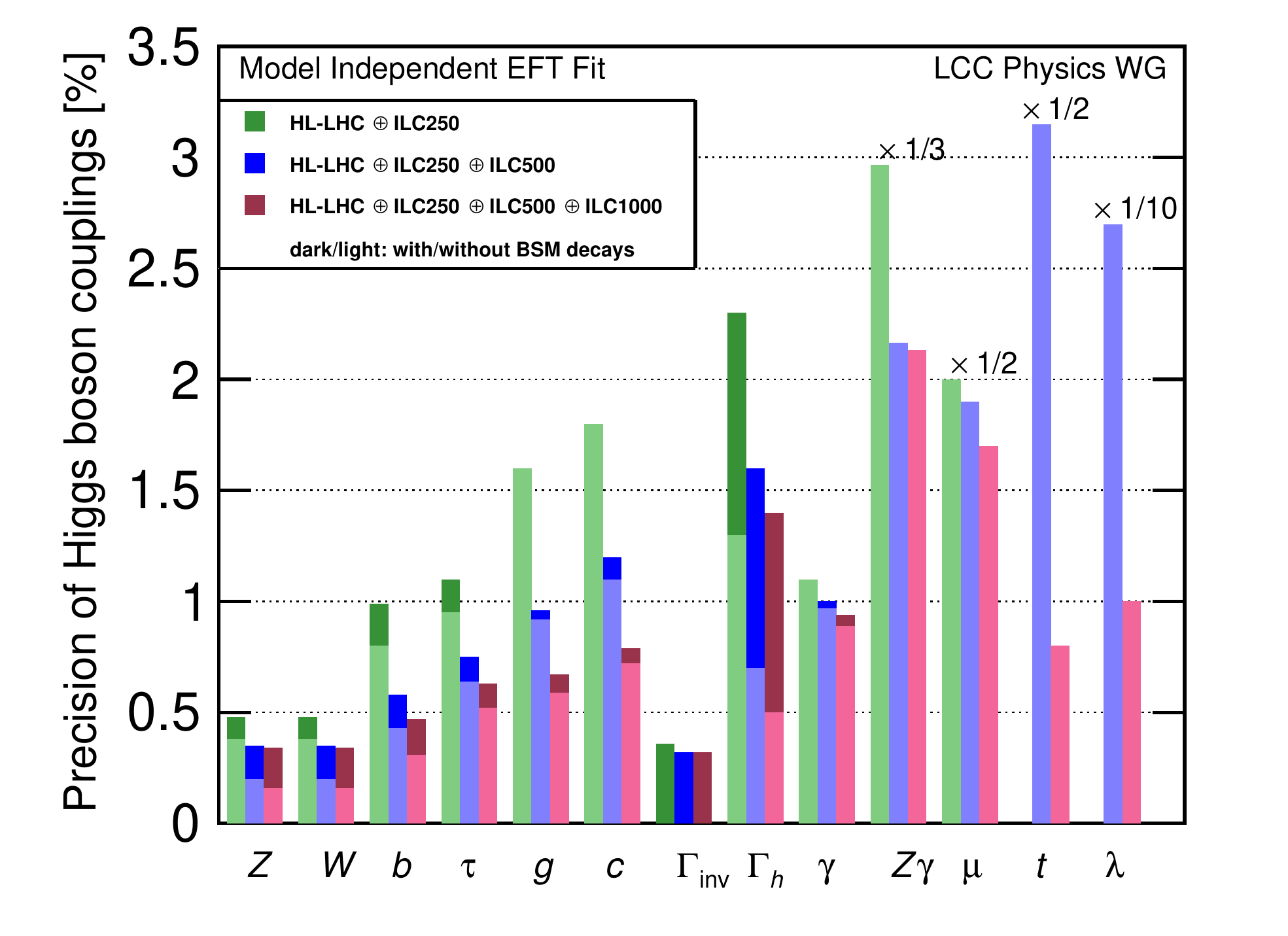} 
\includegraphics[scale=0.38]{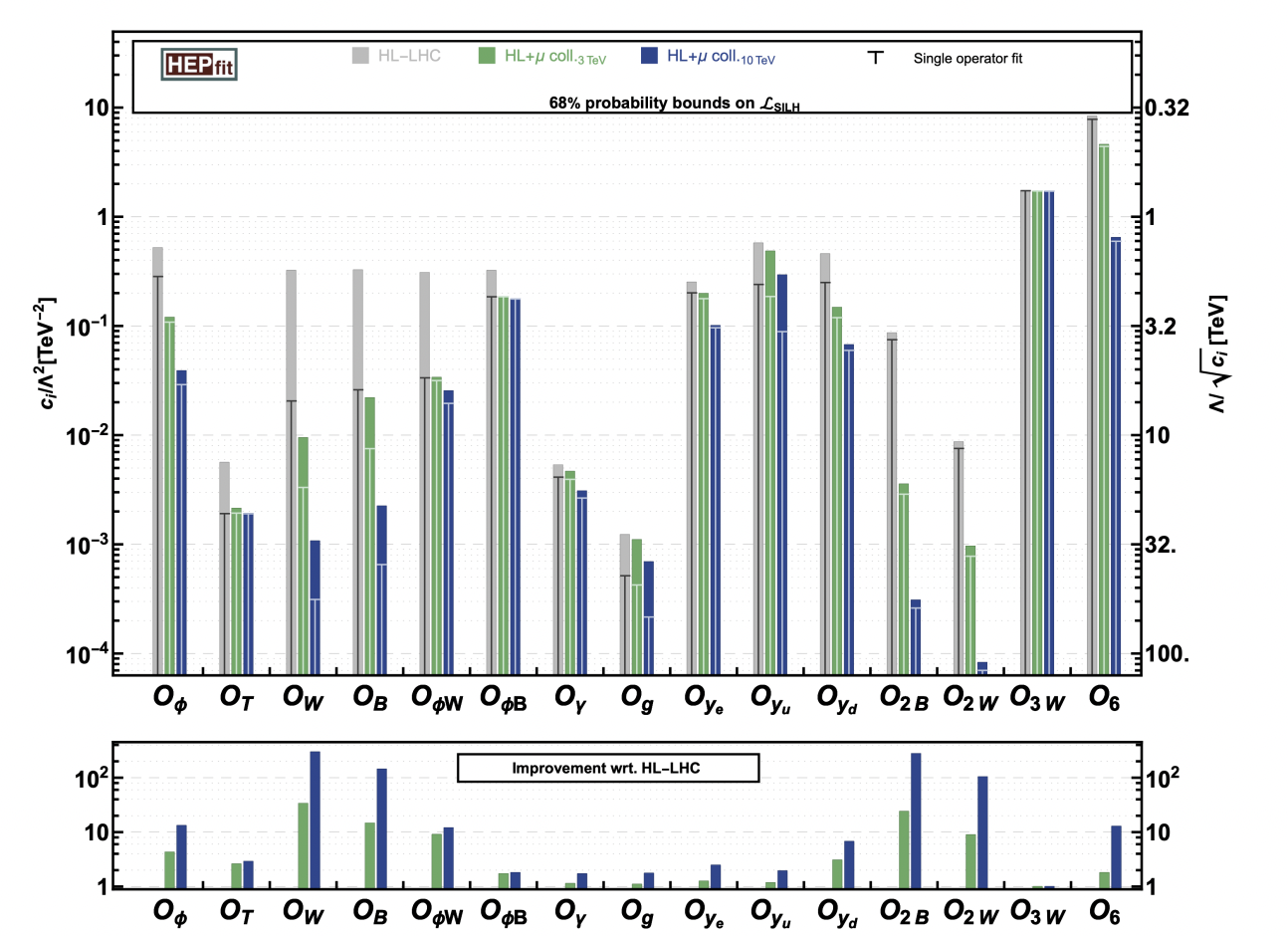} 
\par\end{centering}
\caption{Left, SMEFT projected ILC fit to Higgs, electroweak precision and diboson data.  The thin (fat) lines allow (do not allow) for beyond the Standard Model decays of the Higgs boson. ~\cite{ILCInternationalDevelopmentTeam:2022izu}. Right,  projected SMEFT fit to operators contributing to Higgs production and decay at a muon collider. The reach of the vertical “T” lines indicate the results assuming only the corresponding operator is generated by the new
physics~\cite{MuonCollider:2022xlm}.}
\label{fig:ilc_smeft}
\end{figure}

We note, however, that using a SMEFT approach, the Higgs couplings are determined more precisely than in the $\kappa$ approach due to the inclusion of data outside the Higgs sector. This is especially apparent when considering the $W$ and $Z$ Higgs interactions.   This is illustrated by comparing the results of the ILC  SMEFT fit of the LHS of Figure~\ref{fig:ilc_smeft} with the Higgs only fit shown in Figure~\ref{fig:higgs_sig_comb_fig}.    This figure also demonstrates the effects of allowing for beyond the standard model decays in the fits
and the slight relaxation of the limits in this case seen here is a general feature\cite{deBlas:2019rxi}. On the RHS of  Figure~\ref{fig:ilc_smeft}, we show
a projection of a fit to SMEFT coefficients in the Warsaw basis\cite{Buchmuller:1985jz} at a muon collider. It is clear that there is more than an order of magnitude variation in the obtained precision for the different coefficients.

\begin{figure}
\begin{centering}
\includegraphics[scale=0.65]{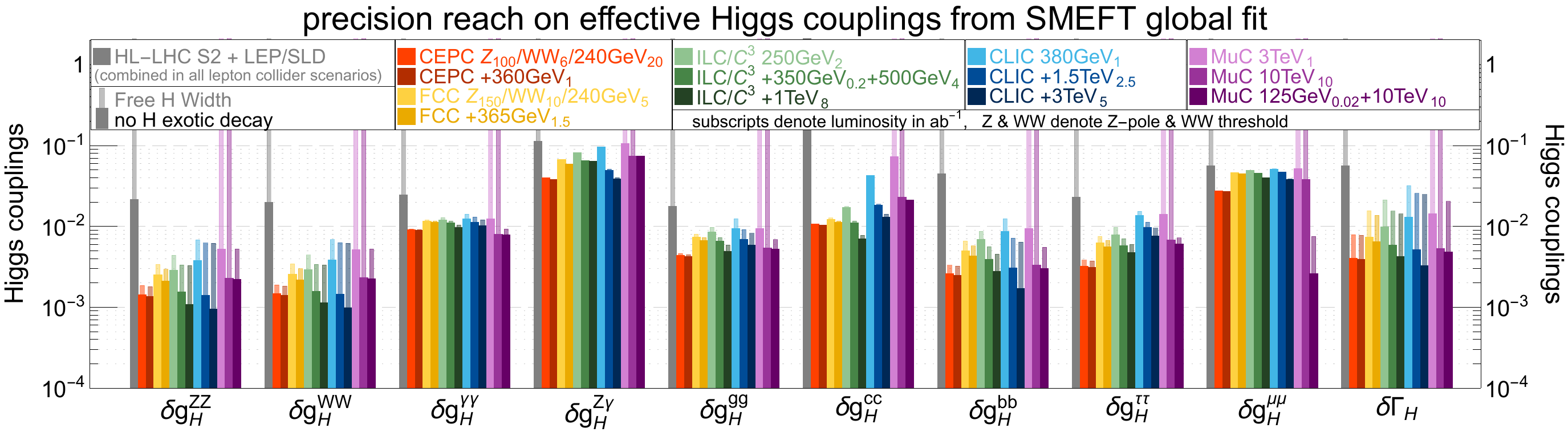} 
\par\end{centering}
\caption{SMEFT  fit to Higgs, electroweak precision and diboson data for future colliders
\cite{deBlas:2022ofj}. }
\label{fig:ef04}
\end{figure}

A comparison of future collider capabilities including Higgs, diboson, and Giga-Z measurements in the SMEFT framework is shown in Fig. \ref{fig:ef04}.  The inclusion of the diboson and Giga-Z data greatly improves the precision of the $\hsm WW$ and $\hsm ZZ$ couplings.  Each of the results includes HL-LHC, along with earlier running of each collider. The fit is done both with a constrained $\Gamma_\hsm$ assumption (no beyond the Standard Model decays) and allowing the Higgs width to float, which requires a model independent measurement from the Higgs recoil technique at the $e^+e^-$ colliders. 

%% file: Tex/cp.tex
\begin{table}[ht]
\renewcommand{\arraystretch}{1.5}
\vspace{-0.4cm}
\begin{center}
\begin{tabular}{|l|ccccccccccc|c|}
\hline\hline
Collider                     &   $pp$       &   $pp$    &   $pp$    &   $e^+e^-$    &   $e^+e^-$    &  $e^+e^-$    &  $e^+e^-$    & $e^-p$ & $\gamma\gamma$ &  $\mu^+\mu^-$  & $\mu^+\mu^-$  & target \\
E (GeV)                     &   14,000   &   14,000  &   100,000     &  250          &  350                 & 500             &   1,000       & 1300    &       125               &    125          & 3000                       &   (theory) \\
${\cal L}$ (fb$^{-1}$) & 300  & 3,000  & 30,000 & 250          &  350                  & 500              &    1,000       & 1000           &        250        &   20                               & 1000  &     \\ 
\hline
\hline
$\hsm ZZ/\hsm WW$ &  $4\!\cdot\!10^{-5}$ &  $2.5\!\cdot\!10^{-6}$ & \checkmark & $3.9\!\cdot\!10^{-5}$ & $2.9\! \cdot\!10^{-5}$ & $1.3\!\cdot \!10^{-5}$ & $3.0\!\cdot\!10^{-6}$  & \checkmark  & \checkmark  &  \checkmark  &  \checkmark  & $<10^{-5}$  \\
\hline
\hline
  $\hsm\gamma\gamma$ & --  & 0.50  & \checkmark & -- & -- & -- & -- & -- & 0.06 & -- & --  & $<10^{-2}$  \\
\hline
  $\hsm Z\gamma$ & --  & $\sim\!1$  & \checkmark &-- & -- & -- & $\sim\!1$   & -- & -- & --  & --  & $<10^{-2}$   \\
\hline
  $\hsm$gg & $0.12$ & $0.011$  & \checkmark & -- & -- & -- & -- & -- & -- & -- &  -- & $<10^{-2}$   \\
\hline
\hline
 $\hsm t\bar{t}$ & 0.24  & 0.05  & \checkmark & -- & -- & 0.29 & 0.08  & \checkmark  & -- & --  & \checkmark & $<10^{-2}$  \\
\hline
 $\hsm\tau\tau$ &  0.07 & 0.008  & \checkmark & $0.01$ & $0.01$  &  ${0.02}$  & ${0.06}$  &  -- & \checkmark & \checkmark & \checkmark & $<10^{-2}$  \\
\hline
 $\hsm\mu\mu$ & --  & -- & -- & -- & -- & -- & --  & -- & -- & \checkmark & -- & $<10^{-2}$   \\
\hline
\hline
\end{tabular}
\end{center}
\caption{
List of expected precision (at 68\% C.L.) of ${CP}$-sensitive measurements of the parameters $f_{CP}^{HX}$ defined in Eq.~(\ref{eq:fCP}).
Numerical values are given where reliable estimates are provided, 
$\checkmark$ mark indicates that feasibility of such a measurement could be considered\cite{Gritsan:2022php}. The $e^{+} e^{-} \rightarrow Zh$ projections are performed with $Z\rightarrow \ell\ell$ but scaled to a ten times larger luminosity to account for $Z\rightarrow q\bar{q}$.}
\label{table-cpscenarios}
\end{table}

The search for $CP$ violation is an important research direction of future experiments in particle 
physics, as $CP$ violation is required for baryogengesis and cannot be sufficiently explained with present knowledge.
$CP$ violation can be searched for in interactions 
of the Higgs boson  with either fermions or bosons at  current and future proposed facilities.
The amount of CP violation is characterized by the quantity,
\begin{equation}
f_{CP}^{\hsm X}\equiv
\frac{\Gamma^{CP\rm\,odd}_{\hsm\to X}
}{\Gamma^{CP\rm\,odd}_{\hsm\to X}
+\Gamma^{CP\rm\,even}_{\hsm \to X}}
\,.
\label{eq:fCP} 
\end{equation}
The dedicated $CP$-sensitive measurements of the $\hsm$ provide simple but 
reliable benchmarks that are compared between proton, electron-positron, photon, and muon 
colliders in Table~\ref{table-cpscenarios}.

Hadron colliders provide essentially the full spectrum of possible measurements sensitive to $CP$ violation 
in the $\hsm$ boson interactions accessible in the collider experiments, with the exception of interactions with 
light fermions, such as $\hsm\mu\mu$.
The $CP$ structure of the $\hsm$ boson couplings to gluons cannot be easily measured at a lepton collider, 
because the decay to two gluons does not allow easy access to gluon polarization. 
On the other hand, most other processes could be studied at an $e^+e^-$ collider, especially with the beam 
energy above the $t\bar{t}\hsm$ threshold. Future $e^+e^-$ colliders are expected to provide comparable $CP$
sensitivity to HL-LHC in $\hsm f{\overline{f}}$ couplings, 
such as $\hsm t\bar{t}$ and $\hsm\tau\tau$, and $\hsm ZZ/\hsm WW$ couplings. 

A muon collider operating at the $\hsm $ boson pole gives access to the $CP$ structure of the $\hsm \mu\mu$ 
vertex using the beam polarization. It is not possible to study the $CP$ structure in the decay because the muon 
polarization is not accessible. At a muon collider operating both at the $\hsm$ boson pole and at higher energy, 
analysis of the $\hsm$ boson decays is also possible. However, this analysis is similar to the studies performed 
at other facilities and depends critically on the number of the $\hsm$ bosons produced and their purity.
A photon collider operating at the $\hsm $ boson pole allows measurement of the $CP$ structure of the 
$\hsm \gamma\gamma$ vertex using the beam polarization. Otherwise, the measurement of $CP$ in both 
$\hsm \gamma\gamma$ and $\hsm Z\gamma$ interactions is challenging and requires high statistics of
$\hsm$ boson decays with virtual photons, which would require a production rate beyond that of the HL-LHC for 
sensitive measurements. 

Measurements of the electric dipole moments of atoms and molecules set stringent constraints 
on $CP$-violating interactions beyond the SM from Higgs bosons appearing in loop calculations. 
Assuming only one $CP$-odd $\hsm$ boson coupling is nonzero at a time, EDM constraints can 
be interpreted as limits on $CP$ violation in the $\hsm$ boson interactions.
Such constraints are either tighter or expected to be tighter with EDM measurements projected 
in the next two decades when compared to $CP$ violation measurements in direct $\hsm$ boson 
interactions at colliders. However, resolving all constraints simultaneously will require direct 
measurements of the $\hsm$ boson couplings in combination with EDM measurements. Moreover, 
it has not been experimentally established whether the $\hsm$ boson couples to the first-family fermions,
and if such couplings are absent or suppressed, EDM measurements provide no constraints on 
$CP$ violation in $\hsm$ boson interactions.

We conclude that the various collider and low-energy experiments provide complementary 
$CP$-sensitive measurements of the $\hsm$ boson interactions. The HL-LHC provides the widest spectrum 
of direct measurements in the $\hsm$ boson interactions and is unique in measuring couplings to gluons, 
but it lacks the ability to set precise constraints on interactions with photons and muons. Such constraints 
may become possible with either photon or muon colliders operating at the $\hsm$ boson pole. 
The electron-positron collider may allow constraints similar to HL-LHC in couplings to fermions 
and the $W$ and $Z$ bosons. Given the coverage provided by HL-LHC, we expect that a future $pp$
collider, such as FCC-hh or SPPC, will surpass HL-LHC and allow the furthest reach in $CP$-sensitive 
measurements of the $\hsm$ boson interactions among the collider experiments. 


%% file: Tex/hh_future.tex
\begin{table}[tb]
\centering
{\small
\begin{tabular}{ c  c  c  c }
\hline 
collider       &  Indirect-$\hsm$ &  $\hsm\hsm$   & combined   \\
\hline
HL-LHC~\cite{ATL-PHYS-PUB-2022-005}         &  100-200\%    & {50\%}    & 50\%   \\  \hline
ILC$_{250}$/C$^3$-250~\cite{ILCInternationalDevelopmentTeam:2022izu,Dasu:2022nux}    &  49\%    & $-$   &49\%    \\
ILC$_{500}$/C$^3$-550~\cite{ILCInternationalDevelopmentTeam:2022izu,Dasu:2022nux}     &  38\%      & 20\%   &20\%  \\
CLIC$_{380}$ ~\cite{Robson:2018zje}  & 50\%     & $-$    &50\%  \\
CLIC$_{1500}$~\cite{Robson:2018zje}   &  49\%   & 36\%  & 29\%  \\
CLIC$_{3000}$~\cite{Robson:2018zje}   &  49\%  & 9\% &  9\% \\
FCC-ee~\cite{Bernardi:2022hny}   &  33\%   & $-$    &33\%   \\
%
FCC-ee (4 IPs)~\cite{Bernardi:2022hny} &  24\%   & $-$    &24\%  \\  %
FCC-hh  ~\cite{Mangano:2020sao}       &  -  & 3.4-7.8\%  & 3.4-7.8\%  \\
{$\mu$(3 TeV)}~\cite{MuonCollider:2022xlm} &  - & 15-30\% & 15-30\% \\
{$\mu$(10 TeV)}~\cite{MuonCollider:2022xlm} &  - & 4\% &  4\% \\
\hline
\end{tabular}
}
\caption{\label{tab:h3} Sensitivity at 68\% probability on the Higgs cubic self-coupling at the various future colliders. Values for indirect extractions of the Higgs self-coupling from single  Higgs  determinations below the first line are taken from \cite{deBlas:2019rxi}.  The values quoted here are combined with an independent determination of the self-coupling with uncertainty 50\% from the HL-LHC. }
\label{tab:future_comp}
\end{table}

By the end of Run 3 in 2024, the LHC will have collected, by combining the ATLAS and CMS dataset, around 600~$\fbi$ of integrated luminosity. A naive extrapolation of the most recent Run 2 results indicates that double Higgs production, as predicted by the Standard Model, will not be observed even with the Run 3 dataset. Assuming current detector performance, it will be possible to set an upper limit on the di-Higgs production cross-section closer to the SM value at 95 \% CL at best. but a measurement of the Higgs self-coupling is thus out of reach of Run 3 and requires either a larger dataset, or/and a higher collision energy.

The HL-LHC will collide protons at 14~TeV (which constitutes a moderate, although non-negligible, increase in center of mass energy with respect to 13 TeV at the current LHC), and is expected to produce an integrated luminosity of 3~$\mathrm{ab}^{-1}$ per interaction point. Such a large increase in the luminosity will allow for the milestone observation of double Higgs production at 5$\sigma$.  This would correspond to observation at the 95\% CL that the Higgs self-coupling is non-zero. Still, the corresponding precision on the Higgs self-coupling will be only of order 50\%. This measurement will be largely driven by the measurement of \hh production. 




The goal for future machines beyond the HL-LHC should be to probe the Higgs potential quantitatively.  Such a level of precision is achievable through the measurement of $\ee$ production at lepton machines at energies above 500 GeV and at hadron machines (FCC-hh). 

The proposed $\ee$ Higgs factories---CEPC, ILC, \CCC, CLIC, and FCC-ee---can access the Higgs self-coupling through analysis of single Higgs measurements.  This relies on the fact that these colliders will measure a large number of individual single Higgs reactions with high precision, allowing an
indirect analysis of possible new physics contributions to the self coupling through loop effects.   It will  be important to have data at two different center of mass energies to increase the level of precision and this requires reaching the second stage of a staged run plan.  

The values for the indirect Higgs  measurement of the self-coupling given in Table~\ref{tab:future_comp} are combined with a HL-LHC projected error of 50\%~ \cite{deBlas:2019rxi,DiMicco:2019ngk}.  Thus, only values well below 50\% represent a significant improvement.  The various estimates are computed using different assumptions on the inclusion of SMEFT parameters representing other new physics effects. On the other hand, many of the values quoted for \hh production are derived from fits including the single parameter $\klambda$ only. At $\ee$ colliders it is more straightforward to simulate the relevant backgrounds, but there is less experience with the high-energy regime studied here.  The uncertainties in the direct determinations at $\ee$ colliders are computed using full-simulation analyses based on current analysis methods.   These have much room for improvement when the actual data is available.  The analyses at hadron colliders are based on estimates of the achievable  detector performance in the presence of very high pileup.  These are extrapolations, but the estimates are consistent with the improvements in analysis methods that we have seen already at the LHC. 

\begin{figure}
\begin{centering}
\includegraphics[scale=0.4]{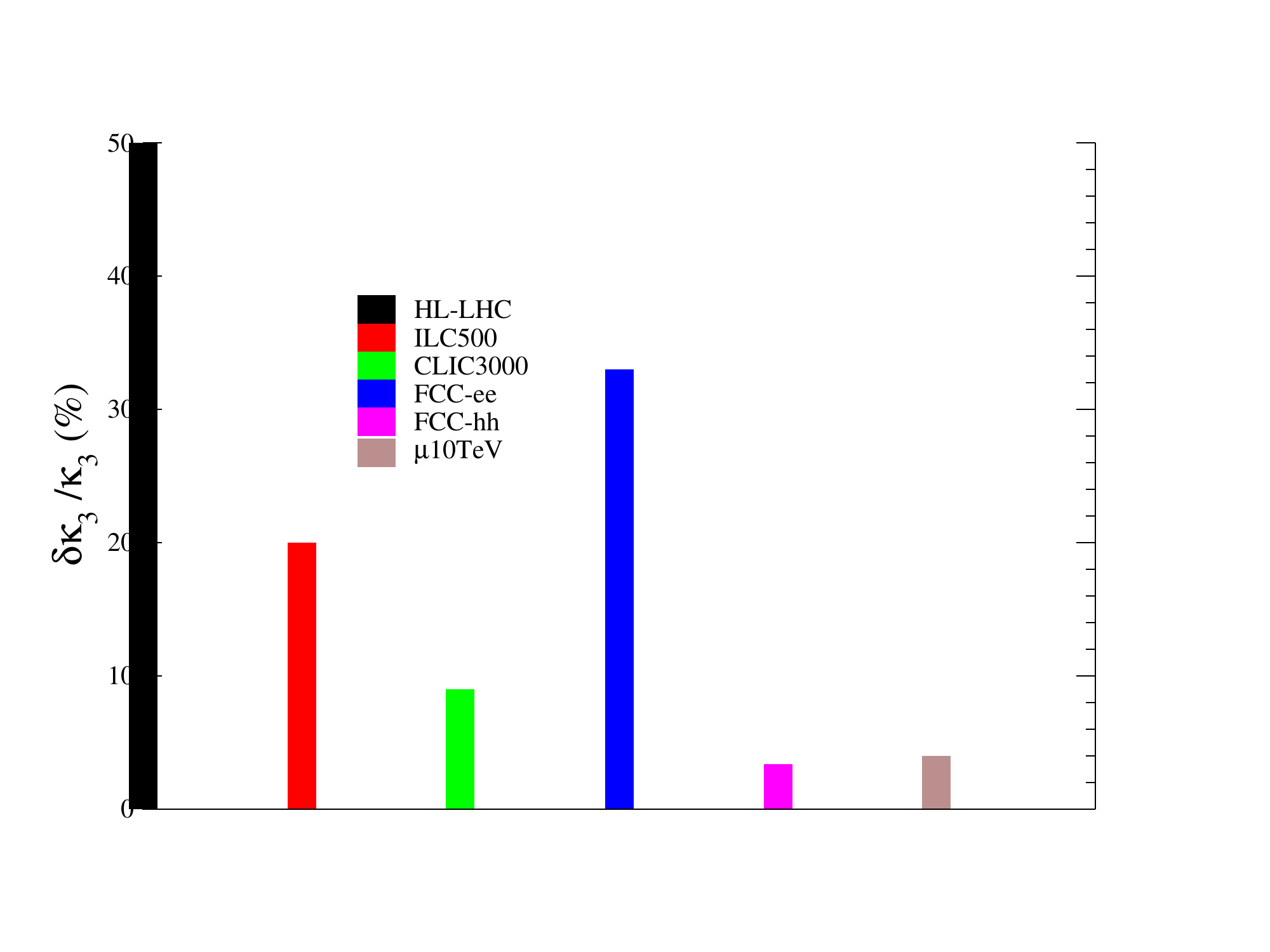}
\par\end{centering}
\caption{Limits on the Higgs self-coupling at future machines. }
\label{fig:hhgraph}
\end{figure}

The projected sensitivities to the Higgs boson self-coupling at the various future colliders are presented in Table~\ref{tab:future_comp} and shown graphically in Fig. \ref{fig:hhgraph}.  
A measurement with $\mathcal{O}$(20\%)  on the Higgs self-coupling would allow to exclude/demonstrate at 5$\sigma$ some models of electroweak baryogenesis as discussed in Section ~\ref{sec:BSM}. 

\begin{figure}
\begin{centering}
\includegraphics[scale=0.3]{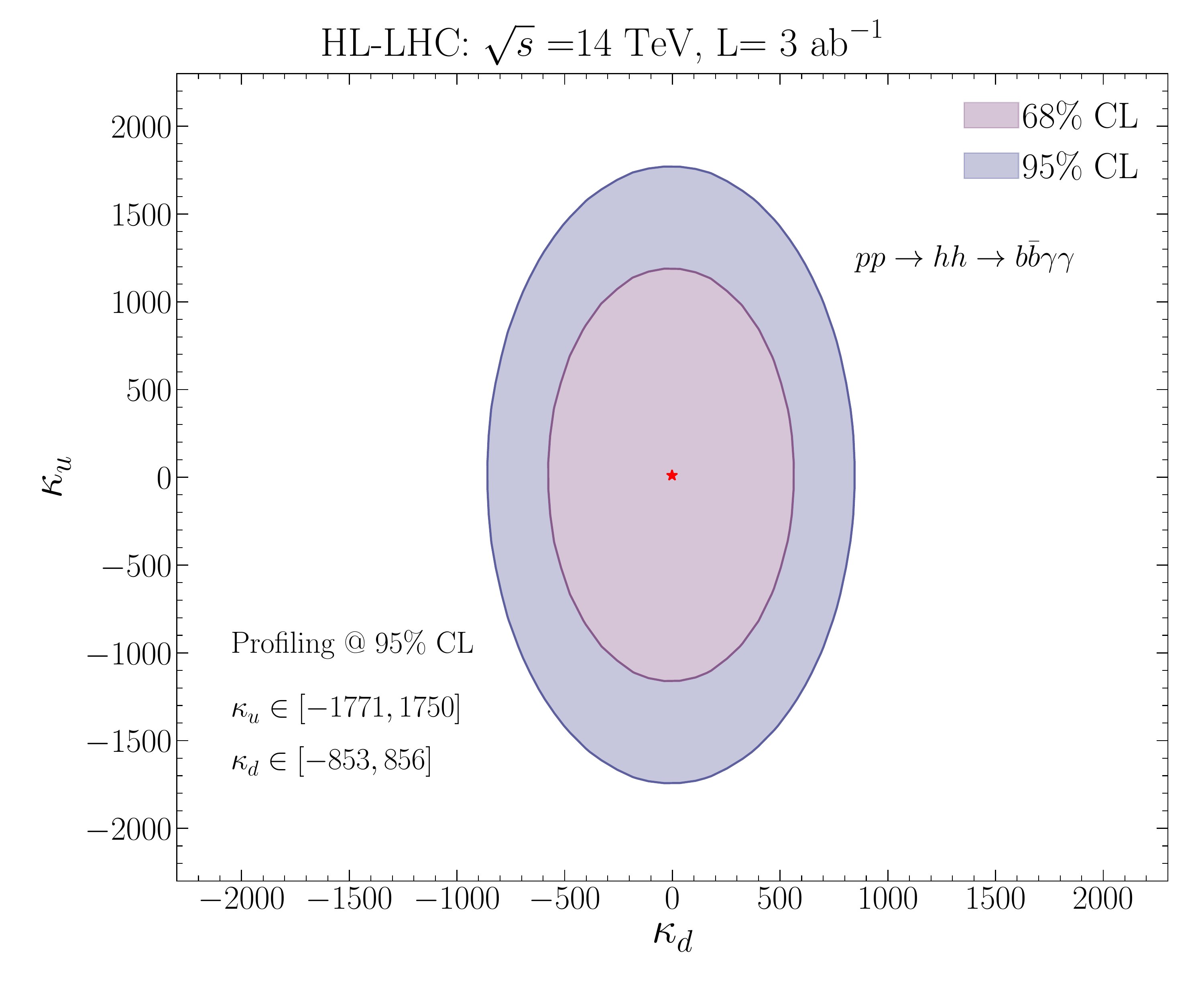}
\par\end{centering}
\caption{Limits on first generation quark Yukawa couplings from di-Higgs production at the HL-LHC\cite{Alasfar:2019wby}.}
\label{fig:firstgen}
\end{figure}

Light quarks contribute to the gluon fusion production of di-Higgs through loop effects and can be used to place limits on $\kappa_f$\cite{Alasfar:2019wby}.  The resulting limits on $\kappa_c$ and $\kappa_b$ do not improve on limits from single Higgs production.  Di-Higgs production at the HL-LHC does, however, provide some limits on the first generation Yukawa couplings as shown in Figure \ref{fig:firstgen}.  Without a UV model these large values of the  first generation Yukawa couplings would be hard to reconcile with other measurements.  However, in Section~\ref{sec:flavor} we discuss how there is a new mechanism that can easily accommodate shifts in the first and second generation Yukawa couplings without being conflict with experimental data.

\begin{figure}[htbp]
    \centering
    \includegraphics[width=0.46\textwidth]{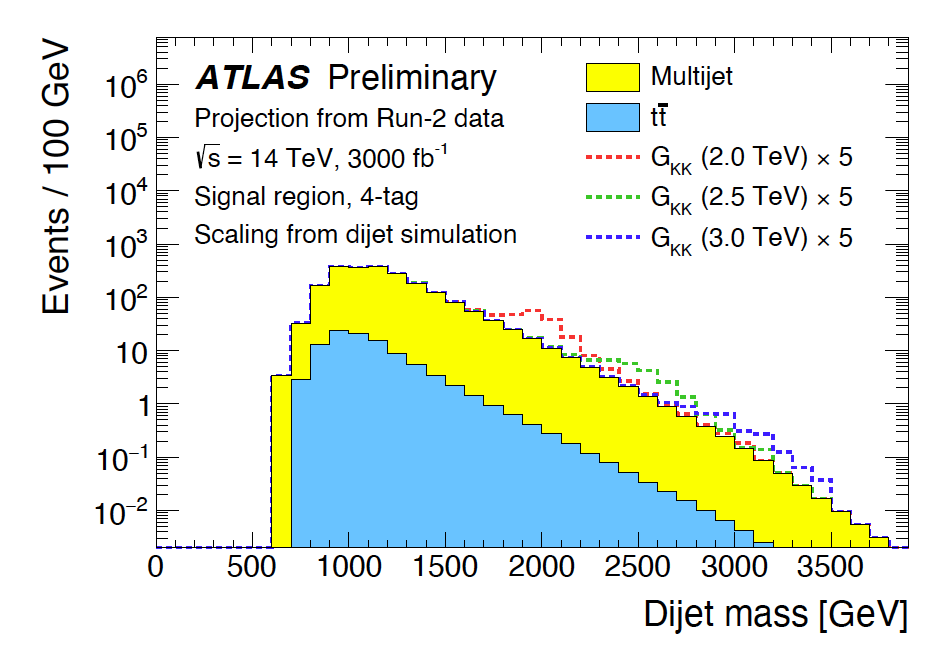}
    \includegraphics[width=0.44\textwidth]{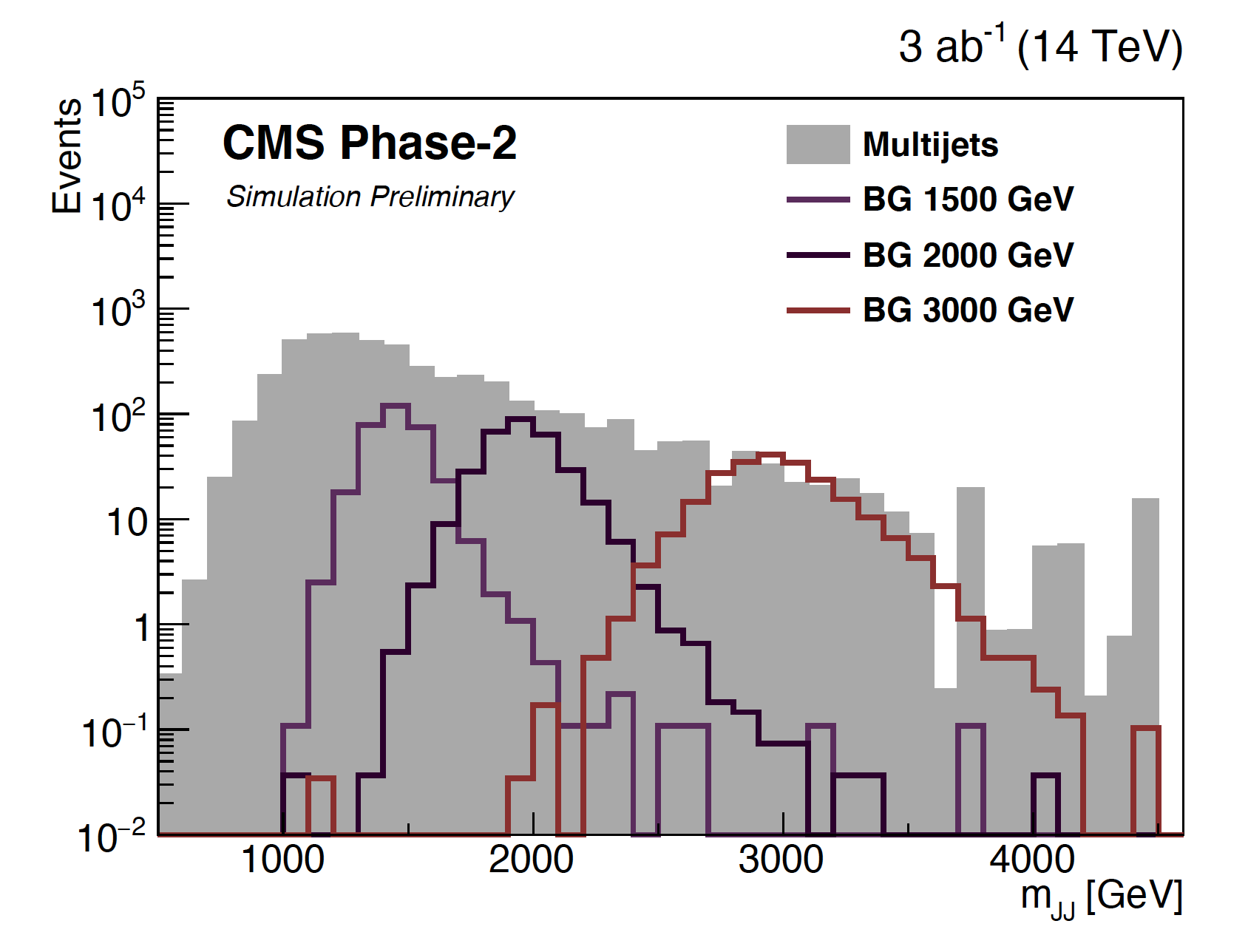}
    \caption{Dijet mass distributions from the truth-level analysis of (left) the 4b subjet 4-tag (signal and background processes are stacked) and (right) 3-tag (signal and background processes are overlaid) events in the signal regions projected for 3 ab$^{-1}$ at $\sqrt{S} =14$ TeV at HL-LHC, in the searches for heavy resonances decaying to a pair of Higgs bosons in the 4$b$ final state via the gluon fusion and VBF production modes, by (left) the ATLAS and (right) CMS Collaboration, respectively. The signal processes shown are spin-2 bulk gravitons.}
    \label{fig:HHRes4B}
\end{figure}

A variety of beyond the Standard Model scenarios predict new resonances decaying to a pair of Higgs bosons. The ATLAS and CMS Collaborations have projected the sensitivity of the searches for the gluon fusion and VBF production modes of new spin-0 and spin-2 particles at the HL-LHC  using the $\hh \rightarrow 4b$ channel, where both Higgs bosons decaying to a pair of b-quarks are highly Lorentz-boosted and the hadronization products of the two bottom quarks are reconstructed as a single large-radius jet. This gives access to new BSM particles of masses up to a few TeV as shown in Figure~\ref{fig:HHRes4B}.
The experimental reach at the HL-LHC is expected to be expanded with improved boosted $\hbb$ tagging capability due to future detector upgrades and improvements of the reconstruction methods~\cite{CMS-DP-2020-002,ATL-PHYS-PUB-2020-019,CMS-DP-2021-017}.

The measurement of the Higgs quartic coupling is extremely challenging due to the small rate for triple Higgs production. Even with 20~$\abi$, the HL-LHC will only observe this channel at $2\sigma$ with no measurement possible.  The FCC-hh  with 10~$\abi$ projects a $1\sigma$ constraint for the quartic coupling of (-2.3,+4.3)\cite{DiMicco:2019ngk}.  A 10 TeV muon collider could potentially obtain a $2\sigma$ constraint of (-.7, +.8) with 20~$\abi$\cite{Chiesa:2020awd}.

%% file: Tex/bsmintro.tex
The ultimate goal of precision Higgs physics is to learn about new physics at high scales.  As discussed in the introduction, the generic scale associated with precision Higgs physics at future colliders typically extends up to a few TeV.  While this was discussed in the context of different UV physics models that can generate Wilson coefficients of a SMEFT approach in the limit that the new physics is very heavy, similar arguments can be made from even more general principles. 

For example, the gauge invariant structure of the SM at the amplitude level accounts for numerous cancellations of contributions to amplitudes that would grow with energy.  This in fact led to the famous argument~\cite{Lee:1977eg}, that the SM Higgs mass could not be arbitrarily large without violating perturbative unitarity in $VV\rightarrow VV$ ($V=W,Z$) scattering.  If one were to allow for arbitrary changes of SM Higgs couplings without preserving gauge invariance, there would be a multitude of amplitudes that would eventually saturate perturbative unitarity.  The leading contribution yields bounds on the energy scale that saturate as $\Lambda\sim \delta \eta_{SM}^{-1/2}$, and therefore scales in the same way as does the EFT in Eq.~\ref{eqn:EFT0}, although
the bounds from unitarity tend to be at the $10-100$~TeV scale for 1\% level measurements\cite{Abu-Ajamieh:2022dtm}.
Note, that even though it scales in the same way as the SMEFT estimate, the amplitudes are proportional to the SM couplings and therefore there would be a wide range of upper limits on the scale of new physics, i.e. the bound from shifts in the muon Yukawa sets a much larger scale than shifts in the top Yukawa, even if the precision on the measurements is similar.

\begin{figure}
\begin{centering}
\includegraphics[scale=0.5]{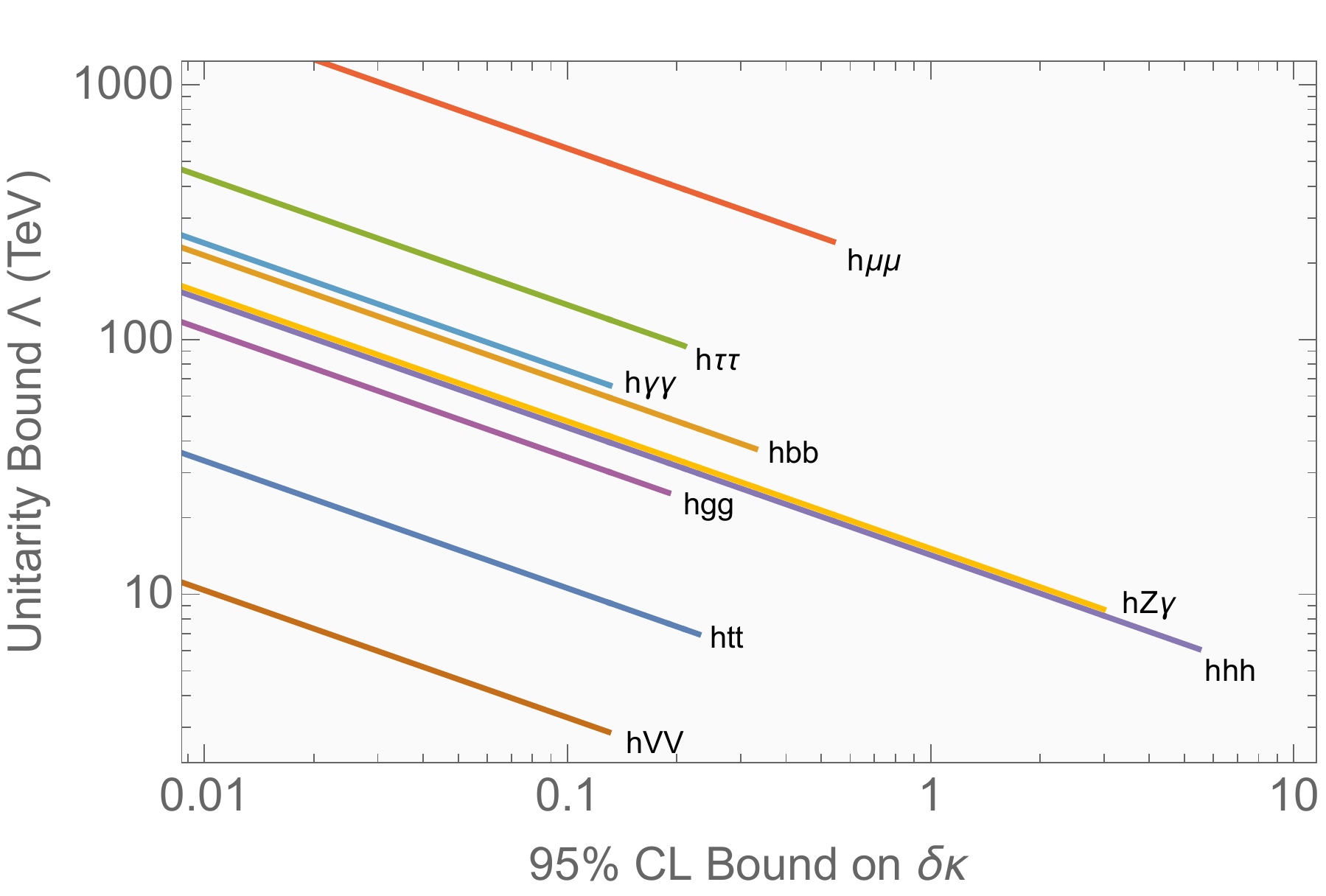} 
\par\end{centering}
\caption{Scale where unitarity is violated as a function of the precision of Higgs coupling measurements.  
The bound is typically only saturated in strongly interacting scenarios and in specific models tends to be significantly lower.\cite{Abu-Ajamieh:2022dtm}.}
\label{fig:unitarity}
\end{figure}

Nevertheless, both the EFT and the perturbative unitarity scales presented are tied to the ultimate precision reached at colliders, with the assumption that a deviation is not seen.  Larger deviations that could be measured at future colliders or the HL-LHC generally imply lower scales for new physics, and thus it is important to understand the types of models that can generate deviations.  Moreover, instead of precision Higgs physics viewed agnostically in all channels simultaneously, physics that causes deviations implies {\em patterns} of correlated deviations or other observables that matter.  Unfortunately the model phase space relevant to all types of new physics cannot be fully covered.  In particular, there is a bifurcation in thinking about how to organize new physics studies.  From the bottoms up point of view, one can think about whether new physics couples predominantly at loop-level
or through tree-level mixing with the Higgs, and then the representations under the SM symmetries that such particles carry.  This is the spirit of Figure~\ref{fig:higgscentral3}.  In this line of reasoning, we can think of tree-level mixing extensions first, and for example investigate the simplest non-trivial representations that could couple to the SM Higgs, a scalar singlet $S$ and a second Higgs doublet $H_2$
(i.e. the full space of 2HDM models).  At loop level we can again categorize into SM singlets, with the simplest example being a 
$\mathbb{Z}_2$ symmetric scalar singlet extension, 
while for non-trivial SM representations that generate loop level effects, the dominant effects would occur in processes that start at loop level in the SM.  This in principle allows for a more general set of spins and representations that could affect the $\hsm gg$ coupling, $\hsm\gamma\gamma$ and $\hsm\gamma Z$ couplings. 
This organization, although not all encompassing, is very pragmatic for illustrating examples of what Higgs precision can test while also being able to examine complementary observables and constraints.  

Unfortunately, organizing by type of particle does {\em not} allow for a characterization of the 
sort shown in Figure~\ref{fig:higgscentral} or Figure~\ref{fig:higgscentral2}.  This is the nature of the Higgs Inverse problem, and it is possible that a generic new particle might naturally live in a narrow part of a parameter space in order to address one of our fundamental questions. For example, one might choose 
to make estimates based in the context of more general questions such as naturalness or whether there was an electroweak phase transition 
that differs from the SM expectation.    This is clearly the more exciting direction for categorization, but nevertheless it is more difficult to come up with a systematic exploration of BSM models.  In the context of naturalness this was attempted to be roughly classified in~\cite{EuropeanStrategyforParticlePhysicsPreparatoryGroup:2019qin}.  To do this one defines a tuning or naturalness parameter $\epsilon$ quantifying the contributions to the Higgs mass parameter from particles at higher scales $\Delta \mh^2$ by
\begin{equation}
    \epsilon = \frac{\mh^2}{\Delta \mh^2},
\end{equation}
where if $\epsilon \ll 1$ the theory is more tuned and $\Delta\mh^2$ is the contribution to the Higgs mass from new physics.  Then the question remains as to what is the size of $\Delta \mh^2$ and how does it correlate to Higgs precision $\delta \eta_{SM}$ and direct searches? This is of course a difficult question to answer systematically, but one can argue~\cite{EuropeanStrategyforParticlePhysicsPreparatoryGroup:2019qin} that
\begin{equation}
    \frac{\delta \eta_{SM}}{\eta_{SM}} \sim c \epsilon
\end{equation}
where $c\sim \mathcal{O}(1)$ in many models. One can then of course directly read off how natural the SM is based on how well Higgs couplings can be measured, i.e. a 1\% tuning corresponds roughly to a 1\% deviation in Higgs couplings.  However, this doesn't allow for a correlation to the direct searches until one defines the form of $\Delta \mh^2$.  In~\cite{EuropeanStrategyforParticlePhysicsPreparatoryGroup:2019qin} different classes of natural models were characterized based on the value of $\Delta \mh^2$ as Soft, SuperSoft, HyperSoft and then correlated to the direct searches.  For example in models like supersymmetry that fall in the Soft category for a large range of parameter space, direct searches for natural supersymmetry at the LHC already go beyond or at least are compatible with the full parameter space explored at $\ee$ Higgs factories.  If a high energy collider such as a muon collider or FCC-hh is built, then the parameter space for all types of natural models with this scaling can be explored to the 1\% level or much better with direct searches, so the complementary nature of precision Higgs physics and energy frontier probes is clear.  While these parametric arguments are useful, full models of naturalness often have many moving parts and particles and therefore a systematic exploration is much more difficult.  For example in models of supersymmetry, new Higgs bosons, electro-weakinos, and scalar stop particles all can alter the Higgs couplings,  while having a spectrum that is far from degenerate.

Another fundamental question often used to motivate Higgs physics studies is the thermal history of the universe.  While the SM thermal history is known theoretically, it relies heavily upon knowing that the Higgs potential is exactly as predicted by the SM {\em and} there are no particles near the electroweak scale that couple to the Higgs that would modify the thermal potential.  Although current and proposed colliders are not directly testing the thermal potential, since all measurements are done at zero temperature and only derivatives of the potential locally are accessible, measurements of the triple Higgs coupling $\lambda_3$ and the quartic Higgs coupling $\lambda_4$ are often taken as good proxies for this question. 

\begin{figure}
\begin{centering}
\includegraphics[scale=0.25]{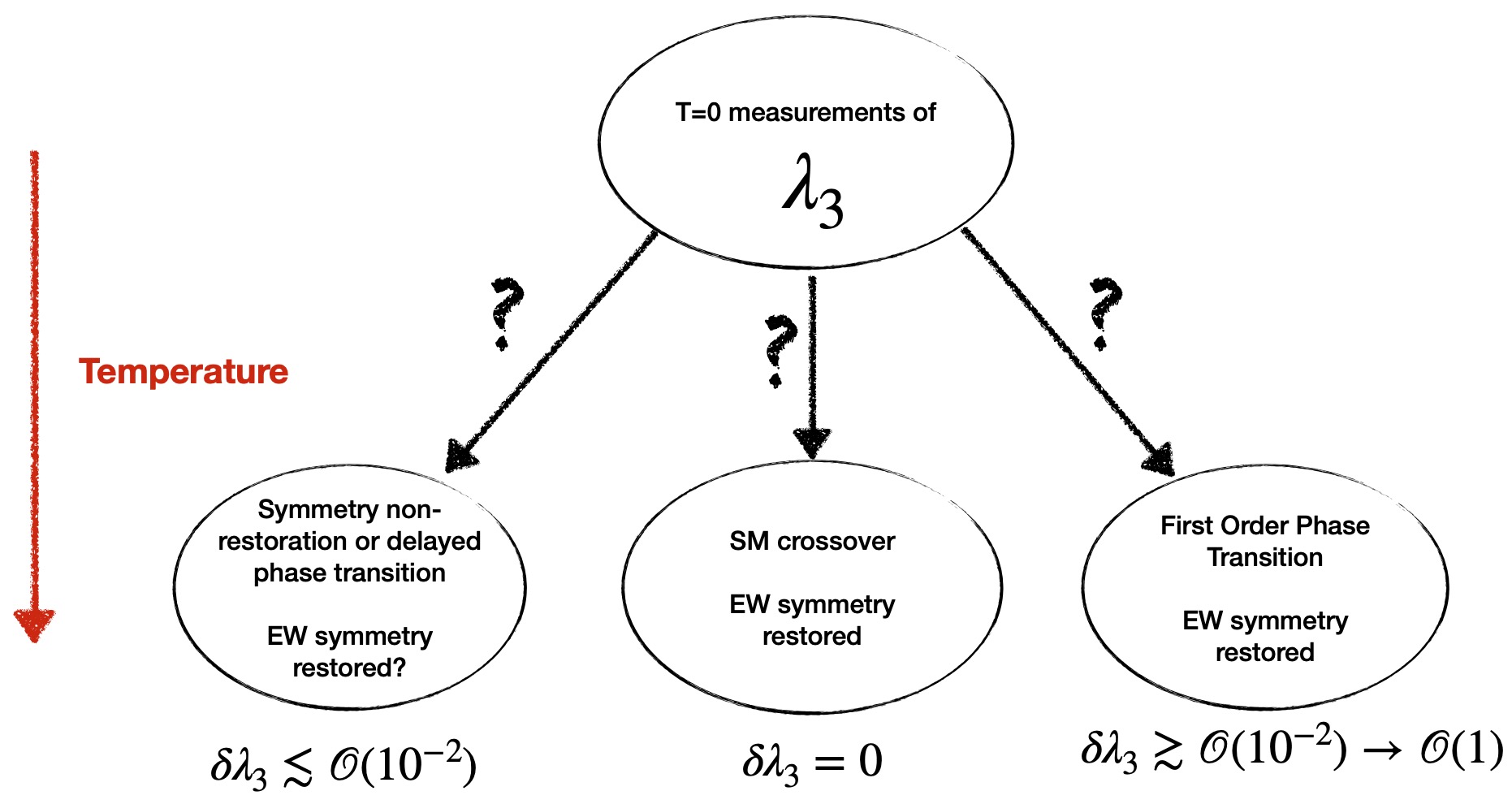} 
\par\end{centering}
\caption{The Higgs boson as the keystone of the Standard Model is connected to numerous fundamental questions that can be investigated by studying it in detail.}
\label{fig:triplehiggsphase}
\end{figure}

Historically, one of the reasons so much emphasis has been put on the precision of the triple Higgs coupling measurement, is a folk theorem that the strength of the Electroweak Phase transition is proportional to the size of the deviation seen in the triple Higgs coupling.  Recent theoretical advances since the last Snowmass have shown counter-examples to this, and moreover the phase diagram of the EW symmetry is now understood to be potentially even more complicated~\cite{Meade:2018saz,Baldes:2018nel,Glioti:2018roy}. This of course does not render the measurement of the triple and quartic Higgs self interactions any less interesting, it is just no longer a benchmark with a binary answer about the EW phase transition.  Given even the best triple Higgs coupling precision projected, currently by a high energy collider, if no deviation is found, there {\em still} could be a first order EW phase transition and possibly EW baryogenesis.  However {\em seeing} a deviation with a precision of down to $\mathcal{O}(1)\%$ would likely cover the most difficult cases where a first order phase transition occurs at the EW scale~\cite{Curtin:2014jma}.  Nevertheless, even if shifts in the Higgs self interactions from their SM values are uncorrelated to the EW phase transition, a measurement of a deviation is still  a profoundly deep answer that can shed light on the origin of electroweak symmetry breaking, the stability of our universe and simply the shape of the Higgs potential experimentally.  Additionally, the triple Higgs coupling is not the only potential observable correlated with a strong EW phase transition.  If new physics that enhances the phase transition is sufficiently light, then there are signatures from exotic Higgs boson decays as discussed in~\cite{Carena:2022yvx}.

\begin{figure}
\begin{centering}
\includegraphics[scale=0.5]{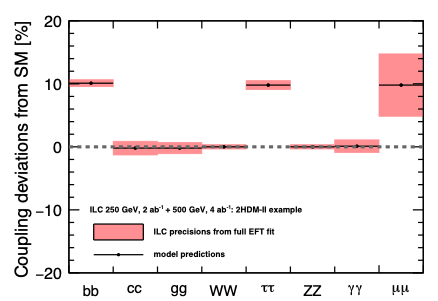}\includegraphics[scale=0.5]{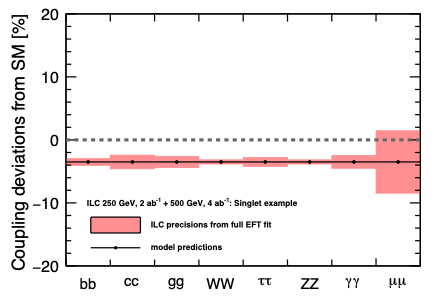}  
\par\end{centering}
\caption{An example from~\cite{ILCInternationalDevelopmentTeam:2022izu} demonstrating different patterns of Higgs deviations from different classes of models, in this case a 2HDM example with a mass scale, $M_A=600$~GeV, and a scalar singlet model with a heavy scalar of mass $2.8$ TeV and the largest mixing angle currently allowed by LHC Higgs couplings measurements.}
\label{fig:ilcinverse}
\end{figure}

As seen with our discussion of naturalness and the EW phase transition, it is difficult to completely organize the parameter space covered by a specific fundamental physics question in terms of Higgs properties and direct searches.  Therefore it is of course even more difficult to disentangle what is the driving BSM question as illustrated in Figure~\ref{fig:higgscentral2}.  This is generically lumped into the aforementioned ``Higgs Inverse" problem of how to map from observables to new physics.  Of course inherent in any discussion of the inverse problem is that there are signals of new physics to disentangle, which of course makes it a good problem to have.  The reason that this problem is potentially tractable is that it isn't just a parametric one based on the overall size of deviations.  Models of new physics tend to induce patterns of deviations.  For example as shown in Figure~\ref{fig:ilcinverse}, adapted from the ILC whitepaper~\cite{ILCInternationalDevelopmentTeam:2022izu}, the pattern of deviations associated with a particular parameter point in a 2HDM model is quite different from that of a SM singlet model. 

The stark difference between models shown in Figure~\ref{fig:ilcinverse}, stems from the fact that a SM singlet inherently affects Higgs couplings universally since it carries no distinguishing quantum numbers, while a 2HDM does.  Therefore one can potentially distinguish certain classes of models with precision measurements at Higgs factories.  However, it should also be noted that the particular points shown in Figure~\ref{fig:ilcinverse} correspond to a 2HDM with a 600 GeV mass scale and a singlet with a 2.8 TeV scalar.  Both of these are clearly out of the direct search reach of circular $\ee$ Higgs factories despite having the precision to test them via Higgs couplings. However, only a 10 TeV muon collider or FCC-hh among the proposed future machines would be able to both reach this level of Higgs precision {\em and} directly discover the new physics states  of the benchmark collider scenarios considered, as even a 3 TeV CLIC would be insufficient~\cite{Brunner:2022usy}.  While this represents just one small corner of the Higgs Inverse problem, it does illustrate the complementary nature of Higgs precision measurements and high energy collider measurements.  In the EF04 topical report where EFT fits are considered in detail, there is additional discussion about the general inverse problem of relating patterns of EFT coefficients to new physics.
In  specific models, it is possible to correlate the deviations in the Higgs couplings with a high scale mass occurring in the model, with some examples given in \cite{https://doi.org/10.48550/arxiv.2209.03303}.

Another possible direction to organize BSM models is purely through the type of signatures they manifest. An example of this which is less studied is the production of triple-Higgs or even quad-Higgs final states at HL-LHC and future colliders.  These are often thought of as  processes that are too rare to observe, in the case of probing the SM quartic coupling, or that have no appreciable rates at the LHC regardless of physics case.   However, there are now viable models with enhanced triple-Higgs and quartic production at the HL-LHC and beyond~\cite{Low:2020iua,Egana-Ugrinovic:2021uew,Chen:2022vac}.  The measurement of the quartic coupling should  be considered a standard part of beyond the Standard Model Higgs phenomenology and triple Higgs and quad Higgs measurements should be pursued at future colliders.

Given that the mapping of fundamental physics questions to Higgs direct and indirect observables is difficult to fully organize comprehensively, for the rest of this section we will instead focus on specific types of models.  While this is a more bottoms up approach it does allow for a concrete grouping of ideas.  Moreover within the classes of models considered here, they will contribute to the numerous fundamental physics questions shown in Figure~\ref{fig:higgscentral}.  In each class of model we will therefore give examples of how they relate to the bigger picture issues.  We will organize based on the characteristic size of effects mentioned previously and outlined in Figure~\ref{fig:higgscentral3}.  In Section~\ref{sec:singlets}, we first discuss models of Higgs singlets.  These are of course the simplest addition to the SM Higgs sector, nevertheless they offer a wide variety of phenomenology and connections to bigger picture issues such as the EW phase transition and the Higgs as a portal to hidden sectors.   In Section~\ref{sec:doublets} we discuss the next simplest extension, additional Higgs doublets, specifically the 2 Higgs doublet models (2HDMs).  These offer an even wider range of observables, and importantly they can allow for differences in flavor associated to the SM Higgs boson that will be specifically highlighted in Section~\ref{sec:flavor} and new targets for CP violation. In Section~\ref{sec:bsmloops}, we then discuss models where the dominant effects on Higgs precision can occur at loop level, for example colored or EW charged states of possibly different spins.   Finally, one must also consider the fact that simple discussions of single Higgs precision is incomplete if our Higgs has new branching fractions into BSM physics.  This of course is difficult in the context of ``kappa" fits or especially an EFT fit given that what the decay mode is would have to be specified to understand the full range of physics.  Typically this is taken into account by modifying Higgs precision fit scenarios, e.g. allowing the Higgs width to vary in an EFT, or allowing the Higgs width to vary and specifying particular modes such as Higgs to invisible in a ``kappa" framework.  From the point of view of BSM physics the phenomenology associated with Higgs exotic decays is wide, so we investigate this separately in Section~\ref{sec:exoticdecays}.  This overview of how BSM physics intersects with Higgs physics is of course far from complete, but hopefully  serves as a set of useful examples for illustrating the abilities of various collider options.  There are a variety of submitted white papers to Snowmass related to BSM Higgs physics that we can not do justice to  in detail, so we highly encourage the reader to look at the individual contributions.

%% file: Tex/singlet.tex
An extension of the Higgs sector of the SM with an additional gauge singlet scalar $S$ represents the simplest BSM physics possibility.  However, despite the simplicity this class of models displays a wide range of phenomenology and connections to fundamental physics questions.  For example with a single degree of freedom  from a real scalar, one can connect to the EW phase transition and thereby models of baryogenesis\cite{Curtin:2014jma,Carena:2022yvx,Papaefstathiou:2022oyi}.  The singlet scalar  can represent the most relevant Higgs portal operators
\begin{equation}\label{eqn:higgsportal}
 \mathcal{L}\supset \lambda_{\hsm \hsm S} \phi^2 S+\lambda_{\phi S} \phi^2 S^2,
\end{equation}
in which if  an additional $Z_2$ symmetry is assumed to be preserved, only the second term applies ($\phi$ is the SU(2) doublet in the unbroken phase).  This Higgs portal scenario can then be connected to dark sectors and dark matter, or can be viewed as a proxy for models of neutral naturalness.  The existence of a new scalar then also applies to the question of whether the Higgs is unique and if it modifies the Higgs potential.  This, of course, can then have implications for the stability of our universe.  Almost the only question that the simplest singlet extension doesn't directly apply to is Higgs flavor at the renormalizable level.  Therefore by studying the singlet whose parameter space is limited, one can connect to a myriad of fundamental questions.  More pragmatically since the parameter space for visible effects can be mapped effectively into a coupling of the singlet to the SM and its mass, for the simplest singlet extensions,  it is much easier to understand the reach and complementary measurements that can be made than in more complicated models.  

We start with the example of a single real singlet extension of the SM.  One can impose an additional ${Z}_2$ symmetry on $S$ such that only a subset of terms are allowed.  This is a useful symmetry to think of the singlet as a portal to a dark sector, since then the singlet (and dark sector) can be odd under this ${Z}_2$ while the Higgs is even, but the $\lambda_{\hsm S}$ coupling in Eq~\ref{eqn:higgsportal} can still exist.  Then there is the option depending on the full potential $V(\phi,S)$ of whether the ${Z}_2$ symmetry is spontaneously broken which would induce an effective $\lambda_{\hsm \hsm S}$ as well or if it is preserved.  In the case where the ${Z}_2$ symmetry is not spontaneously broken, the phenomenology is quite different, since the leading effects on Higgs precision only occur through loops of the singlet.  Nevertheless an $\hsm \hsm S$ coupling is induced after the Higgs gets a vacuum expectation value from the $\lambda{\hsm S}$ coupling in Eq.~\ref{eqn:higgsportal}, so you can pair produce the singlet directly even if Higgs precision effects are highly suppressed. If the symmetry is spontaneously broken, then the effects on Higgs physics can be much larger because then the $\lambda_{\hsm \hsm S}$ coupling induces a mixing of the SM Higgs and singlet states.  This mixing effectively dilutes all SM Higgs couplings as was previously shown in Figure~\ref{fig:ilcinverse}.  One of course could also consider the case without a ${Z}_2$ symmetry which would naively have the same parameters as the spontaneously broken ${Z}_2$ case, but then there are subtle differences because of the symmetry such as whether a shift in the SM triple Higgs coupling is generated.  Additional effects can occur depending on if the singlet mass is below $\mh/2$, then in all symmetry cases the SM Higgs can decays into a pair of singlets.  The phenomenology of the decays can be different of course depending on the symmetry due to whether the singlet can then decay into other SM particles.  This will be discussed further in Section~\ref{sec:exoticdecays}, for the rest of this section we will focus on the case where the mass is above $\mh/2$.



\begin{figure}
\begin{centering}
\includegraphics[scale=0.35]{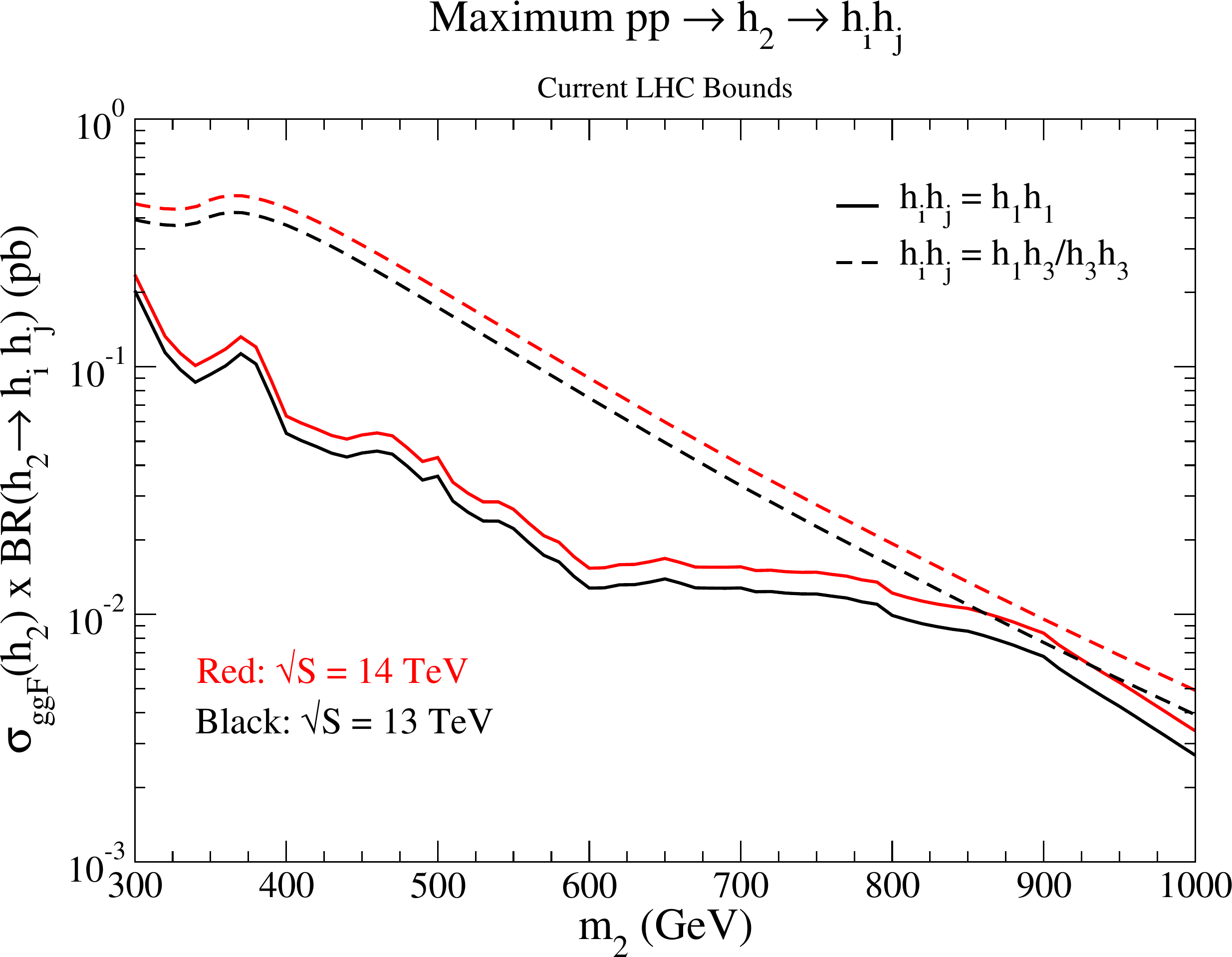} 
\par\end{centering}
\caption{Production of a pair of Higgs bosons in the 
complex singlet model.  $h_1$ is the SM Higgs boson,
and $h_2$, $h_3$ are new gauge singlet scalars  with $m_{2}>2m_{3}$.
The maximum rates allowed by current LHC data are shown\cite{Adhikari:2022yaa}.}
\label{fig:singlet_hh}
\end{figure}

As we have seen, the rich phenomenology of the real singlet scalar is quite extensive.  However, one can add additional scalars, i.e. a singlet complex scalar, or even more.  In these cases the phenomenology can be further complicated and projections onto a two-dimensional plane aren't sufficient.  In particular,  the masses of the various singlets can be different, thus  causing resonance decays to have a much wider range of phenomenology.  Figure \ref{fig:singlet_hh} illustrates this possibility, where we see that a rate larger than that of $\hsm\hsm$ production in the SM is possible\cite{Adhikari:2022yaa}. In such a scenario, there will be new signatures in di-Higgs production where 2 Higgs bosons of different masses can be produced~\cite{Adhikari:2022yaa,Robens:2022lkn,Robens:2022erq}. 

\begin{figure}
\begin{centering}
\includegraphics[scale=0.4]{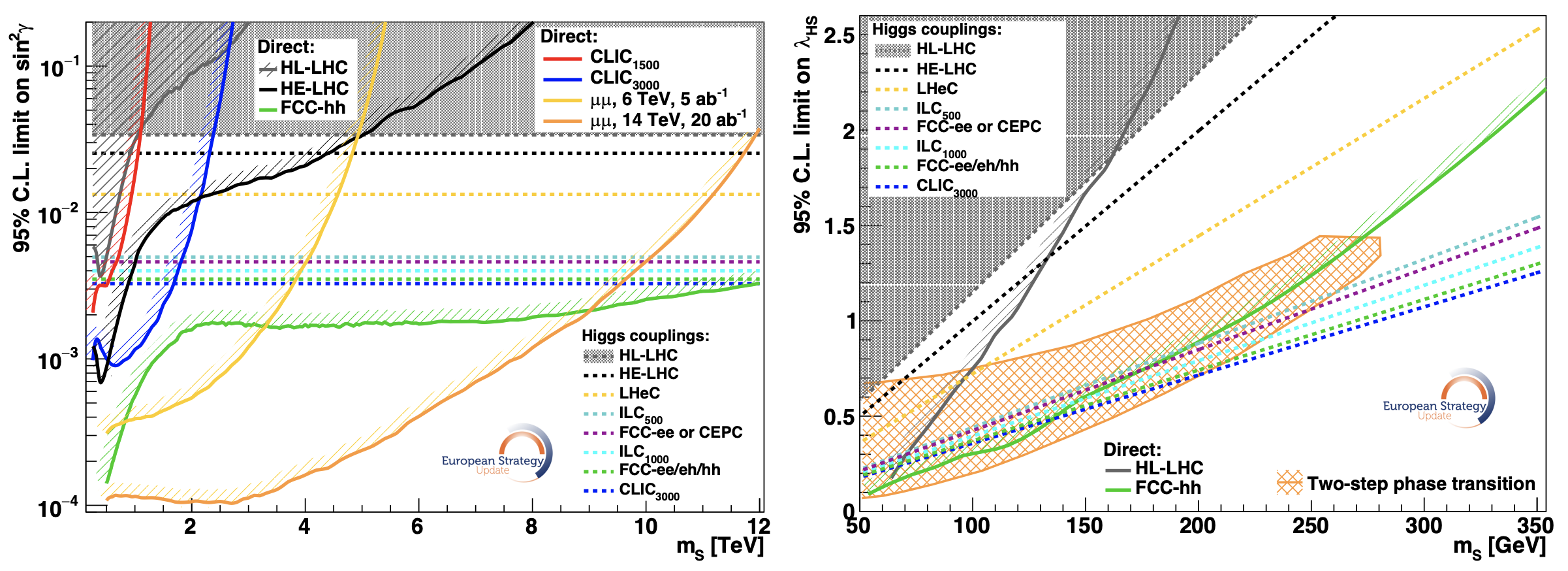} 
\par\end{centering}
\caption{This figure is from~\cite{EuropeanStrategyforParticlePhysicsPreparatoryGroup:2019qin} Figure 8.11, where  the LHS shows the direct and indirect sensitivity to a singlet which mixes with the SM Higgs, while  the RHS shows   the limit of no-mixing,  but  overlaid with regions of parameter space where a strong first-order phase transition is allowed.}
\label{fig:esgsinglet}
\end{figure}

Nevertheless, even though more complicated singlet scalar models can offer new phenomenological signatures to search for at the HL-LHC, the ultimate reach of these models is dictated by directly producing the scalars. An example of how all these observables interplay is shown in Figure~\ref{fig:esgsinglet}~\cite{EuropeanStrategyforParticlePhysicsPreparatoryGroup:2019qin}.  
As shown in Figure~\ref{fig:esgsinglet}, the highest energy probes such as FCC-hh or a muon collider will have the largest possible reach.

%% file: Tex/2hdm.tex
\indent
Two Higgs doublet models (2HDMs) provide the next simplest extension after scalar singlets to the Higgs sector.  They are particularly interesting because they allow for a new Higgs boson 
that can also acquire a vacuum expectation value (VEV) while naturally allowing for small electroweak precision corrections, unlike for instance in generic Higgs triplet models.  
Since the new scalar is an SU(2) doublet, there is a much wider array of phenomenological possibilities than arises  from the model with the SM singlet.  This is because scalar SU(2) doublets can couple to SM particles differently than our Higgs at the renormalizable level unlike a singlet that can only inherit its interactions from the SM Higgs at this level.  2HDMs can share in the wide range of connections to fundamental physics that SM singlets can as well. However, they are particular motivated because in the most compelling solution for naturalness, supersymmetry, they are required by the symmetry structure of the model.  Therefore over time they have received quite a bit of attention and serve as useful benchmarks for the study of Higgs physics for future colliders. 

There are many more states in a 2HDM after going to the mass basis, since there is an entirely new doublet, e.g. the familiar five mass eigenstates: the observed 125 GeV CP-even neutral scalar~$\hsm$, an additional
CP-even neutral scalar H, one CP-odd Higgs boson A, and a pair of charged Higgs bosons
H$^\pm$. Therefore even scanning the phenomenology is quite a bit more complicated than in singlet models, and can often seem daunting.  However, at its core it is important to remember that a 2HDM is just a second copy of {\em our}  SM Higgs.  Therefore, the Lagrangian terms one can write down for the second Higgs with the SM fermions and gauge bosons are identical in structure.  While the gauge symmetry of the SM dictates that the kinetic/gauge interaction terms for ``our" Higgs are identical, differences arise due to the fact that the Higgs potential can be more complicated (as it is a function of both Higgs doublets), and the Yukawa interaction strengths are not fixed by symmetry.  The latter is potentially quite dangerous, as the successful GIM mechanism of the SM could be ruined and new flavor changing neutral current (FCNC) interactions could be introduced in generic 2HDM models.  An idea put forth by Glashow and Weinberg, ``natural flavor conservation" (NFC) was constructed  to avoid FCNCs generically, and is often taken as the organizing symmetry principle of 2HDMs~\cite{PhysRevD.15.1958}.  This imposes a discrete symmetry on the 2HDM which results in the second Higgs doublets Yukawa couplings being proportional to the first.  Imposition of this symmetry results in the standard 4 types of 2HDM models that are often mistaken as the only 2HDM model possibilities(Types I-IV or Types I-II, Type L, and Type F depending on the naming scheme). In fact, this was amusingly pointed out by Georgi as a fallacy by others of confusing sufficient with necessary~\cite{Georgi:1993mps} due to the impressive nature of Glashow and Weinberg who originally wrote down the symmetry condition for NFC.  Nevertheless, given the constraints on flavor one has to address this specifically outside the standard 4 types of NFC 2HDM models, as there is particularly novel phenomenology at future colliders that we will discussion in Section~\ref{sec:flavor}.  Another aspect that we address in Section~\ref{sec:flavor} is the organization of CP violation that can be present in 2HDMs.

Restricting ourselves to the standard types of 2HDM models still allows for an enormous range of phenomenology.   The complications outside the Yukawa sector arise because the potential for the 2HDM, $V(H_1,H_2)$, can allow for both Higgses to acquire VEVs and  quartic terms involving both Higgs doublets in the potential allow for mixing between the 2 Higgs doublets.  Given the ubiquitous nature of NFC 2HDMs, the standard parametrization of the physics is done in terms of a ratio of the VEVs of the 2HDM states, $\tan \beta$, and a mixing angle $\cos (\beta-\alpha)$ as well as the masses of the various eigenstates.  Another way to think of a 2HDM is in the so-called Higgs basis~\cite{Georgi:1978ri,Botella:1994cs}, where one chooses a basis such that the VEV occurs only for the first doublet, $H_1$. The second doublet $H_2$ just has its own set of the usual interactions with the SM, but does not modify the SM Higgs properties at tree level {\em unless} there is a non-trivial mixing, i.e. $\cos (\beta-\alpha)\neq 0$.   In NFC 2HDM models, $\tan \beta$ in the Higgs basis is still useful to parameterize  the effects of the 2HDM in the Yukawa sector and allows for a connection to studies that don't use the Higgs basis.  In the Yukawa sector, which distinguishes the four types of 2HDM, we write separate Yukawas~$\lambda_f^{(1)}$ and $\lambda_f^{(2)}$ as follows, where $1$ refers to the SM Higgs,
\begin{equation}
\label{eq:2HDM-Yukawas}
\lambda_f^{(1)} = \dfrac{\sqrt{2}}{v} m_f,
\qquad
\lambda_f^{(2)} = \dfrac{\eta_f}{\tan\beta} \lambda_f^{(1)},
\end{equation}
and $\eta_f$ dictates the type of 2HDM, given in Table~\ref{tab:2HDMtypes}, and $m_f$ is the mass of fermion type $f$. %
\begin{table}[h!]
\centering
\begin{tabular}
{
@{\hspace{-0.8mm}}
>{\centering}p{1cm}
>{\centering}p{1.8cm}
>{\centering}p{1.8cm}
>{\centering}p{1.8cm}
>{\centering\arraybackslash}p{1.8cm}
@{\hspace{3mm}}
}
& Type-I &  Type-II & Type-L & Type-F \\
\hline
$\eta_{u} $ & 1 & 1 & 1 & 1 \\
$\eta_{d}$ & 1 & $-\tan ^{2} \beta$ & 1 & $-\tan ^{2} \beta$ \\
$\eta_{l}$ & 1 & $-\tan ^{2} \beta$ & $-\tan ^{2} \beta$ & 1 \\
\end{tabular}
\caption{Values of the parameter $\eta_f$ for the different types of 2HDM and for the different types of charged fermions as shown in~\cite{Dawson:2022cmu}.}
\label{tab:2HDMtypes}
\end{table}
\normalsize

\begin{figure}
\begin{centering}
\includegraphics[scale=0.3]{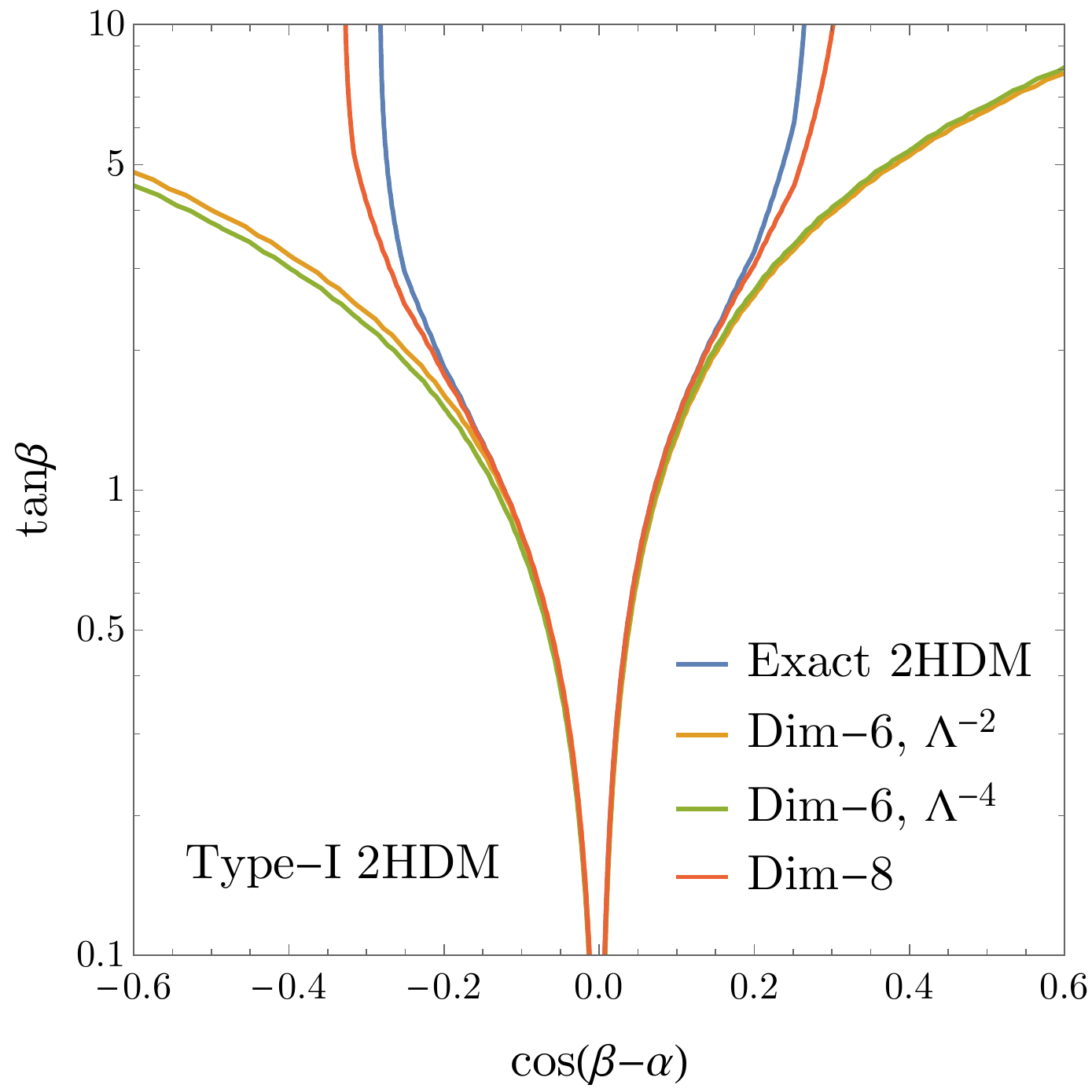} 
\includegraphics[scale=0.3]{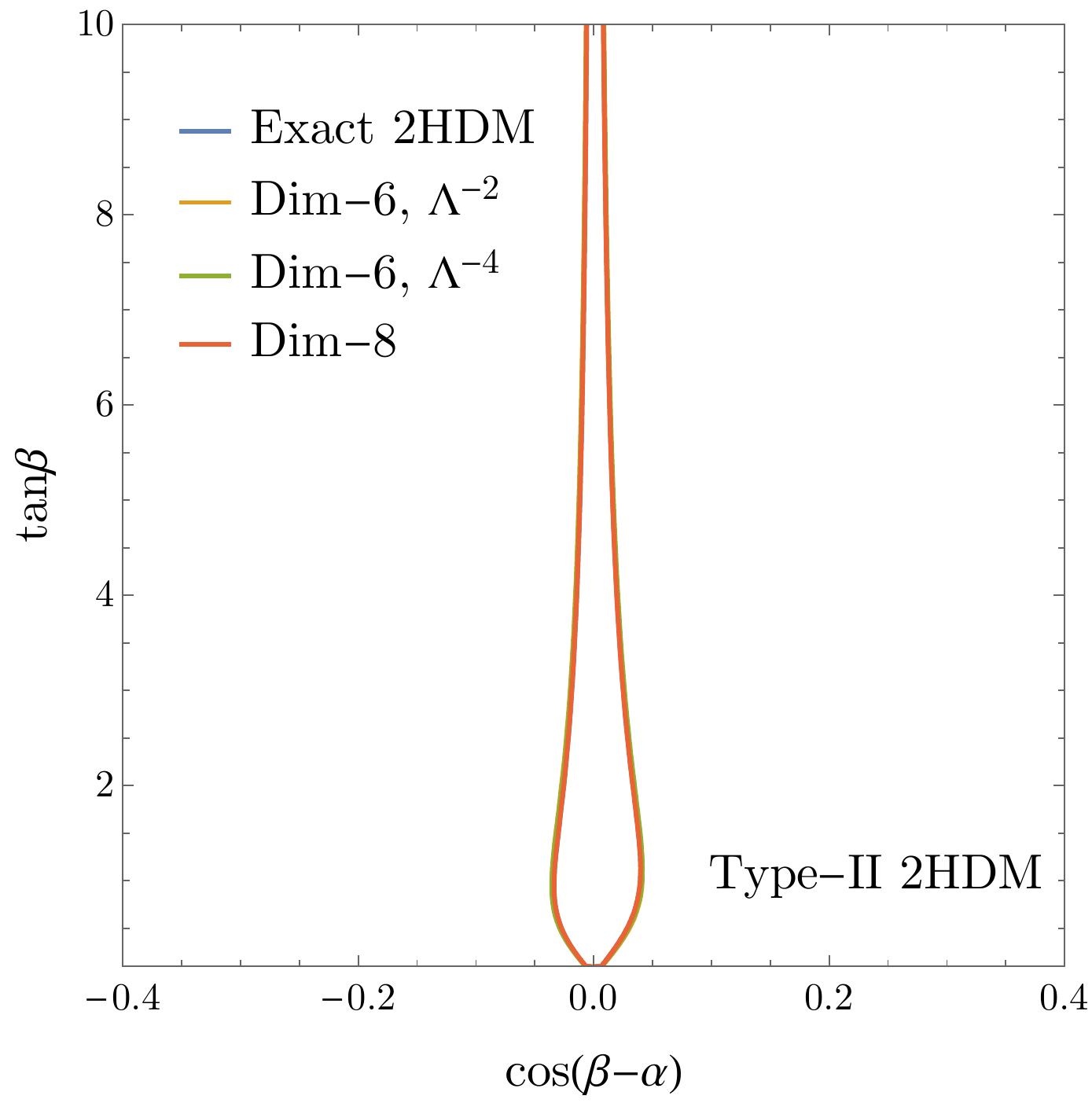} 
\par\end{centering}
\caption{Matching the 2HDM type-I and type-II to the SMEFT at dimension-6 and dimension-8\cite{Dawson:2022cmu}. The allowed regions from Higgs coupling measurements at the LHC are inside the curves centered at $\cos(\beta-\alpha)$. The type-I model is poorly approximated by the dimension-6 SMEFT approximation.}
\label{fig:2hdm_smeft}
\end{figure}

As can be seen from Table~\ref{tab:2HDMtypes}, there is a wide variety of different phenomenology of the second Higgs doublet that does not occur for singlet models.  Moreover it has interesting consequences in terms of matching Higgs precision to an EFT for such a theory.  2HDMs offer two distinct limits where SM Higgs behavior occurs, one is the ``alignment" limit of $\cos(\beta-\alpha)=0$, while one can also decouple the second Higgs doublet by raising its mass terms.  While decoupling implies alignment, it is not a necessary condition for the former.  In the decoupling limit where all new Higgses are heavy, one can obviously map a 2HDM to an EFT.   However, with the inclusion of the couplings that can occur for a generic 2HDM,  one has to be careful about the expansion parameter and how well the lowest dimension operators map to shifts in Higgs precision~\cite{Egana-Ugrinovic:2015vgy,Dawson:2022cmu}.  As shown in Figure~\ref{fig:2hdm_smeft}, the dimension-6 operators are not sufficient to capture the constraints on new physics in the decoupling limit of Type-I 2HDMs, whereas for type-II they are.  This is a good example of how models are crucially important to capture the physics correctly and an EFT inverse problem may lead to incorrect conclusions.

\begin{figure}
\begin{centering}
\includegraphics[scale=0.5]{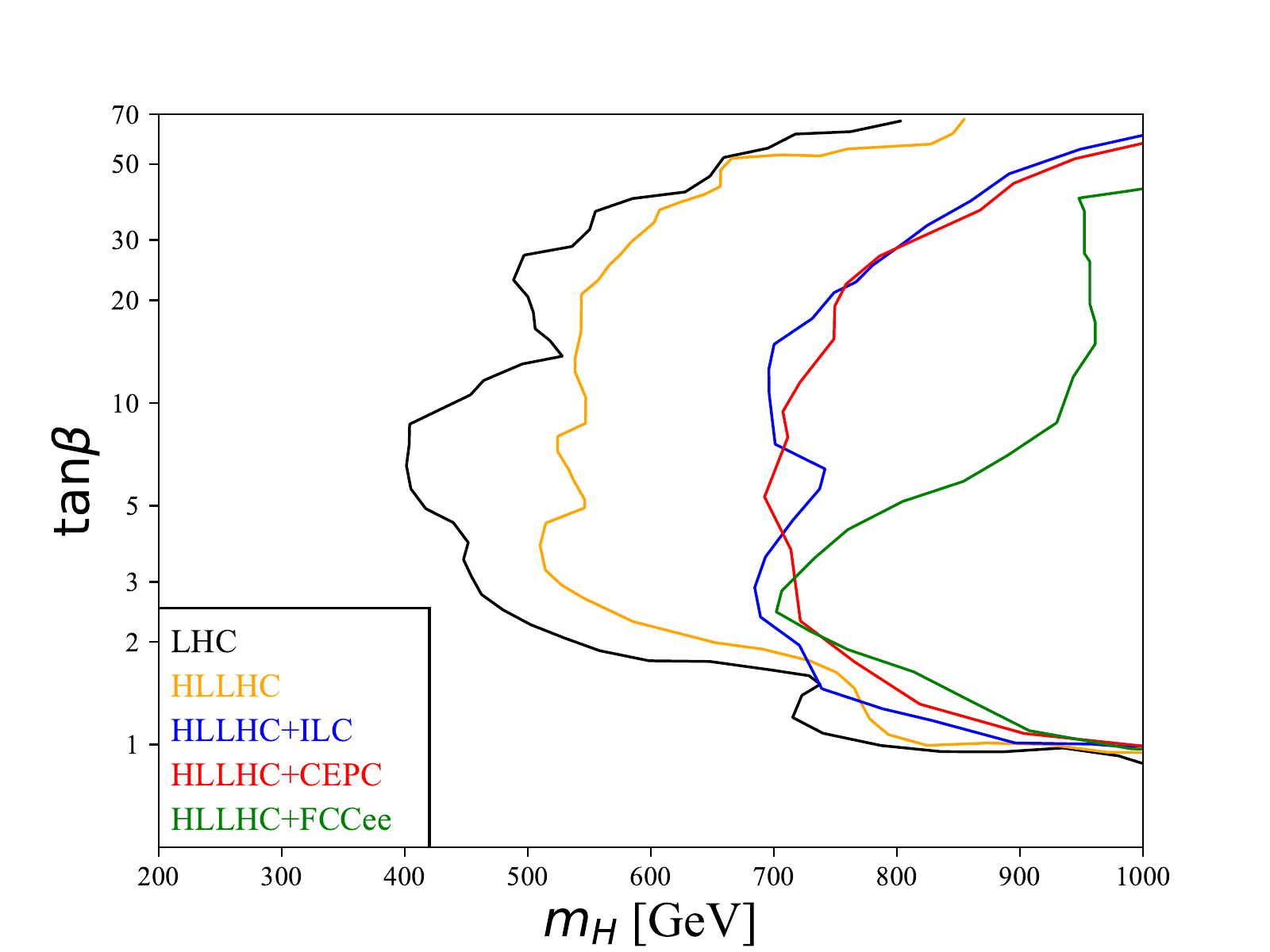} 
\includegraphics[scale=0.3]{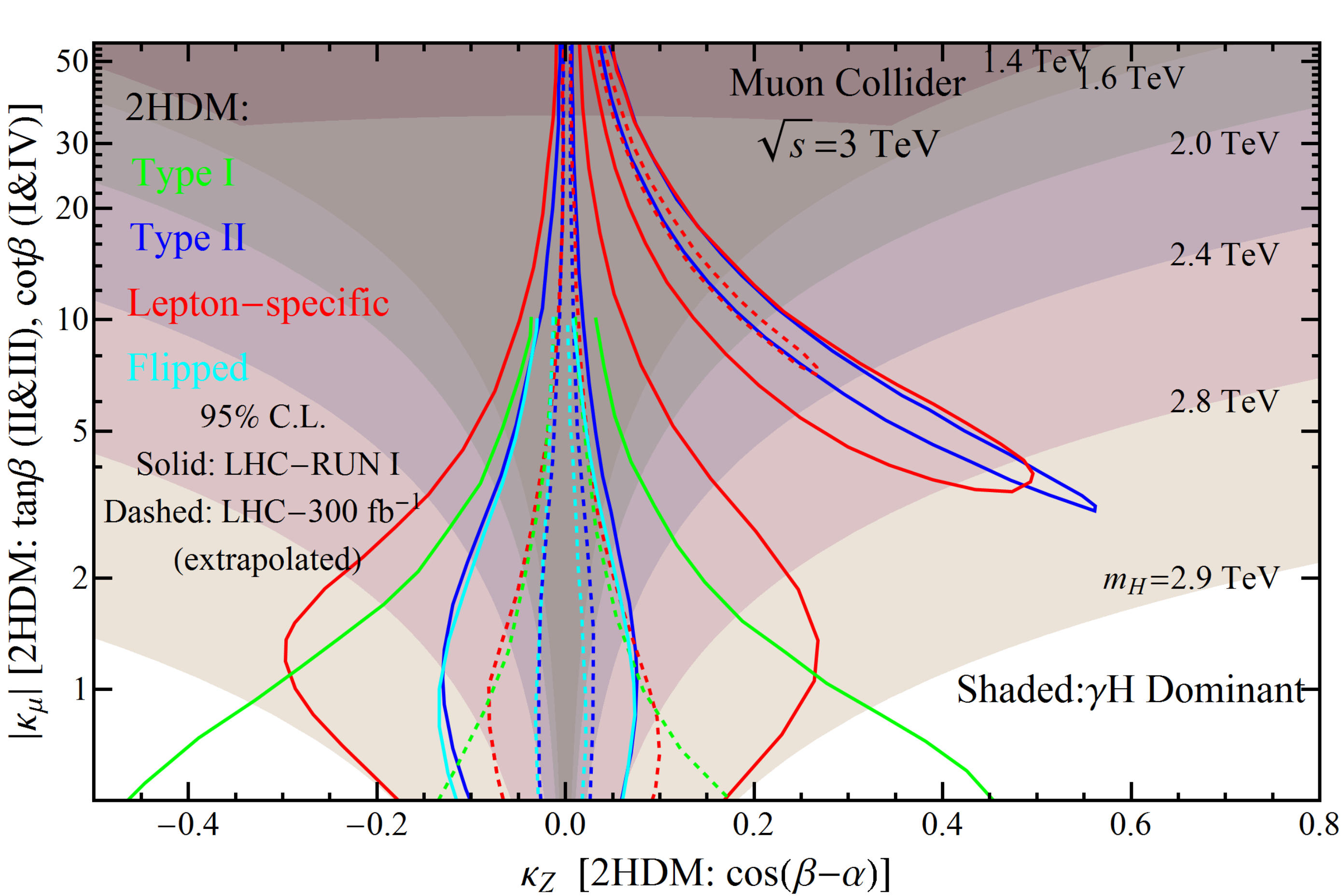} 
\par\end{centering}
\caption{Limits on the parameters of a 2HDM from precision Higgs couplings combined with HL-LHC results. LHS: Limits from future $e^+e^-$ colliders combined with limits from HL-LHC.  The region to the right of the curves is allowed\cite{Bahl:2020kwe}. RHS: Limits from a 3 TeV muon collider.  The mass scales on the right of the figure correspond to the mass of the heavy neutral Higgs boson in the 2HDM\cite{MuonCollider:2022xlm}. }
\label{fig:2hdmfig_com}
\end{figure}

\begin{figure}
\begin{centering}
\includegraphics[scale=0.4]{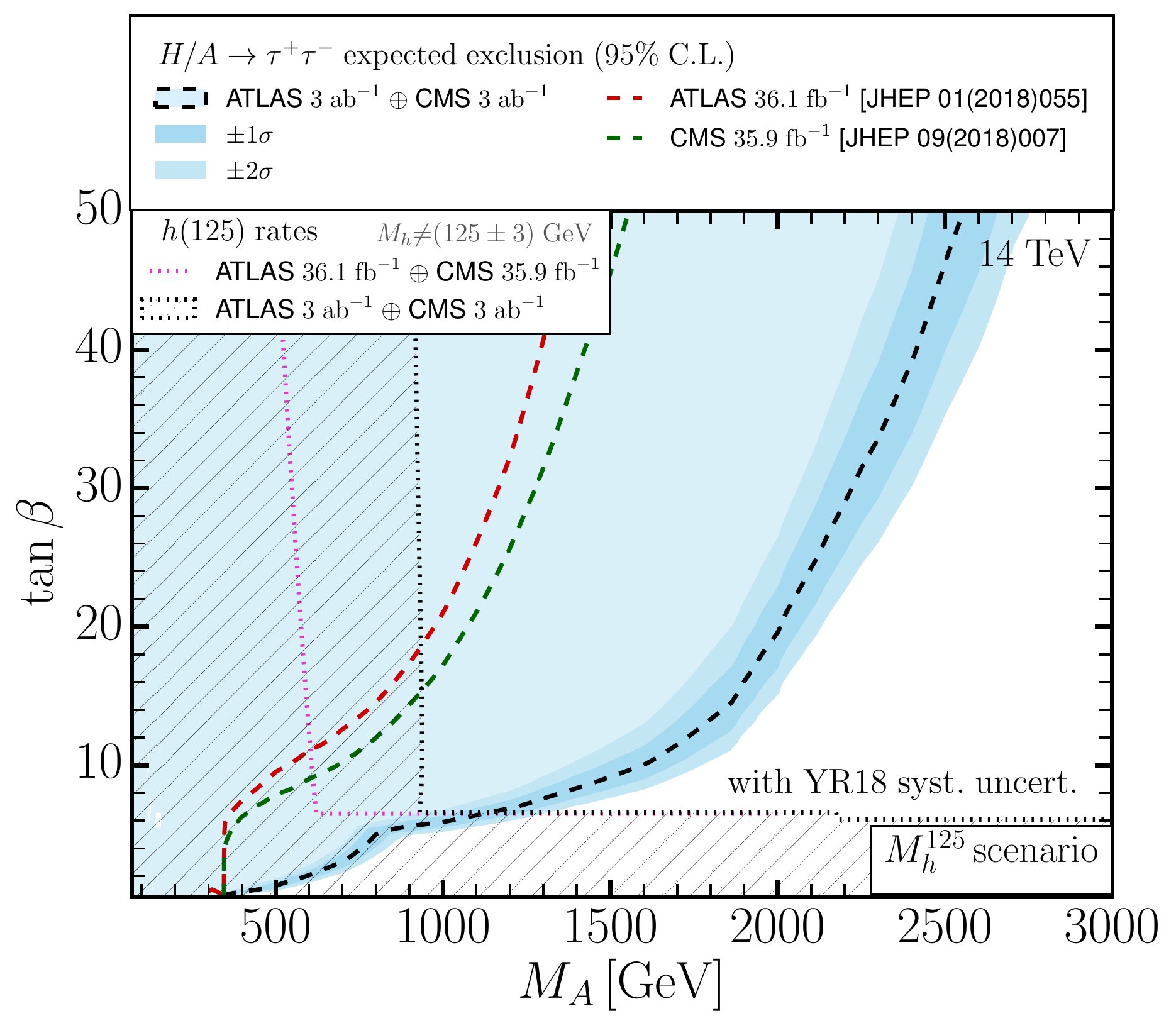}  
\par\end{centering}
\caption{Capability of HL-LHC to probe the scalar sector of the type-II 2HDM.  The cross hatched region corresponds to limits from Higgs coupling measurements\cite{ATL-PHYS-PUB-2022-018}. }
\label{fig:2hdmHL}
\end{figure}

Precision Higgs measurements probe the model parameter space as demonstrated in 
Fig. \ref{fig:2hdmfig_com}\cite{https://doi.org/10.48550/arxiv.2203.07883} and the improvement at lepton colliders for moderate $\tan\beta$ is apparent.  The RHS of Fig. \ref{fig:2hdmfig_com} demonstrates the ability of a high energy muon collider to probe the parameter space of the 2HDM models.  We note that the region of moderate $\tan\beta$ is best probed by $B$ decays. The direct search for the heavier Higgs bosons of the 2HDM is the provenance of the HL-LHC. For high $\tan\beta$, the decay of the heavier Higgs boson to $\tau^+\tau^-$ at HL-LHC will provide a stringent limit, as seen in Fig. \ref{fig:2hdmHL}.  The limits from precision Higgs measurements and direct searches for heavy Higgs bosons probe complementary regions of parameter space.


%% file: Tex/flavor.tex
While 2HDM models have focused heavily on the NFC cases, as explained, the second Higgs doublet could in principle have arbitrary couplings.  Unfortunately this would generically be ruled out by FCNC and other flavor observables.  There have been many attempts to extend 2HDM models with various flavor ansatz that we outline in Table~\ref{t:comparison} including the original NFC models.

\begin{table}[htbp!]
\centering
\begin{tabular}{c | cc}
							& up-type 						& down-type \\ \hline
MFV~\cite{DAmbrosio:2002vsn}				& polynomial of SM Yukawas 				& polynomial of SM Yukawas \\
gFC~\cite{Penuelas:2017ikk,Botella:2018gzy}					& non-universally flavor aligned		& non-universally flavor aligned	 \\
NFC	(types I-IV)~\cite{PhysRevD.15.1958}			& real proportional				& real proportional \\
Aligned 2HDM~\cite{Pich:2009sp,Pich:2010ic}	& complex proportional 	& complex proportional  \\
up-type SFV~\cite{Egana-Ugrinovic:2018znw,Egana-Ugrinovic:2019dqu}			& real proportional				&non-universally flavor aligned	 \\
down-type SFV~\cite{Egana-Ugrinovic:2018znw,Egana-Ugrinovic:2019dqu}		& non-universally flavor aligned	 			& real proportional
\end{tabular}
\caption{Summary of the second doublet Yukawa structure for different 2HDMs which are free from tree-level FCNCs at tree-level in all generations. 
In each column we indicate the relation between the up- and down-type quark Yukawas for the second Higgs doublet and the SM Yukawa matrices. 
Non-universally flavor aligned stands for Yukawas that are flavor-aligned with the SM Yukawas, 
without sharing the SM Yukawa hierarchies. 
Real (complex) proportional stands for proportionality to the corresponding up or down SM Yukawa matrix, with one up- and one down-type real (complex) proportionality coefficient.  This table has been adapted from~\cite{Egana-Ugrinovic:2019dqu}}
\label{t:comparison}
\end{table}

In many cases, the phenomenology of those models shown in Table~\ref{t:comparison} is not particularly different  from the usual type I-IV 2HDM models other than in the 3rd generation. gFC models in principle have arbitrary aligned Yukawa couplings, however general alignment (Aligned Flavor Violation), is not sufficient to avoid FCNCs as discussed in~\cite{Egana-Ugrinovic:2018znw} from the symmetry point of view.  Nevertheless there was a recently introduced concept of Spontaneous Flavor Violation (SFV) which allows for arbitrary diagonal textures in the up-type {\em or} down-type Yukawa couplings of the second Higgs~\cite{Egana-Ugrinovic:2018znw,Egana-Ugrinovic:2019dqu} .  This allows for a wide range of phenomenology that was unexplored previously~\cite{Egana-Ugrinovic:2019dqu,Egana-Ugrinovic:2021uew}.  Furthermore there have been more recent model building efforts in the direction of Flavorful 2HDMs where the SM Higgs is responsible only for third-generation masses, while the second Higgs is responsible for the first two generations ~\cite{Altmannshofer:2016zrn,Altmannshofer:2017uvs}.  These theories provide potential flavor changing targets for future Higgs studies.

\begin{figure}
\begin{centering}
\includegraphics[scale=0.4]{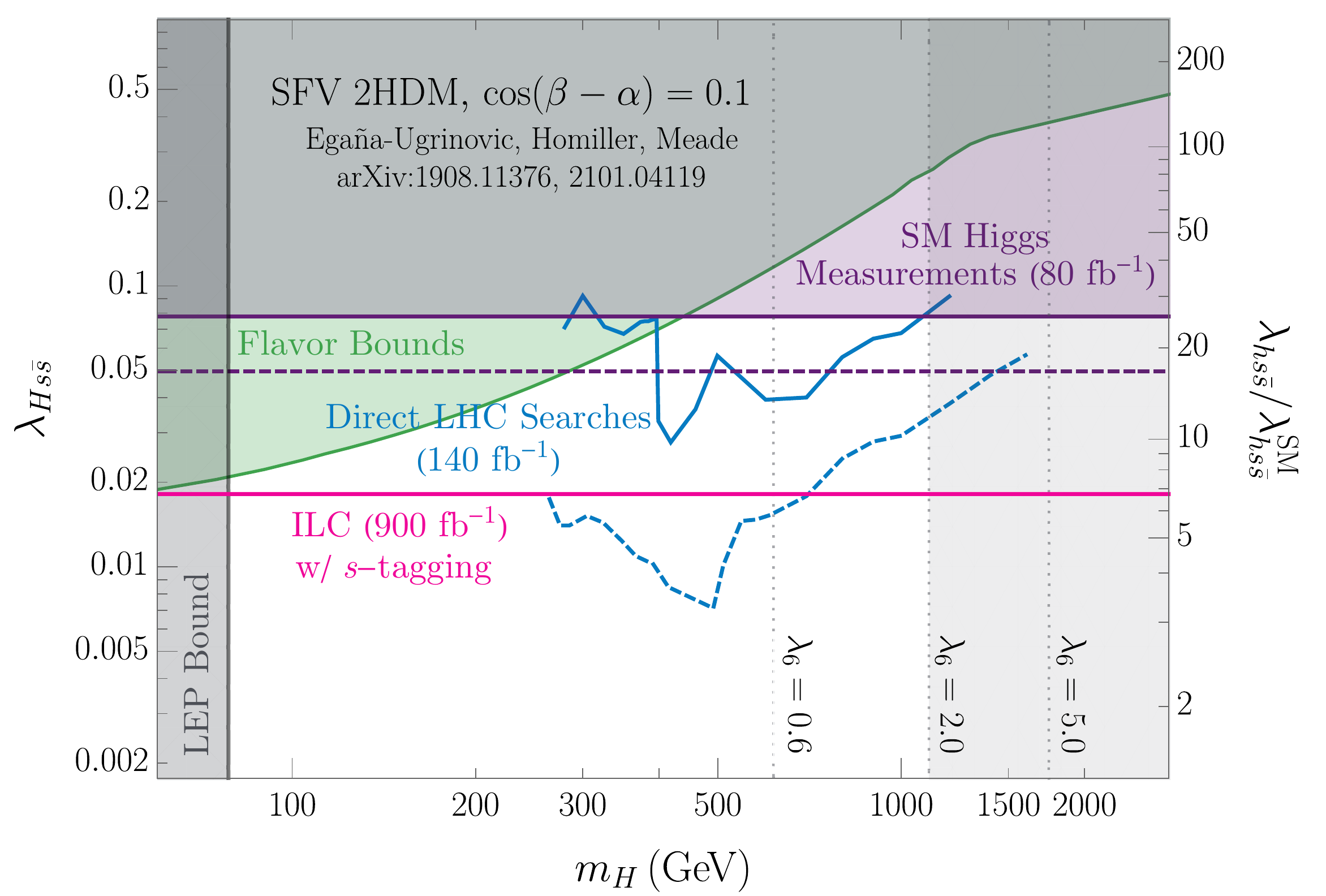} 
\par\end{centering}
\caption{Probes of flavor violation in a 2HDM at future colliders\cite{Albert:2022mpk}.}
\label{fig:flavor_s}
\end{figure}

An example of the interplay of new observables and precision measurements of Higgs couplings is shown in Figure~\ref{fig:flavor_s}. This is for a model of up-type SFV 2HDMs where only the strange coupling is modified.  It can therefore be addressed through direct searches for the extra 2HDM states, flavor bounds, indirect single Higgs bounds, and as shown here direct strange tagging probes at the ILC~\cite{Albert:2022mpk}.

Last but not least, in 2HDM models there can be new sources of CP violation.  While this is a long standing question, the boundaries of where colliders can probe and where EDM measurements constrain are not always appreciated~\cite{Egana-Ugrinovic:2018fpy}.  The organization of what types of CP violation are present is not always well understood either~\cite{Haber:2022gsn}.  In particular, it is important to note that new sources of CP violation that arise in extended Higgs sectors involving neutral scalars fall into two distinct classes: CP-violation from the Yukawa sector and CP-violation from the scalar potential.  While the former is typically discussed in the literature, the latter is more difficult and a discussion of how to find evidence for P-even CP-violation in scalar-mediated processes at future colliders can be found in~\cite{Haber:2022gsn}.   Unfortunately, the energies of ILC and FCC-ee are too low to probe this scenario, but these signals can be searched for at multi-TeV lepton colliders such as CLIC or a future high energy muon collider~\cite{Haber:2022gsn}.

%% file: Tex/bsmloops.tex
In the previous subsections, we have investigated extending the Higgs sector with additional scalar particles that allow for the largest effects on Higgs precision through tree-level mixing.  However, in numerous models, particles in other representations can or must couple to the Higgs boson.  For instance in the context of naturalness, new particles are introduced that are responsible for cancelling the leading quadratic divergences in the SM.  As such these new particles {\em must} couple to the Higgs boson, and their couplings are related through symmetries.  There are multiple types of natural models of this type, which range from supersymmetry, to composite Higgs to models of neutral naturalness.  In the first two cases, there are potentially sizable loop level effects on Higgs precision stemming from the ``top partners" of the SM top which is responsible for the largest contribution to fine tuning when viewed from the Wilsonian point of view. Therefore, one expects in natural models a large contribution to the Higgs-gluon coupling, and one can even test conventional models of naturalness in this way.  In the case of neutral naturalness, at leading loop order the contributions come from SM neutral particles. Therefore the effects are typically smaller than in  the other sorts of models.  However, in all cases the effects on the loops of course decouple with the mass of the partner particles.  For example in the case of stops in supersymmetry, the contribution to gluon fusion scales roughly as $\propto 1/m_{\tilde{t}}^2$~\cite{Ellis:1975ap,Shifman:1979eb} as shown in~\ref{fig:higgscentral3}.  Therefore there is a seesaw effect, where large deviations that can test the most natural models also imply very light states.  This logic  can be extended to models outside of those constructed to solve the hierarchy or little hierarchy problems.  However, the conclusions drawn will be similar.

\begin{figure}
\begin{centering}
\includegraphics[scale=0.5]{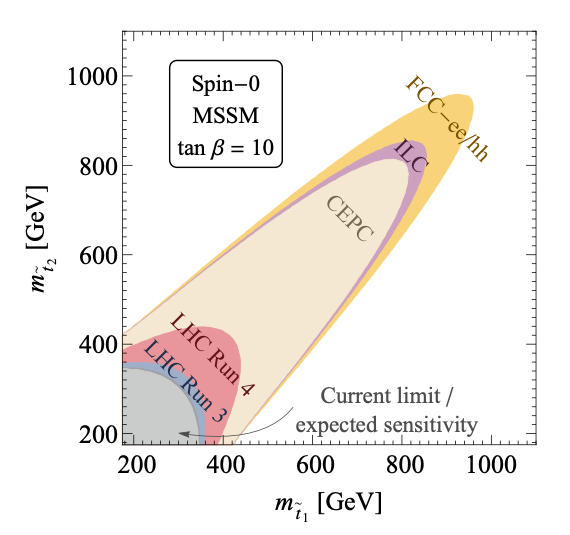}
\par\end{centering}
\caption{The bounds on stop masses in the MSSM for a fixed value of $\tan \beta$, for future $e^+e^-$ colliders and the FCC-hh adapted from~\cite{Essig:2017zwe}.  As can be seen even with the most precise Higgs measurements, the LHC has already probed this space albeit with assuming particular decay modes.}
\label{fig:stopbounds}
\end{figure}

As two useful examples, we consider SUSY models with stops and charginos separately. For stops in the MSSM one can scan over all possible stop masses and mixings to determine the bounds from Higgs precision measurements, for example as done in~\cite{Essig:2017zwe}.  In Figure~\ref{fig:stopbounds}, we demonstrate an example of how stops are constrained from Higgs precision at future colliders. One should note that the estimated size in Figure~\ref{fig:higgscentral3} matches the full models sensitivity.  Additionally, as we see from Figure~\ref{fig:stopbounds}, the LHC has probed this region of stop masses already, so this provides complementary information.  However, the direct search reach from a high energy muon collider or FCC-hh goes significantly beyond what can be obtained via Higgs precision.

%% file: Tex/exotic.tex
As our final example of BSM Higgs scenarios, we discuss exotic Higgs decays, which inherently aren't captured primarily by deviations in SM Higgs precision measurements.  While in principle they do affect SM Higgs precision measurements by changing the total width of the Higgs, the search for exotic Higgs decays can be pushed well beyond this more inclusive test.  Exotic Higgs decays probe new types of BSM physics, in particular hidden sector dynamics arising through a Higgs portal.  In fact, since the largest SM decay mode of the Higgs is set by $y_b$ it isn't particularly difficult to introduce new sizable decay modes of the Higgs~\cite{Curtin:2013fra}.  In fact this program was started to be carried out more systematically soon after the initial discovery of the Higgs boson when large new contributions to the SM Higgs width were allowed.  A bifurcation for exotic decays occurs when considering whether the Higgs boson decays invisibly, or whether there are visible by products in an exotic decay mode.  For example in the context of the Higgs decaying into dark matter one would expect this to give a contribution to the Higgs ``invisible width".  Whereas if there were a moderate Higgs portal coupling to a light new scalar (the low mass limit of Section~\ref{sec:singlets}), the Higgs could decay to $\hsm\rightarrow S S $, and then via the $S-H$ mixing, the $S$ could decay back into SM particles.  These decays could be even more spectacular in certain regions of parameter space where the $S$ could be long lived due to a small coupling and thus displaced decays could occur as part of the phenomenology of exotic Higgs decays.  Lepton colliders can play a key role in the exotic Higgs decay program due to their inherently clean environment.  Moreover at low energy lepton colliders running near the $Z\hsm$ maximum cross section, the energy and momentum are sufficiently well known that one can obtain extra bounds on $\hsm\rightarrow$~anything.  For example using the representative $\hsm\rightarrow S S $ decay, with $S$ a light scalar, we can see that future $e^+e^-$ colliders can significantly improve on HL-LHC limits as shown in Table \ref{tab:higgs_exotic}\cite{Liu:2016zki}.  This is especially the case when the final state has a hadronic component and thus LHC backgrounds are difficult to suppress.  
\begin{table}[ht!]
\centering
\renewcommand{\arraystretch}{1.5}
\begin{tabular}{||c|c|c|c||}
\hline\hline
Channel & HL-LHC& ILC & FCC-ee \\
\hline\hline
$E_T^{miss}$ & 0.056 & $.0025$& $.005$
 \\
\hline
$b {\overline{b}}b {\overline{b}} $ & 0.2 & $9\times 10^{-4}$& $3\times 10^{-4}$\\
\hline
$b {\overline{b}}E_T^{miss}$ & 0.2 & $2\times 10^{-4}$& $5\times 10^{-5}$
\\ \hline
$j j \gamma\gamma $ & $0.01$ & $2\times 10^{-4}$ & $3\times 10^{-5}$\\
\hline\hline
\end{tabular}
\caption{Representative 95$\%$ CL limits on Higgs branching ratios  to a pair of light scalars which then decay to the indicated channels.\cite{Liu:2016zki,ILCInternationalDevelopmentTeam:2022izu}}
\label{tab:higgs_exotic}
\end{table}

However, if a final state exotic decay is dominated by leptons, then the larger number of Higgs produced at a hadron collider can be used to probe lower branching fractions.  Another interesting example is $\hsm\rightarrow \mathrm{invisible}$.  Clearly as seen in Table~\ref{tab:higgs_exotic}, $e^+e^-$ Higgs factories offer an order of magnitude improvement over the HL-LHC.  However, at FCC-hh, the larger production cross sections and range of $p_T$ can be used to constrain the invisible Higgs branching fraction by another order of magnitude down to $\mathcal{O}(10^{-4})$ ~\cite{L.Borgonovi:2642471}.  Regardless of collider, the exotic Higgs program is crucially important for studying the Higgs as a portal to new sectors, since direct or indirect measurements of the Higgs width typically are many orders of magnitude less sensitive at future colliders as discussed in Section~\ref{sec:width}.

\begin{figure}[h]
\begin{centering}
\includegraphics[scale=0.5]{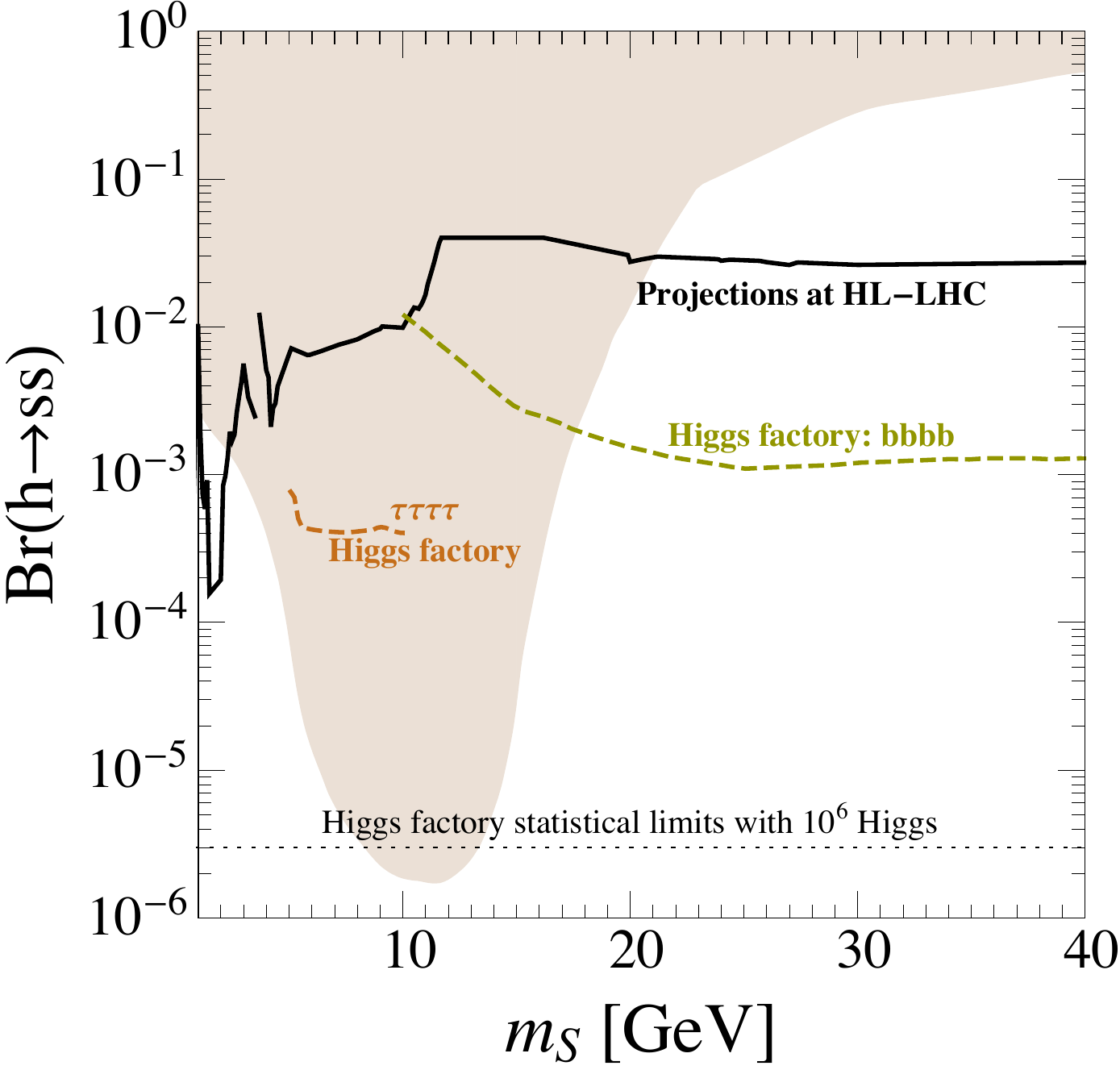} 
\par\end{centering}
\caption{Higgs portal model with $\hsm\rightarrow SS$.  The
shaded region allows for an electroweak phase transition. From Ref \cite{Carena:2022yvx}. See also \cite{https://doi.org/10.48550/arxiv.2203.10184}}
\label{fig:phase}
\end{figure}

Exotic Higgs decays are  not just a test of the Higgs coupling to other sectors.  Given that singlets are a canonical example of the Higgs portal, and that they potentially tie to numerous fundamental questions, exotic Higgs decays can be leveraged in this way as well.  For example the connection to allowed phase transitions is shown in Fig. \ref{fig:phase}, as a function of the light scalar mass and the branching ratio, $BR(\hsm\rightarrow SS)$. Both the HL-LHC and future
Higgs factories can probe the region with an allowed electroweak phase transition\cite{Carena:2022yvx,https://doi.org/10.48550/arxiv.2203.10184,Papaefstathiou:2022oyi}.

%% file: Tex/detector.tex
The Basic Research Needs for High Energy Physics Detector Research \& Development document ~\cite{osti_1659761} compiles a list of requirements on various detector technologies.  The proposed collision energies and data rates of the next generation of energy frontier colliders and the ambitious target precision on various Higgs measurements impose unprecedented requirements on detector technology.

The detectors for the next Higgs Factory must provide excellent precision and efficiency for all basic signatures, i.e. electrons, photons, muon and tau leptons, hadronic jets, and missing energy over an extensive range of momenta. The tracking resolutions should enable high-precision reconstruction of the recoil mass in the $\ee \rightarrow Z\hsm$ process for instance. This translated into a goal for the material profile of less than 20\% X$_0$ for the tracker.  The first sensor layer should be placed within 20 mm of the interaction point to allow very efficient $b$ and $c$ vertex tagging. Future detectors follow the ``particle flow'' methods to fully optimize resolution, where all final-state particles, including neutral particles, are reconstructed by combining corresponding measurements from both the tracking detectors and calorimeters. Particle flow methods benefit from high calorimeter granularity.
These physics goals translate into requirements for transformative and innovative technologies at the next generation of energy frontier experiments focused on precision Higgs and SM physics and searches for BSM phenomena (1) low-mass, highly-granular tracking detectors and (2) highly-granular calorimeters, both with high-precision timing capabilities.

During the Snowmass process, a few key detector technologies R\&D have emerged to enable unprecedented precision on Higgs observables at future $\ee$ colliders. 

\textbf{CMOS Monolithic Active Pixels (MAPs)}~\cite{Hoeferkamp:2022qwg,MAPS} for applications in tracking and electromagnetic sampling calorimetry for future $\ee$ colliders. Larger areas of silicon sensors are needed, several hundred m${^2}$, for the low mass trackers and sampling calorimetry. Trackers require multiple layers, large radii, and micron scale resolution. The CMOS MAPs application presents a promising approach, in which silicon diodes and their readout are combined in the same pixels, and fabricated in a standard CMOS process. CMOS MAPs sensors have several advantages over traditional hybrid technologies with sensors bonded to readout ASICs. These include the sensor and front-end electronics integration, reduced capacitance and resulting noise, lowered signal to noise permitting thinner sensing thickness, very fine readout pitch, and standardized commercial
production. %

\textbf{Low Gain Avalanche Diodes (LGADs)}~\cite{LGAD} offer the possibility of combined fast timing, low mass, and precision spatial resolution. 3D integration is now a standard industry technology, offering dense, heterogeneous, multi-layer integration of sensors and electronics. Advanced process nodes are available as well, lowering power and increasing the density of integration.

\textbf{Dual Readout Calorimetry}~\cite{Pezzotti:2022ndj}, pioneered by the RD52/DREAM/IDEA collaborations,  aims
to provide additional information on the light produced in the sensitive media via, for example, wavelength and polarization, and/or a precision timing measurements, allowing an estimation of the shower-by-shower particle content. This could enable unprecedented
energy resolution for hadronic particles and jets, and new types of particle flow
algorithms.

%% file: Tex/conclusions.tex
The Higgs boson is the central figure in the SM of particle physics.  
Due to its unique nature in the SM, it is connected to some of the most fundamental questions about the universe.  In this report, through a number of concrete examples, we have attempted to give an overview of the interplay of some of these questions and Higgs physics measurements at current and future colliders.  
One of the general themes stressed throughout this report is the connection between precision Higgs physics and the energy scale the measurements  probe. Even though there is no question that the Higgs has been discovered, \textbf{the SM will not be complete until we have enough precision in our knowledge of all its properties to verify that all the  SM predicted couplings exist.  However, that is not the primary goal of the precision Higgs physics program.  If there is a deviation from the SM prediction measured in a Higgs coupling, it must have a {\em cause}, which is what we hope to find and understand.} 

The full run of the HL-LHC will bring increased precision to the measurement of Higgs couplings, the observation of di-Higgs production, and a preliminary measurement of the triple Higgs coupling.  In addition, the search region for additional Higgs bosons will be extended beyond our current knowledge at the HL-LHC.  An $\ee$ Higgs factory is the obvious next step to extend our understanding of the Higgs boson and electroweak symmetry breaking.  
An $\ee$ Higgs factory option can bring unique capabilities, for   instance  a model independent measurement of the Higgs width, and accessibility to invisible and exotic decays of the Higgs, along with increased precision on the Higgs couplings beyond the HL-LHC measurements.   There have been numerous collider options proposed for testing the Higgs sector beyond what can be achieved at the HL-LHC and the improvement in SM quandities is summarized in Table \ref{fig:higgssummaryb}.  The $\ee$ Higgs factory must be pursued with the highest priority by the US particle physics community.\\

Precision Higgs physics at future colliders will  test new physics up to the few TeV scale.  It is important to extend the parameter space beyond what can be covered by the LHC and $\ee$ colliders to the highest energy options, as for instance the FCC-hh or a multi-TeV muon collider. FCC-hh and the muon collider can roughly cover the space of theories responsible for Higgs precision deviations that can be observed at the $\ee$ Higgs factories.  
However, the FCC-hh and muon colliders are less mature concepts and their timelines are currently much longer than those evaluated for the $\ee$ colliders, planning must take this into account as well. It is essential to continue accelerator, magnet and instrumentation research and development {\em now} to preserve accessibility to these technologies when future discoveries are made that can further illuminate the path forward. All in all the Higgs programs outlined provides many compelling options.  It is clear that both a low energy $\ee$ (250 GeV - 1 TeV) and high energy program, such as the muon collider of FCC-hh, would provide a robust coverage of Higgs physics and BSM scenarios. 
